\documentclass[%
 reprint,
superscriptaddress,
 amsmath,amssymb,
pra,
]{revtex4-2}
\usepackage{graphicx}
\usepackage{dcolumn}
\usepackage{cancel}
\usepackage{enumitem}
\usepackage{ragged2e}
\usepackage{blindtext}
\usepackage{subfig}
\usepackage{bm}
\usepackage{xcolor}
\usepackage[normalem]{ulem}
\usepackage{tikz}
\usepackage{soul}
\newcommand{\dd}{%
\begin{tikzpicture}
  \node[inner sep=0pt,outer sep=0pt] (a) {D};
  \draw [line width = 0.2ex](-0.02,0.1) -- (-0.02,-0.1) ;
\end{tikzpicture}
}

\newcommand{\hh}{\hat{h}}

\newcommand{\bob}[1]{\textcolor{black}{#1}}

\begin{document}

\preprint{APS/123-QED}

\title{Quantum Homotopy Algorithm for Solving Nonlinear PDEs and Flow Problems}

\author{Sachin S. Bharadwaj}
\email{sachin.bharadwaj@nyu.edu}
 \affiliation{Department of Mechanical and Aerospace Engineering, New York University, New York 11201 USA}
 \author{Balasubramanya Nadiga}
 \affiliation{Computer, Computational and Statistical Sciences Division, LANL, Los Alamos, New Mexico 87544, USA}
 \author{Stephan Eidenbenz}
 \affiliation{Computer, Computational and Statistical Sciences Division, LANL, Los Alamos, New Mexico 87544, USA}
\author{Katepalli R. Sreenivasan}
\affiliation{Department of Mechanical and Aerospace Engineering, New York University, New York 11201 USA}
\affiliation{Department of Physics and Courant Institute of Mathematical Sciences, New York University, New York, New York 10012, USA}

\date{September 29, 2026}

\begin{abstract}
 \noindent Quantum algorithms to integrate nonlinear PDEs governing flow problems are challenging to discover but critical to enhancing the practical usefulness of quantum computing. We present \textcolor{black}{a} near-optimal, robust, and end-to-end quantum algorithm to solve time-dependent, dissipative, nonlinear PDEs. We embed the PDEs in a truncated, \textcolor{black}{high-dimensional} linear space on the basis of quantum homotopy analysis. The linearized system is discretized and integrated using finite-difference methods \textcolor{black}{with} a compact quantum algorithm. The present approach can adapt its input to the nature of nonlinearity and underlying physics. The complexity estimates improve existing approaches in terms of \textcolor{black}{the time-marching system size, simulation time, accuracy parameters, and post-selection parameters}. We provide a general embedding strategy, bounds on stability criteria, accuracy, gate counts\textcolor{black}{,} and query complexity. A physically motivated measure of nonlinearity is connected to a parameter similar to the flow Reynolds number $Re_{\textrm{H}}$, whose inverse marks the allowed integration window, for given accuracy and complexity. We illustrate the embedding scheme with numerical simulations of \textcolor{black}{Burgers, Fisher--KPP, damped Kuramoto--Sivashinsky, and real Ginzburg--Landau/Allen--Cahn equations}. \textcolor{black}{Together, these examples encompass cubic nonlinearity and saturation, quadratic transport, and fourth-order dissipative stiffness.} This work shows the potential of hybrid quantum algorithms for simulating nonlinear problems on near-term and fault-tolerant quantum devices.

\end{abstract}

\maketitle


\section{Introduction}
\label{sec:Introduction}
\noindent 

{The
  recent acceleration of interest in quantum computing (QC) follows
  demonstrations of its advantages in a small number of specially
  constructed problems. However, for QC to emerge as a versatile tool,
  demonstration of its utility in improving the state-of-the-art in
  challenging scientific and engineering problem settings is
  imperative. In the classical domain, such problems often tend to be
  nonlinear.} For instance, problems in fluid dynamics such as
turbulence, which are central to a wide range of natural phenomena and
engineering applications such as aircraft design, weather prediction
and combustion, are governed by Navier-Stokes equations which are

nonlinear.

Accurate solution of such equations at large enough problem sizes
poses a serious computational challenge to even the biggest
supercomputers. It thus appears useful to explore the new paradigm of
QC and harness the fundamental quantum
advantage. 
In spite of the considerable progress being made in quantum
algorithms, the nonlinearity of governing equations is a critical
bottleneck, since the basic operations of QC are inherently linear and
unitary. In this work we propose a hybrid quantum algorithm that

{works around this
  fundamental linear-nonlinear divide}, while
  preserving an end-to-end quantum advantage, thus 
  adding to the accumulating body of work seeking to
  demonstrate the feasibility of quantum computing in science and
  engineering.

\subsection{\label{subsec:Related Works}Related Work}

Recent years have witnessed various proposals for quantum computation
of fluid dynamics (QCFD) \cite{bharadwaj2025}. They can be classified
into three broad categories 
as listed in \cite{bharadwaj2024compact}. We review them here for completeness as well as to serve as a proper introduction to the present work.

(i) Category A - Quantum Direct
Numerical Simulations (QDNS)
\cite{harrow2009quantum,berry2014high,berry2017quantum,subacsi2019quantum,costa2019quantum,liu2021efficient,childs2021high,costa2022optimal,costa2023further,krovi2023improved,liu2023efficient,an2023linear,fang2023time,bharadwaj2023hybrid,an2023quantum,gonzalez2024quantum,dalzell2024shortcut,lubasch2025quantum,berry2024quantum,brearley2024quantum,leong2022variational,lubasch2020variational,ingelmann2024two,wright2024noisy,pool2024nonlinear},
(ii) Category B -- Lattice Boltzmann and Liouville equation methods
\cite{todorova2020quantum,budinski2021quantum,bakker2023quantum,li2023potential,itani2024quantum,succi2024three,succi2024ensemble,lin2022koopman,schalkers2024efficient,kocherla2023fully,kocherla2024two,penuel2024feasibility}
and (iii) Category C -- Schr\"odinger equation methods
\cite{jin2023quantum,lu2024quantum,meng2023quantum,jin2024quantum}. 
{Even
  though the Lattice Boltzmann and Liouville methods share a common kinetic origin, it appears better to put them in different categories in the context of the distinction between linear and nonlinear systems.}\\

Category A -- This category includes the Quantum Linear Systems Algorithms (QLSA)---on which the present work is based---and Variational Quantum Algorithms (VQA). The VQA approach solves the governing PDE as an optimization problem \cite{leong2022variational,lubasch2020variational,ingelmann2024two,kocher2024numerical,pool2024nonlinear,bravo2023variational}, where a parameterized velocity field is deemed to represent the desired solution when these parameters correspond to the minimum of a suitably defined cost function, after an iterative optimization. This approach has advantages such as relatively shallow circuits \cite{cerezo2021variational,ingelmann2024two,kocher2024numerical}, the possibility of encoding nonlinearity and minimal measurements (single-valued cost function output) \cite{lubasch2020variational,pool2024nonlinear}. However, they have certain limitations such as --- slow convergence \cite{ingelmann2024two}, the problem of barren-plateaus which can worsen with noise \cite{wang2021}, over-parameterization ansatzes \cite{cerezo2021variational,holmes2022connecting,pool2024nonlinear}, as well as the lack of complexity guarantees on quantum advantage \cite{ingelmann2024two,leong2022variational}. 

QLSA methods are best suited for solving linear system of equations \cite{bharadwaj2023hybrid,bharadwaj2024compact,morales2024quantum} and offer complexity guarantees on quantum advantage and are devoid of barren-plateaus and heuristics from optimization. \bob{QLSA can also be used as the central solver for methods proposed in Categories B and C, which would then provide the performance guarantees of the underlying QLSA.  Nonetheless, to solve nonlinear
governing equations with a QLSA-type workflow while preserving an
end-to-end quantum advantage requires innovation on the following two fronts.}

(1) \textit{Linearization}: As the first step, QLSA requires an efficient embedding of the nonlinear PDE, into a linear system of equations spanning a truncated, high-dimensional linear space. Most of the existing quantum proposals rely on techniques such as Carleman and Koopman embedding \cite{joseph2020koopman,liu2021efficient,lin2022koopman,giannakis2022embedding,costa2023further,liu2023efficient,krovi2023improved,gonzalez2024quantum,giannakis2025second,wu2025quantum}. \bob{The first Carleman-based proposals \cite{liu2021efficient,liu2023efficient} showed that it is possible to simulate nonlinear dynamics and compute certain observables efficiently with a quantum algorithm, under dissipativity and nonlinearity assumptions expressed through sufficient conditions such as $R<1$, where $R$ is a measure of nonlinearity, defined via the norms of the matrix operators involved. These restrictive bounds can produce unfavorable dependencies through the Carleman truncation order, solution-norm ratios, post-selection overheads, and accessible time horizon, thereby limiting applications and quantum advantage to weakly nonlinear flows \cite{lin2022koopman}. }\bob{Subsequent works \cite{costa2023further,krovi2023improved} have shown that higher-order methods and rescaling strategies can substantially reduce these costs, including overheads associated with amplitude-decay-dependent factors, while also improving error bounds and removing certain normality constraints.} \textcolor{black}{Yet, Carleman linearization is not an exponentially scaling strategy; but we can, and do, use it as the central benchmark against which to compare the present improvements in terms of embedding, truncation, and physical nonlinearity estimates.}

\bob{A related issue, especially in fluid dynamics problems, is the physical interpretation of the nonlinearity parameter $R$, from the perspective of numerical discretization. A recent work \cite{gonzalez2024efficient} performed a careful analysis revealing that the parameter $R$ may have a spurious dependence on the numerical grid resolution through the discrete derivative norms, even when the underlying physical flow is unchanged.}  \bob{These authors addressed this issue by connecting the required discretization to physically relevant resolution scales such as the Kolmogorov scale and the Reynolds number of a turbulent flow \cite{pope2000turbulent}.  They also delineated physically relevant regimes where Carleman-based quantum algorithms can be useful, and established a connection between $R$ and a physically relevant measure of nonlinearity in fluid flows, which is the well-known Reynolds number, $Re$. } 

\bob{Even with these improvements, a practical use of quantum computers for solving nonlinear problems requires alternative embeddings whose truncation structure admits nonlinear regimes of practically interesting problems. At asymptotic levels of nonlinearity, different linearization approaches could become indistinguishable in performance. Nonetheless, in this work we explore whether, for finite time horizons of evolving the PDEs, there exists a convergent embedding strategy that enables access to higher levels of nonlinearity.} 
In this spirit, parallel efforts have been made using the Koopman generalization of Carleman methods \cite{lin2022koopman,giannakis2022embedding,giannakis2025second,joseph2020koopman}. These functional mappings occur on a continuous Hilbert space, thus making it challenging to find a finite basis set (optimal truncation) that produces an exact mapping of the original nonlinear PDE. A rare exception to this is the Cole-Hopf transformation. In particular, these function spaces have a continuous spectrum that is typically hard to approximate on near-term, gate-based devices, though it might be amenable to photonic devices \cite{lin2022koopman,giannakis2025second}. Without careful topological considerations or data-driven methods \cite{williams2015data,lin2022koopman,giannakis2025second,valva2024physics}, a truncated spectrum would impose constraints on admissible strengths of nonlinearity \cite{lin2022koopman}.\\

In this work we propose an alternative embedding method based on the concept of homotopy \cite{liao2004homotopy}. Within the broad umbrella of homotopy techniques, there exist two primary approaches: (i) Homotopy Analysis Method \cite{liao1992proposed,liao2004homotopy}
 and (ii) Homotopy Perturbation Method \cite{he1999homotopy}. In fact, the latter has been used earlier to build quantum algorithms \cite{xue2021quantum,xue2022quantum}. Despite some merits, it has been shown (even classically) \cite{liao2003beyond,liang2009comparison,liao2012homotopy,gupta2014comparison} that homotopy perturbation methods have certain limitations: weak guarantees on convergence, narrow range of allowed perturbation parameter values, restrictions in terms of accuracy improvements. They are thus typically suitable for weakly nonlinear problems. Several comparative studies \cite{liao2003beyond,liang2009comparison,liao2012homotopy,gupta2014comparison} have shown that the homotopy analysis method is a more general and robust method, which is the choice made in this paper. The details are provided in the remainder of this manuscript. 
 
 The corresponding quantum algorithmic offspring \cite{xue2021quantum,xue2022quantum} also tend to have similar constraints. Some additional restrictions are that the measure of nonlinearity $R$ used in these works is defined similar to ref.~\cite{liu2021efficient}, along with the spurious variation with respect to grid resolution, unrelated to the physical length and time scales; the bound on $K$ is more stringent than Carleman-based methods, given as $K<2/\sqrt{2}$; the matrix operations are restricted to be normal operators;  At the same time, we had independently begun exploring \cite{bharadwaj2023quantum} an alternative homotopy analysis method, leading to the present work which aims to overcome existing bottlenecks. We now briefly review the state-of-the-art QLSA methods to solve the resulting linear system of equations. \\

(2) \textit{Near-term optimality}: Starting from the Harrow-Hassidim-Lloyd (HHL) algorithm \cite{harrow2009quantum}, QLSA approaches have witnessed considerable evolution \cite{berry2014high,berry2017quantum,subacsi2019quantum,liu2021efficient,childs2021high,krovi2023improved,an2023linear,bharadwaj2023hybrid,fang2023time,an2023quantum,berry2024quantum,costa2022optimal,costa2023further,dalzell2024shortcut,liu2023efficient}, by overcoming several challenges and caveats \cite{aaronson2015read, bharadwaj2023hybrid} of the earlier proposals. This has been aided largely by advances made in improving Hamiltonian simulation algorithms \cite{berry2007efficient,berry2014exponential,berry2015simulating,berry2015hamiltonian,low2017optimal,costa2019quantum,berry2020time,an2021time} as well as the introduction of the concept of Linear Combination of Unitaries (LCU) \cite{childs2012hamiltonian,childs2017quantum,bharadwaj2024compact}. However, existing approaches typically face the following challenges: (i) linear, non-optimal dependence on matrix system parameters such as grid size and number of time steps, matrix sparsity and condition number, as well as the specified accuracy; (ii) query complexity, requiring multiple copies of the initial state and repeated measurements and amplitude amplification of intermediate and final solution states; (iii) specific constraints on the properties of matrix operators such as hermiticity, unitarity, positive-definiteness and the range of eigenvalues; (iv) stability requirements for time marching problems, and (v) the depth of quantum circuits due to quantum phase estimation methods and inefficient LCU decomposition. The more recent works, particularly those based on LCU, have addressed some of these aspects, while some \cite{bharadwaj2024compact,bharadwaj2023hybrid} have also (for fluid dynamics) attempted to render the algorithms end-to-end, by addressing state preparation and the post-processing of quantum information.\par

A few additional comments are useful. Here, by ``optimal" we refer to an algorithmic complexity (qubit, query and two-qubit gate complexities) that scales logarithmically or at most poly-logarithmically in all the
system parameters. The term ``near-optimal" refers to a complexity that is \textit{optimal} in most algorithmic parameters except a few. These parameters and the specific dependence on them is derived from the QLSA used here, which is a
compact time-marching quantum ODE solver~\cite{bharadwaj2024compact}. To be more precise, let $N$ denote the
dimension of the spatial linear system to be marched in time, $\tau$ the
number of simulation time steps, $s$ the sparsity of the finite-difference matrices involved,
$\epsilon_{\rm LCU}$ the LCU expansion parameter $\sim O(1)$, $\varepsilon_U$ the error in solution estimation,
and $\kappa$ the condition number of the scaled matrix used in the
Neumann-series representation. Under the specific assumptions of this algorithm, 
the overall complexity scales as
$
~ O\!\left(s,\epsilon_{\rm LCU},\log(N\tau),\operatorname{polylog}(1/\varepsilon, \kappa)\right).$

Thus, in this restricted sense, the complexity is logarithmic or
polylogarithmic in all parameters except that the
sparsity $s$ and $\epsilon_{\rm LCU}$ remain outside the
logarithm. In the compact time-marching
construction, the usual condition-number burden is not claimed to
disappear universally; rather, it is translated into the success
probability of the LCU/postselection step. For the flow-type
time-marching systems of Ref.~\cite{bharadwaj2024compact},
this postselection factor can be made a controlled prefactor by
rescaling the matrix norm and using Richardson extrapolation, so that it
does not necessarily dominate the asymptotic gate complexity. When such
conditioning is not possible, or when one uses the standard
amplitude-amplification route, a linear condition-number-type overhead \cite{costa2022optimal}
may reappear. Thus, the near-optimality claim is a qualified statement
about the compact time-marching solver in the controlled-postselection
regime.

Category B -- The Lattice Boltzmann Method tracks discrete probability distributions of the fluid elements that evolve in time according to a fixed set of velocity directions. Despite some innate advantages of this approach, they are faced with important challenges \cite{todorova2020quantum,budinski2021quantum,lin2022koopman,bakker2023quantum,li2023potential,kocherla2023fully,itani2024quantum,succi2024three,succi2024ensemble,schalkers2024efficient,kocherla2024two,penuel2024feasibility}, such as errors from inaccurate quantum circuit modelling of nonlinear collision-streaming operations \cite{itani2024quantum} and physical boundary conditions \cite{todorova2020quantum}. Further, their computational advantage tend to be downgraded because (a) qubit complexity scales linearly with problem size (grid size and time steps) when lattice positions are encoded as binary state vectors, leading to non-unitarity of streaming or collision steps \cite{kocherla2023fully,schalkers2024importance}, (b) repeated measurements are needed at every time step \cite{kocherla2024two,budinski2021quantum} and (c) multiple copies are required of the initial state \cite{itani2024quantum}. In spite of the progress being made \cite{wang2025quantum}, a single algorithm surpassing all these bottlenecks remains elusive. An alternative strategy would entail solving the Liouville equation  \cite{lin2022koopman,succi2024ensemble} which is a fundamentally linear formulation for general linear and nonlinear flow problems. On the other hand, this is prone to errors from spurious effects that appear with increasing grid resolution, known as Gibbs oscillations. Either way, it appears that the best computational complexity for the algorithms in this category can be attained by translating the governing equations of these methods into a linear system of equations, which can subsequently be solved  by a Quantum Linear Systems Algorithm (QLSA) \cite{li2023potential,penuel2024feasibility,bakker2023quantum} (category A).\\

Category C -- Here, the governing PDEs are mapped into a Schr\"odinger or a Schr\"odinger-like equation. Algorithms as in \cite{meng2023quantum, meng2024simulating} map the Navier-Stokes equations into an analogous hydrodynamic, nonlinear Schr\"odinger equation by using a Madelung transformation, which is then simulated as a quantum circuit. While this approach has a quantum mechanical context, the overall complexity of the algorithm is linear with the problem size and so the asymptotic computational advantage is elusive. An alternative is to Schr\"odingerise the fluid equations \cite{jin2023quantum,lu2024quantum,hu2024quantum,jin2024quantum}. This builds an exact mapping from a classical PDE into a dilated Schr\"odinger equation having an extra dimension with an effective Hamiltonian, which could offer an improvement in the overall computational complexity, especially on analog quantum computers, if not on gate-based devices. However, such mappings are restricted to a specific set of PDEs rather than to general nonlinear PDEs such as the Navier-Stokes equations \cite{lu2024quantum}; further, there is still a non-optimal, linear dependence (gate complexity) on the solution accuracy and time 
\cite{hu2024quantum} for simulating similar problems to those considered here. 
\bob{\subsection{Algorithm Overview and Current Contributions}}
\label{subsec:Current Contribution}

\bob{\noindent\textbf{Computational task}
It bears repeating that the broad goal of this work is to construct an end-to-end quantum algorithm for nonlinear, dissipative, time-dependent PDEs of the form in Eq.~\eqref{eq:general PDE}. Given boundary data, an initial condition, a physical simulation time $T$, and a target accuracy $\varepsilon$, the algorithm prepares a quantum state proportional to the approximate solution at time $T$, or estimates of that solution.}

\medskip
\bob{\noindent\textbf{Input access model assumptions}
After spatial discretization on $N_{g}$ grid points, the governing PDE yields, yields the quadratic nonlinear class of equation represented by Eq.~\eqref{eq:quadratic PDE}. The quantum algorithm to solve this assumes the following input access model:
\begin{enumerate}[wide, labelwidth=!, labelindent=0pt,itemsep=0pt]
\item Access to an efficient oracle that prepares the elements of the sparse finite-difference operators $\dd_1,\dd_2$ and the source or boundary contribution $\dd_0(t)$;
\item Well-defined boundary and initial condition inputs as well as the base solution $\bar u_0$ for the homotopy construction. An efficient preparation of quantum states proportional to these input states. The present approach offers great flexibility in choosing the input quantum states. Further, flow simulations allow smooth, sparse, random, or analytic initial and base flow inputs, that have efficient state preparations.
\item A prescribed total time $T$ to integrate the PDEs and accuracy tolerance $\varepsilon$. The homotopy truncation order $M$ and convergence-control parameter $\hh$ are then chosen to meet the prescribed accuracy and convergence requirements.
\item Access to the sparse lifted matrices generated by the homotopy construction.
\end{enumerate}
}

\medskip
\bob{\noindent\textbf{Output}}
\bob{The algorithm execution is considered successful when the post-selected solution subspace is the normalized-state, proportional to the truncated-approximation (order $M$) of the solution 
\begin{equation}
|\bar u^{(M)}(T)\rangle
=
\frac{\bar u^{(M)}(T)}{\|\bar u^{(M)}(T)\|},
\end{equation} for the target nonlinear PDE, with a total error tolerance \begin{equation}
\varepsilon_{\rm total}(t)
=
\varepsilon_{\bar\gamma}(t)
+
\varepsilon_{\rm fd}(t)
+
\varepsilon_Q(t)
\leq
\varepsilon,
\qquad
0\leq t\leq T,
\end{equation}
where $\varepsilon_{\bar\gamma}$ is the homotopy truncation error, $\varepsilon_{\rm fd}$ is the finite-difference error, and $\varepsilon_Q$ is the quantum-solver error. The output can also be the expectation value of a linear or nonlinear observable \cite{bharadwaj2023hybrid,bharadwaj2024compact,bharadwaj2024qflows} given by
$
\langle \bar u^{(M)}(T)|O_{\rm obs}|\bar u^{(M)}(T)\rangle.
$
 The success probability for post-selecting the solution subspace determines the query complexity, which is bounded as discussed in Section \ref{sec:Complexity Analysis and Error Analysis}. The full classical reconstruction of all grid values would require additional readout overhead and is not the output model used for the claimed advantage.}

\medskip
\bob{\noindent\textbf{Complexity and end-to-end cost.}
Let $s$ be the sparsity of the homotopy-lifted operators, $\epsilon_{\rm LCU}$ the LCU expansion parameter (see Section \ref{sec:TMCQC}), $\varepsilon_Q$ the core QLSA's accuracy, $\tau$ the number of time steps, and $\kappa$ the condition parameter of the scaled system. Given the bounded query complexity (see Section \ref{sec:Complexity Analysis and Error Analysis}), the end-to-end algorithmic complexity can be estimated as
\begin{equation}
\widetilde O\left[
s\epsilon_{\rm LCU}
\log(N_gM^2\tau)
{\rm polylog}\left(\frac{1}{\varepsilon_Q},\kappa\right)
\right].
\end{equation} Together with efficient state preparation and observable estimation subroutines as in Refs.~\cite{bharadwaj2023hybrid,bharadwaj2024compact}, the end-to-end complexity is near-optimal in the restricted sense described earlier. The corresponding qubit complexity is $O\left(\log(N_gM^2\tau)\right),$
up to a constant number of ancillas.
}

\medskip
\bob{\noindent\textbf{Nonlinearity measures and time horizons}
A main analytical outcome is a practically relevant, homotopy Reynolds number $Re_{\rm H}$ given by \begin{equation}
Re_{\rm H}
:=
\frac{\|\bar u_0\|_\infty}{\bar U}
\left.
\frac{L\|\dd_2\|_\infty}{\|\dd_1\|_\infty}
\right|_{\Delta x\sim \ell_{\rm phys}},
\end{equation}
where \(\ell_{\rm phys}\) is the smallest physical scale that must be resolved. This measures the effective strength of the nonlinearity admissible by the algorithm for the said complexity guarantee.
 It is computed by rescaling algorithmic parameters by physically relevant velocity and length scales. In physical time $t$, the finite-time convergence condition can be written as
\begin{equation}
Re_{\rm H}
<
\frac{2L}{|\hh|\bar U t},
\end{equation}
where $\bar U$ is the velocity scale associated with the base solution $\bar u_0$ and $L$ is the characteristic length. Equivalently, the time-integration window is bounded as
\begin{equation}
t<t_{\rm NS},
\qquad
t_{\rm NS}
=
\frac{L}{\bar U}
\frac{2}{|\hh|Re_{\rm H}}.
\end{equation}
Thus $t_{\rm NS}$ is the physical time up to which the homotopy truncation is certified by this bound. To simulate up to a prescribed final time $T$, the sufficient condition is $T<t_{\rm NS}$. Larger values of $Re_{\rm H}$ can therefore be admitted, but only over shorter physical time windows. For the one-dimensional Burgers scaling derived in Sec.~\ref{sec:length scales}, \(Re_{\rm H}=O(Re)\) for normalized profiles, and therefore \(t_{\rm NS}=O(Re^{-1})\).}


\medskip
\bob{
The quantum advantage is therefore embodied by the asymptotically near-optimal scaling of the algorithmic costs. In addition to this, the underlying compact QLSA is an end-to-end, near-term focused, and near-optimal solver for time-dependent linear systems: it treats input preparation, uses compact hardware-aware LCU circuit constructs, includes matrix rescaling and Richardson extrapolation, and supports quantum post-processing of observables~\cite{bharadwaj2023hybrid,bharadwaj2024compact}. The present work lifts this quantum-advantage-preserving solver to nonlinear PDEs and is guaranteed when the inputs are efficiently preparable, the lifted system remains sparse and the output is an encoded state or a small number of observables rather than full classical reconstruction.}

\medskip
\bob{\textbf{Main advances} The first key advance is the embedding strategy itself: the nonlinear PDE is lifted into a homotopy-generated space built around a physically chosen base solution $\bar u_0$, rather than a full tensor hierarchy. This adapts the embedding to the imposed data and the underlying dissipative physics. The retained variables correspond to deformation and closure generated by the homotopy recursion, so the approximation space is tied to the dynamics rather than to algebraic powers of the state. Once the order of truncation $M$ is fixed, the closure variables are introduced by ensuring that the retained set of deformation equations are fully closed, rather than the result of arbitrary discard of newly generated terms. We provide the general $M$-th order deformation equation and a systematic embedding procedure for constructing the lifted linear system at arbitrary $M$.}

\bob{A second key advance is the connection between homotopy convergence and physically meaningful flow scales. Raw matrix-norm estimates can contain spurious grid dependence because derivative norms change with resolution. We remove this ambiguity by first writing the homotopy nonlinearity estimate in terms of the discrete derivative operators and then evaluating the resulting operator ratio at the smallest physically required resolved scale, leading to the homotopy Reynolds parameter \(Re_{\rm H}\) defined in Sec.~\ref{sec:length scales}. Further, a connection to the physical Reynolds number $Re$ of the flow is also derived, paving the way for viewing the algorithm's capabilities from the perspective of practical applications. This also reveals a finite-time nonlinear convergence window (exponential decay in error with respect to truncation order): for one-dimensional Burgers, the homotopy window scales as $Re^{-1}$, while the corresponding physically rescaled Carleman finite-time window scales as $Re^{-5/2}$ (see Appendix \ref{app:carleman-comparison}). Thus, in the tested Burgers regime, the homotopy window deteriorates more slowly with increasing Reynolds number. The simulations also reach a higher effective nonlinear regime, including $\nu=10^{-3}$ and $Re_{\rm H}=O(10^2)$. Furthermore, a more careful comparison with a physically rescaled Carleman baseline shows (see Appendix \ref{app:carleman-comparison}) that, in the tested near-frontier regimes for the Burgers equation problem, the homotopy embedding reaches the target accuracy with a smaller lifted dimension and better captures the interior-gradient structure of the steep transition layer.}

\bob{The numerical evidence for the proposed method is taken beyond Burgers equation itself. As discussed in Appendix \ref{app:additional_pde_benchmarks}, the same homotopy workflow is tested on Fisher--KPP, damped Kuramoto--Sivashinsky, and real Ginzburg--Landau/Allen--Cahn equations, covering logistic reaction, cubic saturation, quadratic transport, and fourth-order stiffness. For these examples, the numerical simulations have demonstrated the algorithm's performance in producing accurate and convergent solutions.}

\bob{Finally, the embedding is incorporated into an end-to-end quantum algorithm that is near-optimal for the specific cases considered here and built on subroutines that are near-term and hardware-aware \cite{bharadwaj2024compact}. In essence, the outcomes of this work form a significant advancement, pushing the frontiers of quantum
simulations of nonlinear PDE problems that bridge science and engineering domains, for the era of near-term
and fault-tolerant quantum computing.}

\section{\label{sec:Problem Formulation}Problem Statement}

The governing PDE of a general nonlinear dynamical system may be written as
\begin{equation}
    \frac{\partial\mathbf{u}(\mathbf{x},t)}{\partial t} = \mathcal{D} (\mathbf{u}(\mathbf{x},t)) + \bm{f}(\mathbf{x},t),\label{eq:general PDE}
\end{equation}
where $\mathbf{u}(\mathbf{x},t)=(u_{1}(\mathbf{x},t),\cdots u_{d}(\mathbf{x},t))$ is a d-dimensional velocity field varying \bob{continuously} in space and time dimensions given by $\mathbf{x} = (x_{1},\cdots x_{d})\in\mathbb{R}^{d}$ and $t$. $\bm{f}(\mathbf{x},t)$ is a source term and $\dd$ is a general differential operator. 
First, we note that, the quantum PDE solver and the embedding scheme discussed in the remainder of this manuscript is applicable to a general nonlinear PDE represented by eq. (\ref{eq:general PDE}). Flow problems in nature and engineering have quadratic nonlinearity, so we use this as our starting point to develop the algorithmic machinery. \bob{After numerical discretization, a governing PDE with a quadratic nonlinearity would be given by an ODE system of the form}
\begin{equation}
    \frac{d \mathbf{u}(t)}{d t}= \bob{\dd_{1}} \mathbf{u}(t)^{\otimes 2}+ \bob{\dd_{2}} \mathbf{u}(t) + \dd_{0}(t).
    \label{eq:quadratic PDE}
\end{equation} \bob{Here $\dd_1$ and $\dd_2$ denote the first- and second-spatial-derivative
finite-difference operators, respectively, so that the subscripts refer to derivative order. The term $\dd_{0}(t)$ denotes possible source terms, including external forcing, pressure/forcing contributions when they are retained, and any inhomogeneous boundary-condition
terms introduced by the discretization.} The explicit form of the operators $\bob{\dd_{1}},\bob{\dd_{2}}$ and $\dd_{0}$ clearly depend on the specific structure of the \bob{ODE} itself and the numerical formulation used to discretize them. For turbulence in an incompressible fluid, the Navier-Stokes equations take the form
\begin{align}
    \frac{\partial \textbf{u}}{\partial t} + \textbf{u}\cdot\nabla \textbf{u} &= \nu \nabla^{2}\textbf{u} + \tilde{\bm{f}}(\mathbf{x}, t) ~~\textrm{and}
    \label{eq:NS1}\\
    \nabla\cdot\textbf{u} &= 0,
    \label{eq:NS2}
\end{align}
where $\tilde{\bm{f}}(\bm{x}, t)=~ - \frac{1}{\rho}\nabla p + \mathbf{g/\rho}$, with $p$ the pressure, $\mathbf{g}$ the forcing term, and $\nu$ is the kinematic viscosity. \textcolor{black}{We demonstrate the algorithm's performance for the one-dimensional Burgers equation, but consider other problems in Appendices A to F, especially F}.

\bob{\noindent \textbf{One-dimensional Burgers initial-boundary-value problem.} We define the problem on $x\in[0,L]$ and $t\in[0,T]$,} and write

\begin{equation}
\bob{\frac{\partial u}{\partial t}+u\frac{\partial u}{\partial x}=\nu\frac{\partial^2u}{\partial x^2}+\tilde f(x,t),}
\label{eq:burgers_problem1}
\end{equation}
\bob{with viscosity $\nu>0$, domain length $L=1$, Dirichlet boundary conditions $u(0,t)=1$ and $u(L,t)=-1$, and initial condition $u(x,0)=\cos(\pi x)$. We set $\tilde f(x,t)=0$ in the numerical example and complexity estimates below, and relegate a discussion of the role of nonzero source terms to Sec.~VI.}

\bob{From the perspective of eq.~(\ref{eq:quadratic PDE}), we have $\bob{\dd_{1}} = \nabla$, $\bob{\dd_{2}}=\nabla^{2}$ and $\dd_{0} =  \tilde{f}(x,t)$. For the general $d$-dimensional case, these operators may be defined as in \cite{gonzalez2024quantum}. After spatial discretization on $N_g$ grid points with spacing $\Delta x=L/N_g$, the solution is represented by a vector $u(t)\in\mathbb{R}^{N_g}$. The first- and second-derivative matrices $\dd_1$ and $\dd_2$ are assumed to be sparse finite-difference matrices, and the initial vector $u_{\rm in}$ is obtained by sampling $\cos(\pi x)$ on the grid. We assume sparse access to the nonzero entries of $\dd_1$ and $\dd_2$, and efficient preparation of a quantum state proportional to the discretized initial condition and to the known base solution $\bar u_0$.}

\bob{Given $T>0$, an accuracy tolerance $\epsilon>0$, a homotopy truncation order $M$, and a convergence-control parameter $\hh$, the goal is to produce a quantum state proportional to the solution $u(x,t)$ at time T, using a homotopy analysis approximation.}

\section{\label{sec:HAM}Homotopy Analysis Method}

Homotopy is a fundamental concept in topology, dealing with continuous deformations of functions or operators, embedded in respective topological spaces. This idea was initially adopted in \cite{liao2012homotopy} to develop a new semi-analytic approach to solve nonlinear PDEs, called the Homotopy Analysis Method \cite{liao1992proposed,liao2004homotopy}. This has now been used widely to study various applications \cite{liao1997boundary,liao2012homotopy} such as viscous boundary layers \cite{liao1999blasius,liao2002blasius2}, magnetohydrodynamics \cite{liao2003mhd}, porous media and non-Newtonian flows \cite{xu2010porous}, all with non-trivial nonlinearities. We now outline some essential definitions and concepts.

\subsection{Homotopy Analysis}

\textbf{Definition 1:} A \textit{homotopy} \cite{liao2012homotopy}  $\xi: \mathbf{\Phi} \times [0,1] \rightarrow \mathbf{\Phi '}$ between two continuous functions $f(\mathbf{\phi})$ and $g(\mathbf{\phi})$ is a continuous function from a topological product space $\mathbf{\Phi}$ bounded to the interval $[0,1]$ to $\mathbf{\Phi '}$, such that $\forall \mathbf{\phi \in \Phi}$, $\xi(\mathbf{\phi},0) = f(\mathbf{\phi})$ and $\xi(\mathbf{\phi},1) = g(\mathbf{\phi})$.

Let us consider the following example to build some intuition. Consider two continuous functions $f(\phi) = \sin(\pi\phi) $ and $g(\phi) = \phi(\phi - 1)$. We can now construct a homotopy product state $\xi$, with a certain parameter $q \in [0,1]$ such that
\begin{equation}
    \xi(\phi,q) := (1-q)f(\phi) + q(g(\phi)).
    \label{eq:homtopy relation 1}
\end{equation}

\begin{figure*}[htpb!]
        \subfloat[]{
        \includegraphics[trim={0cm 0cm -0.5cm 0cm},clip=true,scale=0.62]{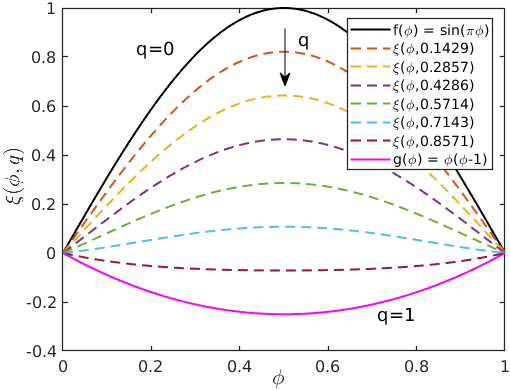}}
        \subfloat[]{\includegraphics[trim={-0.5cm -0.5cm 0cm 0cm},clip=true,scale=0.45]{Figures/time_evol_HAA.jpeg}}
        \caption{\justifying (a) Shows the continuous homotopic deformations of $\xi$ for increasing values of $q$, starting from $f(\phi)$ (solid black line for $q=0$) to $g(\phi)$ (solid magenta line for $q=1$). (b) The left section of this panel, shows the schematic of the time-evolving velocity field of the governing PDE, as computed by a fully classical Direct Numerical Simulations (DNS). The right panel shows the solution as approximated by a Quantum Homotopy Algorithm. The homotopy method decomposes the full velocity solution into (i) $\bar{u}_{0}$ which is the initial guess and solution of the chosen linear PDE known for all $(x,t)$ and (ii) $\sum_{p=1}^{M}\bar{u}_{p}$ which are the higher-order deformation terms that contribute to the nonlinear corrections to the guess solution. The horizontal, red dotted-line indicates a nonlinear time $t_{NS}$, which represents the upper bound on the time horizon up to which the solutions from the quantum homotopy algorithm is maintained within the required accuracy threshold $\varepsilon$ (exponential convergence), while preserving algorithm's complexity. }\label{fig:homtopy example}
\end{figure*}

One can easily see that varying $q$ continuously from 0 to 1 yields a continuous deformation of $\xi$ from $f(\phi)$  (at $q=0$) to $g(\phi)$ (at $q=1$) as shown in figure \ref{fig:homtopy example}.a. Sometimes this is also referred to as the \textit{zeroth order deformation}. This definition can similarly be extended to an operator space, where the solution of (say) a differential operator can be continuously deformed towards the solution of another. \\

\textbf{Definition 2:} The parameter $q \in [0,1]$ that generates a continuous homotopic series of functions is known as the \textit{embedding parameter} while $\xi(\phi,q):= f(\phi) \sim g(\phi)$ is the family of \textit{homotopy functions}.\\

\noindent A typical homotopy-based algorithm to solve nonlinear PDEs works as follows. Consider a nonlinear continuous function $f(\phi)\in C^{\infty}$ with the objective of computing the solution $\phi$, such that $f(\phi)=0$. Let $\phi = \phi_{0}$ be an initial guess of the actual solution. We then construct a homotopy relation given by
\begin{equation}
    \xi(\phi,q) := (1-q)(f(\phi)-f(\phi_{0})) + q(f(\phi)),
\end{equation}
where $q\in [0,1]$ is the embedding parameter. 
We then require the solution $\phi$ to satisfy $\xi(\phi,q) = 0$, such that
\begin{equation}
    (1-q)(f(\phi)-f(\phi_{0})) + q(f(\phi)) = 0.
\end{equation}
Note that, for $q=0$, we have $f(\phi)=f(\phi_{0})$ which implies $\phi = \phi_{0}$ (thus recovering our initial guess). For $q=1$, $\phi$ satisfies the objective equation $f(\phi)=0$. The solution $\phi$ therefore continuously deforms (with $q$) from $\phi_{0}$, towards the solution to our nonlinear equation. This implicitly reveals that $\phi$ itself is a function of $q$ and we therefore replace $\phi$ with $\bar{\phi}(q)$, from here on. Thus when $q=0$,
\begin{equation}
    f(\bar{\phi}(0)) = f(\phi_{0})
    \implies \bar{\phi}(0) = \phi_{0}.
\end{equation}
Similarly, when $q=1$,
\begin{equation}
    f(\bar{\phi}(1)) = 0
    \implies \bar{\phi}(1) = \phi.
\end{equation}

To estimate the solution, we proceed by expanding $\bar{\phi}(q)$ around $q=0$, assuming $\bar{\phi}(q)$ is continuous and analytic around $q=0$. We get
\begin{equation}
    \phi = \phi_{0} + \sum_{p=1}^{\infty}\bar{\phi}_{p}q^{p}, \textrm{~~~~where,}
\end{equation}
\begin{equation}
   \bar{\phi}_{p} = \frac{1}{p!}\frac{d^{p}\bar{\phi}(q)}{dq^{p}}\Big{|}_{q=0}.
    \label{eq:phi_p derivatives}
\end{equation}

Under this assumption, the series would be convergent for $q=1$. Truncating the above series at some $p=M$, for $q=1$, we obtain the homotopy series solution up to a $M$-th order approximation given by
\begin{equation}
    \phi \approx \phi_{0} + \sum_{p=1}^{M}\bar{\phi}_{p},
\end{equation}
also referred to as the \textit{$M$-th order deformation} \cite{liao2012homotopy}. Evaluating this summation requires us to compute each $\phi_{p}$. This is done via eq.~(\ref{eq:phi_p derivatives}), which forms a system of linear recursive equations, each successive term depending on preceding terms. These nonlinear functions can be extended quite straightforwardly to differential operators of a PDE, discretized numerically to obtain a linear \bob{ODE} system of equations, to be solved by an efficient QLSA.

\subsection{Homotopy embedding of nonlinear PDEs}
 
The first step in a broad strategy to extend the above mechanism to nonlinear PDEs is to choose an initial guess of the solution, typically the solution of (say) a well-known linear ODE $\dd_{L}=0$, such that the dynamics of the solution has some similarities to the original PDE, satisfy the boundary and initial conditions, and, in some cases, may also admit an analytical solution. Then we build a homotopy relation that continuously deforms the initial guess towards the solution of the PDE. Concretely, the operator $\dd$ in eq. (\ref{eq:general PDE}), may be decomposed into linear and nonlinear terms as $\dd (\mathbf{u}(\mathbf{x},t)) = \dd_{L} (\mathbf{u}(\mathbf{x},t)) + \dd_{N} (\mathbf{u}(\mathbf{x},t))$. Then one builds a homotopy that continuously deforms the solution from the linear-part of the dynamics (initial guess) towards a solution of the fully nonlinear dynamics (unknown). However, an important point is that, the homotopy relation can be constructed with any linear operator $\dd_{L}$, even when the original equation does not contain an explicit linear part. Now, choosing an initial guess $u_{0}(x,t)$ that satisfies $\dd_{L}(u_{0})=0$, and setting the boundary conditions as $u(0,t)=u_{l}$ and $u(L,t)=u_{r}$, initial conditions as $u(x,0)=u_{\textrm{in}}$, the homotopy relation is written as
\begin{equation}
    (1-q)\big[\dd_{L}\big(u(x,t)-u_{0}(x,t)\big)\big] = \hh q\Bigg[\frac{\partial u}{\partial t} - \dd(u(x,t)) - \tilde{f}(x,t)\Bigg],\label{eq:homotopy for vel}
\end{equation}
where $q\in[0,1]$ and $\hh \neq 0$ is a parameter used to control the stability and radius of convergence of the series solution. As earlier, $u_{0}$ corresponds to $q=0$ and we begin to use the notation $u(x,t)\mapsto \bar{u}(x,t,q)$ to emphasize that $u$ is also indirectly a function of $q$. Following eq.~(\ref{eq:phi_p derivatives}), we can write the $M$-th order series solution as
\begin{align}
    \bar{u} &= \bar{u}_{0} + \sum_{p=1}^{p=M}q^{p}\bar{u}_{p},\\
    \implies \bar{u} &= \bar{u}_{0} + \sum_{p=1}^{p=M}\frac{q^{p}}{p!}\frac{d^{p}\bar{u}}{dq^{p}}\Bigg\vert_{q=0}.\label{eq:deformations}
\end{align}
Additionally we need to specify the initial boundary conditions of $\bob{\bar{u}_{p} =\frac{1}{p!}\left.
\frac{\partial^p\bar u}{\partial q^p}
\right|_{q=0},}$ as follows:
\[
\bar{u}_{p}(x,t) =
\begin{cases}
    u_{l}-u_{0}(x,t) & p=1 ~\&~ x=0 \\
    u_{r}-u_{0}(x,t) & p=1 ~\&~ x=L \\
    0   & p>0 ~\&~ x=\pm L \\
\end{cases}
\] 
\[ \bar{u}_{p}(x,0)  =
\begin{cases}
    u_{\textrm{in}}-u_{0}(x,0) & p=1  \\
    0 & \textrm{else} \\
\end{cases}
\]
Now to compute the sum in eq.\ (\ref{eq:deformations}), we need to compute the deformation equations for each term $\bar{u}_{p}$, by differentiating eq.~(\ref{eq:homotopy for vel}) with respect to $q$, $p$-times. \bob{We first note that,} 
\begin{equation}
   \bob{ \frac{1}{p!}\left. \frac{d^p}{dq^p} \left[(1-q)\dd_L(\bar u-\bar u_0)\right] \right|_{q=0} = \dd_L(\bar u_p)-\chi_p\dd_L(\bar u_{p-1}),}
\end{equation}
\bob{where $\chi_1=0$ and $\chi_p=1$ for $p\geq2$. Using this we obtain the $p-$th order derivative as,}
\bob{
\begin{align}
   &\frac{1}{p!}
\left. \frac{d^p}{dq^p} \left\{ \hh q \left[
\frac{\partial \bar u}{\partial t}
-\dd_L(\bar u)-\dd_N(\bar u)-\tilde f
\right] \right\} \right|_{q=0}
\nonumber\\
&\qquad =
\frac{\hh}{(p-1)!}
\left.
\frac{d^{p-1}}{dq^{p-1}}
\left[
\frac{\partial \bar u}{\partial t}
-\dd_L(\bar u)-\dd_N(\bar u)-\tilde f
\right]
\right|_{q=0}.
\end{align}}
\bob{Since $\tilde f$ is independent of $q$, it contributes only to the
$p=1$ equation. Since $\bar u_0$ is chosen to satisfy the linear part of the dynamics, it can be seen that}
\bob{\begin{equation}
    \frac{\partial \bar u_0}{\partial t}-\dd_L(\bar u_0)-\tilde f=0.
\end{equation}}
\bob{Using the above equations, we obtain the following system of linear ODEs.}

\bob{\begin{align}
    \dd_{L}(\bar{u}_{1})
    &= \bob{-\hh\dd_{N}(\bar{u}_{0})},\\
    \dd_{L}(\bar{u}_{2})
    &= \bob{\dd_{L}(\bar{u}_{1})
    + \hh\Bigg\{\frac{\partial \bar{u}_{1}}{\partial t}
    -\dd_{L}(\bar{u}_{1})-}\nonumber\\&~~~~~~~~~~~~~~~~~~~~~~~\bob{-\frac{d}{dq}\dd_{N}(\bar{u})|_{q=0}\Bigg\}},\\
    \dd_{L}(\bar{u}_{3})&= \bob{\dd_{L}(\bar{u}_{2})
    + \hh\Bigg\{\frac{\partial \bar{u}_{2}}{\partial t}
    -\dd_{L}(\bar{u}_{2})-}\nonumber\\&~~~~~~~~~~~~~~~~~~~~~~~\bob{-\frac{1}{2!}  \frac{d^{2}}{dq^{2}}\dd_{N}(\bar{u})|_{q=0}
    \Bigg\}},\\
    &\vdots \nonumber\\
    \dd_{L}(\bar{u}_{p})&= \bob{\dd_{L}(\bar{u}_{p-1})
    + \hh\Bigg\{\frac{\partial \bar{u}_{p-1}}{\partial t}-\dd_{L}(\bar{u}_{p-1})-}\nonumber\\&~~~~~~~~~~~~~~~~~\bob{-\frac{1}{(p-1)!}
    \frac{d^{p-1}}{dq^{p-1}}\dd_{N}(\bar{u})
    |_{q=0}\Bigg\}},~~ p\geq2 .
\end{align}}

Note that, a given $p$-th order deformation equation is a function of $(\bar{u}_{0},\bar{u}_{1},\cdots,\bar{u}_{p})$, where all the $p-1$ terms $(\bar{u}_{0},\bar{u}_{1},\cdots,\bar{u}_{p-1})$ are known, by successively solving the preceding $p-1$ ODEs. These can be solved numerically by using (say) method of finite differences, which yields a linear system of equations amenable to matrix inversion algorithms. However from a quantum algorithmic perspective, computing each successive $\bar{u}_{p}$ serially entails repeated re-initialization and measurement of the quantum state, which compromises the overall quantum advantage. We therefore propose a general quantum embedding strategy that computes the summation in eq.~(\ref{eq:deformations}) directly by solving for all the terms simultaneously. We discuss this Quantum Homotopy Analysis Algorithm in the following section.
 \vspace{0.5cm}
 
\section{\label{sec:QHAA}Quantum Homotopy Analysis Algorithm}
 
We consider the unsteady Burgers equation described in Section \ref{sec:Problem Formulation} to outline the steps involved.
\vspace{0.5cm}

\subsection{Unsteady Burgers equation}
First we set $\dd_{L}(u) = \partial_{t} u$ and $\dd(u) = \nu \partial^{2}_{x}u-u\partial_{x}u$. The initial guess is the solution to the well-known linear diffusion equation $\partial_{t} \bar{u}_{0} = \bob{\nu}\partial^{2}_{x}\bar{u}_{0}$, solved under the same boundary and initial conditions. The general solution to this is given as (see proof in Appendix \ref{app:analytical solution diffusion})
\begin{align}
    \bar{u}_{0}(x,t) = \Bigg(1-\frac{2x}{L}\Bigg) &+  \sum_{n=1}^{\infty} \Bigg\{\frac{2}{\pi n(4n^{2}-1)} \times \nonumber \\ &\times\exp\Bigg(-\frac{4\pi^{2}n^{2}\nu}{L^{2}}t\Bigg) \sin\Big(\frac{2n\pi x}{L}\Big)\Bigg\}. \label{eq: diffusion solution u0}
\end{align} At $t=0$ and $L=1$, this gives $\bar{u}_{0}=\cos(\pi x)$, is the initial condition. The above choice of operators implies that the time evolution of the estimated solution is a weighted sum of the evolving guess solution $u_{0}$, which is thus deformed towards the nonlinear solution by the evolving deformation terms $\bar{u}_{p}$ as shown in figure \ref{fig:homtopy example}(b).  
For simplicity, we set $\tilde{f}(x,t)=0$ here, but discuss consequences of a finite source term in Section \ref{sec:length scales}. The homotopy is written as

\begin{equation}
    (1-q)\Bigg[\frac{\partial}{\partial t}\big(u(x,t)-u_{0}(x,t)\big)\Bigg] = \hh q\Bigg[\frac{\partial u}{\partial t} + u\frac{\partial u}{\partial x} - \nu \frac{\partial^{2} u}{\partial x^{2}}\Bigg].\label{eq:homotopy for burgers unsteady}
\end{equation}
We now proceed to compute the deformation equations as earlier and obtain

\bob{
\begin{align}
    \frac{\partial \bar{u}_{0}}{\partial t}
    &= \nu \frac{\partial^{2} \bar{u}_{0}}{\partial x^{2}},\\
    \frac{\partial \bar{u}_{1}}{\partial t}
    &= \hh\Bigg[ \bar{u}_{0}\frac{\partial  \bar{u}_{0}}{\partial x}\Bigg],\\
    \frac{\partial \bar{u}_{2}}{\partial t}
    &=
    (1+\hh)\frac{\partial \bar{u}_{1}}{\partial t}
    + \hh\Bigg\{
    \frac{\partial  \bar{u}_{0}\bar{u}_{1}}{\partial x}
    - \nu \frac{\partial^{2} \bar{u}_{1}}{\partial x^{2}}
    \Bigg\},\\
    \frac{\partial \bar{u}_{3}}{\partial t}
    &=
    (1+\hh)\frac{\partial \bar{u}_{2}}{\partial t}
    + \hh\Bigg\{
    \frac{\partial  \bar{u}_{0}\bar{u}_{2}}{\partial x}
    + \frac{1}{2}\frac{\partial  \bar{u}^{2}_{1}}{\partial x}
    - \nu \frac{\partial^{2} \bar{u}_{2}}{\partial x^{2}}
    \Bigg\},\\
    &\vdots \nonumber\\
    \frac{\partial \bar{u}_{M}}{\partial t}
    &=
    (1+\hh)\frac{\partial \bar{u}_{M-1}}{\partial t}
    + \hh\Bigg\{
     \frac{1}{2}\frac{\partial}{\partial x}\sum_{k=0}^{M-1}
    \bar{u}_{k} \bar{u}_{M-1-k}
    -\nonumber\\&~~~~~~~~~~~~~~~~~~~~~~~~~~~~~~~~~~~- \nu \frac{\partial^{2} \bar{u}_{M-1}}{\partial x^{2}}
    \Bigg\}.
    \label{eq:Mth deformation Burgers}
\end{align}}

Equation (\ref{eq:Mth deformation Burgers}) gives the general, $M$-th order deformation equation that allows us to calculate the homotopy solution up to an arbitrary number of terms as required, based on the nonlinearity of the problem. The next step is to eliminate the serial and recursive structure of the above equations. We propose and implement a linear embedding \cite{bharadwaj2024thesis} via dimension dilation that transforms the overall set of equations into a closed, linear system of equations. Unlike ref.~\cite{xue2024quantum}, the present method eliminates the need of auxiliary spaces and derivatives and offers a straightforward approach to construct higher dimensional variables and their equations of motion, which has been elusive in earlier works. We begin with the idea proposed in ref.~\cite{bharadwaj2024thesis} to reduce the newly formed nonlinear terms into functions of $\bar{u}_{0}$ or its derivatives (which are known or given), thus giving a closed system of equations. Building on this, we propose an embedding that suffers no truncation errors and provides an improvement over ref.~\cite{bharadwaj2024thesis}. 
Consider eq.~(\ref{eq:Mth deformation Burgers}) and note that the quadratic nonlinearity is represented by the terms ($ \bar{u}_{k}\bar{u}_{M-k-1}$). First, as in the Carleman method \cite{liu2021efficient}, we define new variables $v_{m}$ given below and derive equations of motion for each. For clarity we consider the $M=4$ case:

\begin{equation}
  \begin{split}
    v_{1} &= \bar{u}_{1}\\
    v_{2} &= \bar{u}_{2}\\
    v_{3} &= \bar{u}_{3}\\
    v_{4} &= \bar{u}_{4}
  \end{split}~~~
  \begin{split}
      v_{5} &= \bar{u}^{2}_{1}\\
    v_{6} &= \bar{u}_{1}\bar{u}_{2}\\
    v_{7} &= (\partial_{x}\bar{u}_{1})^{2}\\
    \alpha_{0} &= \bar{u}_{0} = v_{0}
  \end{split}~~~
  \begin{split}
    &\alpha_{1} = \partial_{x}\bar{u}_{0} \\ 
    &\alpha_{2} = \partial_{xx}\bar{u}_{0} \\ 
    &\alpha_{3} =\partial_{x}(\bar{u}^{2}_{0}) \\ 
    &\alpha_{4} = \partial_{xx}(\bar{u}^{2}_{0}) 
  \end{split}
\end{equation}

Given $\bar{u}_{0}$, all $\alpha_{i}$ are known as well. Also note the we will use $\alpha_{0}$ and $v_{0}$ interchangeably, both representing $u_{0}$. Now the hierarchy of the equations of motion, for each $v_{i}$ and their embedding in the dilated space is done as follows:

\begingroup\color{black}
\begin{align}
    \frac{\partial v_{0}}{\partial t}
    &= \frac{\partial \bar{u}_{0}}{\partial t}
    = \nu\dd_{2}v_{0},
    \\[1ex]
    \frac{\partial v_{1}}{\partial t}
    &= \frac{\partial \bar{u}_{1}}{\partial t}
    = \hh\,\bar u_{0}\frac{\partial\bar u_{0}}{\partial x}
    = \frac{\hh}{2}\frac{\partial \bar u_{0}^{2}}{\partial x}
    \nonumber\\
    &=
    \hh\alpha_{0}\dd_{1}(v_{0}),
    \\[1ex]
    \frac{\partial v_{2}}{\partial t}
    &= \frac{\partial \bar{u}_{2}}{\partial t}
    \nonumber\\
    &=
    (1+\hh)\frac{\partial \bar u_{1}}{\partial t}
    +\hh\left[
    \frac{\partial(\bar u_{0}\bar u_{1})}{\partial x}
    -\nu\frac{\partial^{2}\bar u_{1}}{\partial x^{2}}
    \right]
    \nonumber\\
    &=
    (1+\hh)\frac{\hh}{2}\alpha_{3}
    +\hh\left[
    \bar u_{0}\frac{\partial\bar u_{1}}{\partial x}
    +\frac{\partial\bar u_{0}}{\partial x}\bar u_{1}
    -\nu\frac{\partial^{2}\bar u_{1}}{\partial x^{2}}
    \right]
    \nonumber\\
    &=
    (1+\hh)\frac{\hh}{2}\alpha_{3}
    +\hh\left[
    (\alpha_{0}\dd_{1}+\alpha_{1}-\nu\dd_{2})v_{1}
    \right] \nonumber \\ &:=\mathcal H_{2},
    \end{align}
    \begin{align}
    \frac{\partial v_{3}}{\partial t}
    &= \frac{\partial \bar{u}_{3}}{\partial t}
    \nonumber\\
    &=
    (1+\hh)\frac{\partial \bar u_{2}}{\partial t}
    +\hh\left[
    \frac{\partial(\bar u_{0}\bar u_{2})}{\partial x}
    +\frac{1}{2}\frac{\partial(\bar u_{1}^{2})}{\partial x}
    -\nu\frac{\partial^{2}\bar u_{2}}{\partial x^{2}}
    \right]
    \nonumber\\
    &=
    (1+\hh)\boxed{\mathcal H_{2}}
    +\hh\left[
    \bar u_{0}\frac{\partial\bar u_{2}}{\partial x}
    +\frac{\partial\bar u_{0}}{\partial x}\bar u_{2}
    +\frac{1}{2}\frac{\partial v_{5}}{\partial x}
    -\nu\frac{\partial^{2}\bar u_{2}}{\partial x^{2}}
    \right]
    \nonumber\\
    &=
    (1+\hh)\boxed{\mathcal H_{2}}
    +\hh\left[
    (\alpha_{0}\dd_{1}+\alpha_{1}-\nu\dd_{2})v_{2}
    +\frac{1}{2}\dd_{1}v_{5}
    \right]\nonumber \\ &:=\mathcal H_{3},
     \end{align}
    \begin{align}
    \frac{\partial v_{4}}{\partial t}
    &= \frac{\partial \bar{u}_{4}}{\partial t}
    \nonumber\\
    &=
    (1+\hh)\frac{\partial \bar u_{3}}{\partial t}
    +\hh\left[
    \frac{\partial(\bar u_{0}\bar u_{3})}{\partial x}
    +\frac{\partial(\bar u_{1}\bar u_{2})}{\partial x}
    -\nu\frac{\partial^{2}\bar u_{3}}{\partial x^{2}}
    \right]
    \nonumber\\
    &=
    (1+\hh)\boxed{\mathcal H_{3}}
    +\hh\left[
    \bar u_{0}\frac{\partial\bar u_{3}}{\partial x}
    +\frac{\partial\bar u_{0}}{\partial x}\bar u_{3}
    +\frac{\partial v_{6}}{\partial x}
    -\nu\frac{\partial^{2}\bar u_{3}}{\partial x^{2}}
    \right]
    \nonumber\\
    &=
    (1+\hh)\boxed{\mathcal H_{3}}
    +\hh\left[
    (\alpha_{0}\dd_{1}+\alpha_{1}-\nu\dd_{2})v_{3}
    +\dd_{1}v_{6}
    \right]\nonumber \\ &:=\mathcal H_{4},
     \end{align}
    \begin{align}
    \frac{\partial v_{5}}{\partial t}
    &= \frac{\partial(\bar u_{1}^{2})}{\partial t}
    =2\bar u_{1}\frac{\partial\bar u_{1}}{\partial t}
    =2v_{1}\left(\frac{\hh}{2}\alpha_{3}\right)
    =\hh\alpha_{3}v_{1},
        \end{align}
    \begin{align}
    \frac{\partial v_{6}}{\partial t}
    &= \frac{\partial(\bar u_{1}\bar u_{2})}{\partial t}
    =
    \frac{\partial\bar u_{1}}{\partial t}\bar u_{2}
    +\bar u_{1}\frac{\partial\bar u_{2}}{\partial t}
    \nonumber\\
    &=
    \frac{\hh}{2}\alpha_{3}v_{2}
    +v_{1}\left[
    (1+\hh)\frac{\hh}{2}\alpha_{3}
    +\hh(\alpha_{0}\dd_{1}+\alpha_{1}-\nu\dd_{2})v_{1}
    \right]
    \nonumber\\
    &=
    \frac{\hh}{2}\alpha_{3}v_{2}
    +(1+\hh)\frac{\hh}{2}\alpha_{3}v_{1}
    +\hh[
    \alpha_{0}v_{1}\dd_{1}v_{1} +
    \nonumber \\ &~~~~~~~~~~~~~~~~~~~~~~~~~~~~~~~~~~~~~~~~~~~~~+\alpha_{1}v_{1}^{2}
    -\nu v_{1}\dd_{2}v_{1}
   ]
    \nonumber\\
    &=
    \frac{\hh}{2}\alpha_{3}v_{2}
    +(1+\hh)\frac{\hh}{2}\alpha_{3}v_{1}
    +\hh\Bigg[
    \frac{\alpha_{0}}{2}\dd_{1}v_{5}
    +
    \nonumber \\ &~~~~~~~~~~~~~~~~~~~~~~~~~~~~~~~~~~+\alpha_{1}v_{5}
    -\nu\left\{
    \frac{1}{2}\dd_{2}v_{5}-v_{7}
    \right\}
    \Bigg]
    \nonumber\\
    &=
    \frac{\hh}{2}\alpha_{3}v_{2}
    +(1+\hh)\frac{\hh}{2}\alpha_{3}v_{1}
    +\hh\bigg[
    \frac{\alpha_{0}}{2}\dd_{1}v_{5}
    +
    \nonumber \\ &~~~~~~~~~~~~~~~~~~~~~~~~~~~~~~~~~~+\alpha_{1}v_{5}
    -\frac{\nu}{2}\dd_{2}v_{5}
    +\nu v_{7}
    \bigg],
        \end{align}
    \begin{align}
    \frac{\partial v_{7}}{\partial t}
    &=
    \frac{\partial}{\partial t}
    \left[
    \left(
    \frac{\partial\bar u_{1}}{\partial x}
    \right)^{2}
    \right]
    \nonumber\\
    &=
    2\left(\frac{\partial\bar u_{1}}{\partial x}\right)
    \frac{\partial}{\partial t}
    \left(
    \frac{\partial\bar u_{1}}{\partial x}
    \right)=
    2\left(\frac{\partial\bar u_{1}}{\partial x}\right)
    \frac{\partial}{\partial x}
    \left(
    \frac{\partial\bar u_{1}}{\partial t}
    \right)
    \nonumber\\
    &=
    2\dd_{1}v_{1}\,
    \dd_{1}\left(\frac{\hh}{2}\alpha_{3}\right)
    =
    \hh\alpha_{4}\dd_{1}v_{1}.
\end{align}
\endgroup

We now have a closed system of equations in the embedded space spanned by $\{v_{i}\}$. Note that, as we introduce new terms, the resulting set of equations have no additional truncation or associated errors, as would be the case in Carleman or Koopman type embeddings. \bob{Upon discretization, the right-hand side of the above equations can be
written as a finite-dimensional linear matrix operator. Since $v_{0}$ is
retained in the lifted state, the $\alpha_i$-dependent terms are represented
as linear couplings from $v_0$, giving a homogeneous system for the Burgers
example. For the present example, the coefficients are fixed by the chosen base profile on the considered time window, and the resulting lifted operator have known constant entries.}
\bob{\begin{equation}
    \frac{d\mathbf{v}}{dt} = A\mathbf{v} 
    \label{eq:linearized ODE}
\end{equation}}

Here, $A$ is the constant coefficient matrix obtained after discretizing the
operators in the lifted hierarchy.  The above time-dependent ODE can be solved numerically using an Euler method to march forward in time, using a time marching quantum algorithm \cite{bharadwaj2024compact} (see Section \ref{sec:TMCQC}). {Some essential details of this algorithm are repeated here (from [2]) in Appendices B-D for completeness.} The above embedding strategy can be generalized similarly, up to any arbitrary order of truncation, $M$. For the current example, the matrix equation would be assembled as
\bob{\begin{equation}
   \frac{d}{dt}\begin{bmatrix}
v_{0}\\v_{1}\\v_{2}\\v_{3}\\v_{4}\\v_{5}\\v_{6}\\v_{7}
   \end{bmatrix}
=
\begin{bmatrix}
a_{00} & 0      & 0      & 0      & 0 & 0      & 0      & 0\\
a_{10} & 0      & 0      & 0      & 0 & 0      & 0      & 0\\
a_{20} & a_{21} & 0      & 0      & 0 & 0      & 0      & 0\\
a_{30} & a_{31} & a_{32} & 0      & 0 & a_{35} & 0      & 0\\
a_{40} & a_{41} & a_{42} & a_{43} & 0 & a_{45} & a_{46} & 0\\
0      & a_{51} & 0      & 0      & 0 & 0      & 0      & 0\\
0      & a_{61} & a_{62} & 0      & 0 & a_{65} & 0      & a_{67}\\
0      & a_{71} & 0      & 0      & 0 & 0      & 0      & 0
\end{bmatrix}
\begin{bmatrix}
v_{0}\\v_{1}\\v_{2}\\v_{3}\\v_{4}\\v_{5}\\v_{6}\\v_{7}
\end{bmatrix}.
\label{eq:lifted_matrix_M4}
\end{equation}}

where each non-zero $a_{ij}$ is the finite difference matrix operator of $\bob{\dd_{1}}$ and $\bob{\dd_{2}}$, scaled appropriately by a combination of factors $(\hh,\alpha_{k},\nu)$. \bob{The non-zero entries of the lifted matrix are listed explicitly in Table~\ref{tab:lifted_matrix_entries}.} The exact dependence of these blocks on the spatial grid and numerical formulation, as well as the corresponding stability restriction are discussed in Sec.~\ref{subsec:Numerical Setup}.

\begin{table}[htpb!]
\begingroup\color{black}
\centering
\renewcommand{\arraystretch}{1.45}
\setlength{\tabcolsep}{4pt}
\begin{tabular}{p{0.47\columnwidth}p{0.47\columnwidth}}
\hline
$a_{00}=\nu\dd_{2}$
&
$a_{42}=(1+\hh)\hh L$
\\
$a_{10}=\hh\alpha_{0}\dd_{1}$
&
$a_{43}=\hh L$
\\
$a_{20}=(1+\hh)\hh\alpha_{0}\dd_{1}$
&
$a_{45}=(1+\hh)\dfrac{\hh}{2}\dd_{1}$
\\
$a_{21}=\hh L$
&
$a_{46}=\hh\dd_{1}$
\\
$a_{30}=(1+\hh)^{2}\hh\alpha_{0}\dd_{1}$
&
$a_{51}=\hh\alpha_{3}$
\\
$a_{31}=(1+\hh)\hh L$
&
$a_{61}=(1+\hh)\dfrac{\hh}{2}\alpha_{3}$
\\
$a_{32}=\hh L$
&
$a_{62}=\dfrac{\hh}{2}\alpha_{3}$
\\
$a_{35}=\dfrac{\hh}{2}\dd_{1}$
&
$\begin{aligned}
a_{65}
&=\hh\Big(
\dfrac{\alpha_{0}}{2}\dd_{1}
+\alpha_{1}
-\dfrac{\nu}{2}\dd_{2}
\Big)
\end{aligned}$
\\
$a_{40}=(1+\hh)^{3}\hh\alpha_{0}\dd_{1}$
&
$a_{67}=\hh\nu$
\\
$a_{41}=(1+\hh)^{2}\hh L$
&
$a_{71}=\hh\alpha_{4}\dd_{1}$
\\
\hline
\end{tabular}
\endgroup
\caption{\justifying \bob{Explicit nonzero entries of the lifted homogeneous matrix $A$ for
the $M=4$ Burgers embedding. Here
$L=\alpha_{0}D_{1}+\alpha_{1}-\nu D_{2}$. The entries
$a_{20}$, $a_{30}$, and $a_{40}$ arise because $v_{0}=\bar u_{0}$ is retained
in the lifted state, so the $\alpha_{3}$-dependent driving terms are written
as linear couplings from $v_{0}$ rather than as a separate forcing vector.}}
\label{tab:lifted_matrix_entries}
\end{table}

\subsection{\label{subsec:Numerical Setup}Numerical Setup and Stability criteria}
\bob{
We now discuss the numerical discretization of the homogeneous lifted system
in Eq.~\eqref{eq:linearized ODE}. For a spatial grid with $N_g$ points and
spacing $\Delta x=L/N_g$, each lifted variable $v_i$ is represented by an
$N_g$-dimensional vector. Therefore, for the $M=4$ example in
Eq.~\eqref{eq:lifted_matrix_M4}, the full lifted vector has dimension
$8N_g$ and the matrix $A$ has dimension $8N_g\times 8N_g$. The dependence of $A$ on the grid enters through the finite-difference
matrices $\dd_1,\dd_2\in\mathbb{R}^{N_g\times N_g}$ and through diagonal
coefficient matrices obtained by sampling the known base solution and its
derivatives on the grid. Specifically, for interior grid points,}
\bob{
\begin{align}
    \dd_1(v_i)_j&=\frac{(v_i)_{j+1}-(v_i)_{j-1}}{2\Delta x},\\
\dd_2(v_i)_j&=\frac{(v_i)_{j+1}-2(v_i)_j+(v_i)_{j-1}}{(\Delta x)^2}.
\end{align}}

\bob{Thus the entries of $\dd_1$ scale as $O(\Delta x^{-1})$, while those of
$\dd_2$ scale as $O(\Delta x^{-2})$. Coefficients $\alpha_i$ in
Table~\ref{tab:lifted_matrix_entries} are diagonal matrices, $\alpha_i=\mathrm{diag}\{\alpha_i(x_1),\alpha_i(x_2),\ldots,\alpha_i(x_{N_g})\}.$
For example, $\alpha_0\dd_1$ denotes pointwise multiplication by the sampled
base profile followed by the first-derivative matrix. Once the grid and base
profile are fixed, every block $a_{ij}$ in Eq.~\eqref{eq:lifted_matrix_M4}
is therefore an explicit sparse $N_g\times N_g$ matrix, and the full lifted
matrix has $O(N_g)$ nonzero entries for the fixed $M=4$ embedding.
}\par
\bob{
The time domain $t\in[0,T]$ is discretized into $\tau=T/\Delta t$ time
steps using explicit or implicit Euler schemes,
}
\begin{align}
\bob{
\mathbf v^{j+1}
}
&\bob{
=
(I+\Delta t A)\mathbf v^j
=
A_E\mathbf v^j,
}
\\
\bob{
\mathbf v^{j+1}
}
&\bob{
=
(I-\Delta t A)^{-1}\mathbf v^j
=
A_I\mathbf v^j.
}
\end{align}
\bob{The implicit scheme is unconditionally stable for the dissipative linear
part, whereas the explicit scheme requires a timestep restriction. For the
Dirichlet finite-difference Laplacian, the eigenvalues of $\dd_2$ are
}
\begin{equation}
\bob{
\lambda_j(\dd_2)
=
-\frac{4}{(\Delta x)^2}
\sin^2\left(\frac{j\pi}{2(N_g+1)}\right),
\qquad
j=1,\ldots,N_g,
}
\end{equation}
\bob{
up to the conventional adjustment depending on whether boundary or interior
points are counted. Hence the diffusive block $\nu\dd_2$ has strictly
negative eigenvalues. The full lifted matrix, however, also contains
auxiliary variables and lower-order couplings. After a permutation of the
lifted variables, the matrix is block lower triangular with diagonal blocks
$\nu\dd_2$ and zero blocks. Therefore,} $\bob{\sigma(A)=\sigma(\nu\dd_2)\cup\{0\}.}$
\bob{The zero eigenvalues arise from the auxiliary lifted variables. Thus the
correct statement is that the lifted matrix has no eigenvalues with positive
real part; it is not necessary, nor generally correct, to require all
eigenvalues of $A$ to have strictly negative real part.
}\bob{For a normal dissipative operator with $\mathrm{Re}(\lambda_k)<0$, the
explicit Euler stability condition may be written as
}
\begin{equation}
\bob{\Delta t
<
\min_{\mathrm{Re}(\lambda_k)<0}
\left(
\frac{-2\mathrm{Re}(\lambda_k)}{|\lambda_k|^2}
\right).
}
\end{equation}
\bob{Since the lifted matrix is generally non-normal and contains nilpotent
lower-order couplings, we use the conservative row-sum restriction
}
\begin{equation}
\bob{\Delta t \lesssim \|A\|_\infty^{-1},
\qquad
\|A\|_\infty
=
\max_i\sum_j\|a_{ij}\|_\infty,
}
\end{equation}

\bob{
with the nonzero blocks $a_{ij}$ given explicitly in
Table~\ref{tab:lifted_matrix_entries}. This gives the practical CFL-type
restriction used for the explicit scheme.
}

Furthermore, these time stepping schemes can be implemented either iteratively or as a one-shot method to compute solution $\forall t$ \cite{bharadwaj2024compact}. This is discussed in Appendix B.  The goal is to now implement these time marching operations using an efficient quantum algorithm to solve the governing equation defined \textcolor{black}{for the Burgers problem, which we discuss in Section \ref{sec:TMCQC} (relegating other problems to appendices A-F).}

\subsection{\label{subsec:homotopy convergence}Homotopy Convergence and Truncation Error} We now discuss the convergence of the Homotopy series solution and estimate the error introduced due to truncation \cite{liao2004homotopy,odibat2010study,liao2012homotopy}.\\

\noindent \textbf{Lemma 1:} (Convergence \cite{liao2004homotopy,odibat2010study,liao2012homotopy}) \textit{Consider the homotopy series $\bar{u} = \bar{u}_{0} + \sum_{p=1}^{\infty}q^{p}\bar{u}_{p},$ with $q\in[0,1]$. $\bar{u}_{p}$ belongs to a Banach space equipped with $\mathcal{L}^{k}$ Lebesgue measures for $k\in[1,\infty]$. The series is convergent if there exists a $\gamma_{\hh} = \frac{\|\bar{u}_{p+1}\|}{\|\bar{u}_{p}\|}$ such that $0<\gamma_{\hh} < 1$ $\forall p>\tilde{p}$, where $\tilde{p}\in \mathbb{Z}^{0+}$.}

\vspace{0.5cm}
\bob{Proof: Let us denote the series summation up to $M$ terms as}

\begin{equation}
\bob{S_M=\sum_{p=0}^{M}q^p\bar u_p.}
\label{eq:SM_def}
\end{equation}

\bob{From the ratio condition in Lemma~1, for all $p\geq\tilde p$,}

\begin{equation}
\bob{\|\bar u_{p+1}\|\leq \gamma_{\hh}\|\bar u_p\|,\qquad 0<\gamma_{\hh}<1.}
\label{eq:lemma1_ratio}
\end{equation}

\bob{For $M\geq\tilde p$, the difference of two successive partial sums satisfies}
\bob{
\begin{align}
\|S_{M+1}-S_M\|  &= |q|^{M+1}\|\bar u_{M+1}\|\nonumber \\&\leq \|\bar u_{M+1}\|\leq \gamma_{\hh}\|\bar u_M\|\leq \gamma_{\hh}^2\|\bar u_{M-1}\|\cdots\leq \gamma_{\hh}^{M-\tilde p+1}\|\bar u_{\tilde p}\|.
\label{eq:successive_partial_sum}
\end{align}}

\bob{Now consider two partial sums $S_M$ and $S_{M'}$ with $M>M'\geq\tilde p$. Using Eq.~\eqref{eq:SM_def}, the difference can be written as a telescoping sum.}

\begin{widetext}
\begin{align}
\bob{\|S_M-S_{M'}\|} &\bob{= \|(S_M-S_{M-1})+(S_{M-1}-S_{M-2})+\cdots+(S_{M'+1}-S_{M'})\|} \nonumber\\
&\bob{\leq \|S_M-S_{M-1}\|+\|S_{M-1}-S_{M-2}\|+\cdots+\|S_{M'+1}-S_{M'}\|} \nonumber\\
&\bob{\leq \gamma_{\hh}^{M-\tilde p}\|\bar u_{\tilde p}\|+\gamma_{\hh}^{M-\tilde p-1}\|\bar u_{\tilde p}\|+\cdots+\gamma_{\hh}^{M'-\tilde p+1}\|\bar u_{\tilde p}\|} \nonumber\\
&\bob{= \frac{\gamma_{\hh}^{M'-\tilde p+1}\left(1-\gamma_{\hh}^{M-M'}\right)}{1-\gamma_{\hh}}\|\bar u_{\tilde p}\|.}
\label{eq:convergence_ineq}
\end{align}
\end{widetext}

\noindent \textbf{Lemma 2:} (Truncation error \cite{liao2004homotopy,odibat2010study,liao2012homotopy}) \textit{Consider a truncated version of the homotopy series in Theorem 1, computed up to $M$ terms, with $M\in \mathbb{Z^{+}}$, given by $\bar{u} = \bar{u}_{0} + \sum_{p=1}^{p=M}q^{p}\bar{u}_{p},$ and $q\in[0,1]$. Let the series be convergent for a $\gamma_{\hh} = \frac{\|\bar{u}_{p+1}\|}{\|\bar{u}_{p}\|}$ such that $0<\gamma_{\hh} < 1$ $\forall p>\tilde{p}$, where $\tilde{p}\in \mathbb{Z}^{0+}$. The truncation error $\varepsilon_{\gamma}$ in estimating the solution $u$ relative to the fully asymptotic series can be provided by the upper bound 
\begin{equation}
    \varepsilon_{\gamma}=\Big|\Big|u -\sum_{p=0}^{p=M}q^{p}\bar{u}_{p} \Big|\Big| \leq \frac{\gamma^{M+1}_{\hh}}{1-\gamma_{\hh}}\|\bar{u}_{0}\|
\end{equation}}

\noindent \textbf{Proof:} To estimate the truncation error, we begin by considering eq.~(\ref{eq:convergence_ineq}),  \begin{equation}
    \|S_{M} - S_{M'}\| \leq \bob{ \frac{\gamma_{\hh}^{M'-\tilde p+1}\left(1-\gamma_{\hh}^{M-M'}\right)}{1-\gamma_{\hh}}\|\bar u_{\tilde p}\|.}
\end{equation} We now let $\tilde{p}=0,$ $\bob{M}\rightarrow \infty$, and relabeling $M'\rightarrow M$, we can note at once that
\begin{equation}
    \varepsilon_{\gamma} = \Big|\Big|u -\sum_{p=0}^{p=M}q^{p}\bar{u}_{p} \Big|\Big|
    \leq \frac{\gamma^{M+1}_{\hh}}{1-\gamma_{\hh}}\|\bar{u}_{0}\|.
\end{equation}

We use \textbf{Lemmas 1} and \textbf{2} to assert the convergence of the homotopy series solution for \textbf{Problem 1} and thus estimate the truncation errors. We consider here both analytical and numerical proofs. The analytical proof refers to the case of the computation of the norms using continuous integral based (analytical) solution using the homotopy process. In the latter case, the finite difference operator used to compute the numerical solution takes the form of discretized non-analytic integration.\\

\noindent \textbf{Theorem 1:} \textit{\bob{Consider a truncated homotopy series solution for Problem~1, computed up to $M$ terms, with $M\in\mathbb{Z}^{+}$, given by $\bar u=\bar u_0+\sum_{p=1}^{M}q^p\bar u_p$, and $q\in[0,1]$. In the controlled-cancellation regime described in the proof below, the analytical homotopy coefficients satisfy}}

\begin{equation}
\bob{\gamma_{\hh}:=\frac{\|\bar u_{p+1}\|}{\|\bar u_p\|}=O(|\hh|t).}
\end{equation}

\textit{\bob{Thus, for sufficiently small $|\hh|t$, the analytical series is convergent. If $0<\gamma_{\hh}<1$, the truncation error is bounded by}}

\begin{equation}
\bob{\epsilon_\gamma\leq \frac{\gamma_{\hh}^{M+1}}{1-\gamma_{\hh}}\|\bar u_0\|.}
\end{equation}

\textit{\bob{Consequently, a prescribed accuracy $\epsilon_\gamma$ is obtained by choosing}}

\begin{equation}
\bob{M\geq \left\lceil\frac{\log\left(\epsilon_\gamma(1-\gamma_{\hh})/\|\bar u_0\|\right)}{\log(\gamma_{\hh})}-1\right\rceil.}
\end{equation}

\textit{\bob{For the finite-difference computation, the corresponding ratio satisfies}}

\begin{equation}
\bob{\bar\gamma_{\hh}=O\!\left(|\hh|\|\dd_1\|\|\bar u_0\|t\right).}
\end{equation}

\textit{\bob{Thus, for sufficiently small $|\hh|\|\dd_1\|\|\bar u_0\|t$, the finite-difference homotopy series is convergent. If $0<\bar\gamma_{\hh}<1$, the truncation error is bounded by}}

\begin{equation}
\bob{\epsilon_{\bar\gamma}\leq \frac{\bar\gamma_{\hh}^{M+1}}{1-\bar\gamma_{\hh}}\|\bar u_0\|.}
\end{equation}

\textit{\bob{Consequently, the finite-difference computation requires}}

\begin{equation}
\bob{\bar M\geq \left\lceil\frac{\log\left(\epsilon_{\bar\gamma}(1-\bar\gamma_{\hh})/\|\bar u_0\|\right)}{\log(\bar\gamma_{\hh})}-1\right\rceil.}
\end{equation}
\noindent \textbf{Proof:} For this, we invoke the method of mathematical induction. First we estimate the upper bound for $\gamma_{\hh}$, for which we consider the homotopy terms $\bar{u}_{p}$, whose equations of motion are given by eq.~(\ref{eq:Mth deformation Burgers}). For this, we consider the max-norm of successive terms given by $\|\bar{u}_{p}(x,t)\| = \sup_{x\in[0,1]}|\bar{u}_{p}(x,t)|$ for $L=1$.\\

\noindent \textbf{Case ($p=0$):} This corresponds to $\bar{u}_{0}$ given by eq.~(\ref{eq: diffusion solution u0}) and its norm 
\begin{align}
\|\bar{u}_{0}(x,t)\| & = \Big|\Big|(1-2x)+  \sum_{n=1}^{\infty}  \Big\{\frac{2e^{-4\pi^{2}n^{2}\nu t}}{\pi n(4n^{2}-1)} \sin(2n\pi x)\Big\}\Big|\Big|
\end{align}
First note that for the solution $|\bar{u}_{0}(x,t)|$ varies from $|\cos(\pi x)|_{t=0}$ to $|1-2x|_{t=\infty}$. In both cases the $\|\bar{u}_{0}(x,t)\|=1$. Also observe that, $|\cos(\pi x)|\geq|1-2x|$ and that the solution varies continuously between $t\in[0,\infty)$ with $\exp(-4\pi^{2}n^{2}\nu t)$ being a decreasing function.  These observations, along with the fact that there exists a wide range of functions and great flexibility in choosing the initial conditions,  without loss of generality, one can state $\|\bar{u}_{0}(x,t)\|\leq1$. \bob{Moreover, $\bar u_0$ is a smooth prescribed base profile on the finite time interval considered here. Hence the finite set of spatial derivatives needed in the estimates below is bounded by constants independent of the homotopy order $p$ and truncation order $M$. In particular, on the normalized domain $L=1$,}
$
\bob{\|\partial_x^r\bar u_0(\cdot,t)\|_\infty=O(1),~~~\|\partial_x(\bar u_0^2)(\cdot,t)\|_\infty=O(1).}
$\\

\noindent \textbf{Case ($p=1$):} Proceeding to the next term, we consider the analytical and numerical terms separately for clarity.\\

\noindent (a) Analytical -- First we note that $\exp(-4\pi^{2}n^{2}\nu t)$ attains a maximum at $t=0$, therefore we compute the norms of $\frac{\partial^{k} \bar{u}_{0}}{\partial x^{k}}$ at $t=0$, which for odd and even values of $k$, give $|\pi^{k}\sin(\pi x)|$ and $|\pi^{k}\cos(\pi x)|$, respectively. So if $c_{0}(t) = \sup_{0\leq t'\leq t}\|\partial_{x}(\bar{u}_{0}(\cdot,t')^2)\|_\infty$For $\bar{u}_{1}$, we have
\begin{align}
   \|\bar{u}_{1}(x,t)\|  &= \frac{\hh}{2}\int_{0}^{t}\Bigg|\Bigg|2\bar{u}_{0}\frac{\partial \bar{u}_{0}}{\partial x}dt\Bigg|\Bigg|,\\
     & \leq \frac{\bob{|\hh|}}{2}\int_{0}^{t}\|-2\pi\cos(\pi x)\sin(\pi x)\|dt \nonumber \\&\leq \frac{|\hh|}{2}c_{0}(t)t=\mathcal{O}\left(\bob{|\hh|} t\right).
\end{align}

This gives us our first ratio, bounded as follows $ \frac{\|\bar{u}_{1}\|}{\|\bar{u}_{0}\|} = O(\bob{|\hh|} t)\leq 1 $. 
\noindent (b) Numerical -- The above equations may be rewritten as follows in terms of the norms of the differential operators.
\begin{align}
   \|\bar{u}_{1}(x,t)\|&= \frac{\hh}{2}\int_{0}^{t}\Bigg\| \frac{\partial \bar{u}_{0}^{2}}{\partial x}\Bigg\| dt \\
   &=\frac{\hh}{2}\int_{0}^{t}\Big\|  \bob{\dd_{1}}\bar{u}_{0}^{2}\Big\| dt \\
   &\leq \frac{\bob{|\hh|}}{2} \|  \bob{\dd_{1}} \|  \|\bar{u}_{0} \|^{2}t
\end{align}
Therefore using operator norms this ratio is required to be bounded as, $\frac{\|\bar{u}_{1}\|}{\|\bar{u}_{0}\|} =  O(\bob{|\hh|} \|  \bob{\dd_{1}} \|  \|\bar{u}_{0} \|t) \leq 1$.\\

\noindent \textbf{Case ($p=2$):} Employing a similar calculation to $p=2$ we obtain, the following.
\vspace{0.2cm}
\bob{(a) Analytical --}

\begin{equation}
\bob{\|\bar u_2(x,t)\|=\left\|\int_0^t\left[(1+\hh)\frac{\partial \bar u_1}{\partial t}+\hh\left\{\frac{\partial \bar u_0\bar u_1}{\partial x}-\nu\frac{\partial^2\bar u_1}{\partial x^2}\right\}\right]dt\right\|.}
\end{equation}
\begin{equation}
\bob{\|\bar u_2(x,t)\|\leq \int_0^t\left[\underbrace{|1+\hh|\left\|\frac{\partial \bar u_1}{\partial t}\right\|}_{\underset{\mathbf{{\Huge 1}}}{\downarrow}}+\underbrace{|\hh|\left(\left\|\frac{\partial \bar u_0\bar u_1}{\partial x}\right\|+\nu\left\|\frac{\partial^2\bar u_1}{\partial x^2}\right\|\right)}_{\underset{\mathbf{{\Huge 2}}}{\downarrow}}\right]dt.}
\end{equation}
\bob{Let $\rho_{\hh}:=\pi|\hh|t/2$. The first underbraced term contributes $O(|1+\hh|\rho_{\hh})$, while the second contributes $O((2+\nu\pi)\rho_{\hh}^2)$. Hence}
\begin{equation}
\bob{\|\bar u_2\|\leq O\!\left(|1+\hh|\rho_{\hh}\right)+O\!\left((2+\nu\pi)\rho_{\hh}^{2}\right).}
\end{equation}
\bob{The linear inherited term cannot, in general, be absorbed into the quadratic term for arbitrarily small $t$. It is suppressed only when the convergence-control parameter is chosen so that $|1+\hh|\lesssim \rho_{\hh}$; in particular, it vanishes exactly at $\hh=-1$. In this controlled-cancellation regime,}
\begin{equation}
\bob{\|\bar u_2\|\leq O(\rho_{\hh}^{2}).}
\end{equation}
\bob{Since the $p=1$ estimate gives $\|\bar u_1\|\leq O(\rho_{\hh})$, we obtain}
\begin{equation}
\bob{\frac{\|\bar u_2\|}{\|\bar u_1\|}\leq O(\rho_{\hh})=O\!\left(|\hh|t\right)\leq 1.}
\end{equation}
\bob{\noindent (b) Numerical --}
\begin{equation}
\bob{\|\bar u_2(x,t)\|=\left\|\int_0^t\left[(1+\hh)\frac{\partial \bar u_1}{\partial t}+\hh\left\{D_1(\bar u_0\bar u_1)-\nu D_2\bar u_1\right\}\right]dt\right\|.}
\end{equation}
\bob{\begin{align}
\|\bar u_2(x,t)\|\leq &|1+\hh|\frac{|\hh|}{2}\|D_1\|\|\bar u_0\|^2t+\nonumber \\ &|\hh|\left\{\frac{|\hh|t^2}{2}\|D_1\|^2\|\bar u_0\|^3+\frac{|\hh|\nu t^2}{2}\|D_2\|\|D_1\|\|\bar u_0\|^2\right\}.
\end{align}}
\bob{Let $\bar\rho_{\hh}:=|\hh|\|D_1\|\|\bar u_0\|t/2$. Then we obtain the following leading contributions,}
\begin{equation}
\bob{\|\bar u_2\|\leq O\!\left(|1+\hh|\bar\rho_{\hh}\|\bar u_0\|\right)+O\!\left[\left(1+\frac{\nu\|D_2\|}{\|D_1\|\|\bar u_0\|}\right)\bar\rho_{\hh}^{2}\|\bar u_0\|\right].}
\end{equation}
\bob{Under the controlled-cancellation condition $|1+\hh|\lesssim \bar\rho_{\hh}$ and the resolution scaling used in Sec.~VI, $\nu\|D_2\|/(\|D_1\|\|\bar u_0\|)=O(1)$, so}
\begin{equation}
\bob{\frac{\|\bar u_2\|}{\|\bar u_1\|}\leq O(\bar\rho_{\hh})=O\!\left(|\hh|\|D_1\|\|\bar u_0\|t\right)\leq 1.}
\end{equation}

\bob{Case $(p=M+1)$: Now assume that $\gamma_{\hh}\leq O(\pi|\hh|t)\leq 1$ holds up to order $p=M$. We show that the same bound holds for the $(M+1)$-st deformation coefficient. Using Eq.~\eqref{eq:Mth deformation Burgers}, define}

\begin{equation}
\bob{\Phi_M:=(\alpha_0\dd_1+\alpha_1-\nu\dd_2)\bar u_M+\frac{1}{2}\dd_1\sum_{k=1}^{M-1}\bar u_k\bar u_{M-k}.}
\end{equation}
\bob{The additional $\alpha_i$ coefficients that appear when the lifted system is written explicitly at higher order do not introduce new homotopy orders. They are fixed functions of the prescribed base profile $\bar u_0$ and its spatial derivatives. For the coefficient recursion in Eq.~\eqref{eq:Mth deformation Burgers}, the base-profile dependence entering the $(M+1)$-st equation is contained in $\alpha_0=\bar u_0$ and $\alpha_1=\partial_x\bar u_0$. Higher $\alpha_i$'s generated by the lifted closure affect only fixed pre-factors in the matrix representation, not the geometric dependence on $\rho_{\hh}$.}
\bob{Then the $(M+1)$-th deformation equation is}
\begin{equation}
\bob{\frac{\partial\bar u_{M+1}}{\partial t}=(1+\hh)\frac{\partial\bar u_M}{\partial t}+\hh\Phi_M.}
\end{equation}
\bob{Integrating in time and using $\bar u_{M+1}(x,0)=0$ gives}
\begin{equation}
\bob{\|\bar u_{M+1}\|\leq |1+\hh|\,\|\bar u_M\|+|\hh|\int_0^t\|\Phi_M\|\,dt.}
\end{equation}
\bob{Let $\rho_{\hh}:=\pi|\hh|t/2$. By the induction hypothesis, $\|\bar u_k\|\leq O(\rho_{\hh}^k)$ for $1\leq k\leq M$. The first term therefore gives $O(|1+\hh|\rho_{\hh}^M)$. For the second term, the linear part of $\Phi_M$ contributes one additional factor of $\rho_{\hh}$, while each convolution product $\bar u_k\bar u_{M-k}$ has order $O(\rho_{\hh}^{M})$ and gains one additional factor of $\rho_{\hh}$ after the spatial derivative, multiplication by $|\hh|$, and integration in time. Hence}
\begin{equation}
\bob{|\hh|\int_0^t\|\Phi_M\|\,dt\leq O(\rho_{\hh}^{M+1}).}
\end{equation}
\bob{Consequently,}
\begin{equation}
\bob{\|\bar u_{M+1}\|\leq O(|1+\hh|\rho_{\hh}^{M})+O(\rho_{\hh}^{M+1}).}
\end{equation}
\bob{The inherited contribution proportional to $1+\hh$ is suppressed when $|1+\hh|\lesssim\rho_{\hh}$, and vanishes exactly at $\hh=-1$. In this controlled-cancellation regime,}
\begin{equation}
\bob{\|\bar u_{M+1}\|\leq O(\rho_{\hh}^{M+1}).}
\end{equation}
\bob{A similar calculation for the $M$-th coefficient gives $\|\bar u_M\|\leq O(\rho_{\hh}^{M})$. Therefore}
\begin{equation}
\bob{\frac{\|\bar u_{M+1}\|}{\|\bar u_M\|}\leq O(\rho_{\hh})=O\!\left(|\hh|t\right)\leq 1.}
\end{equation}
\bob{Thus, by induction, $\gamma_{\hh}\leq O(\pi|\hh|t)\leq 1$ in the controlled-cancellation regime. For the finite-difference computation, the same argument with $\bar\rho_{\hh}:=|\hh|\|\dd_1\|\|\bar u_0\|t/2$ gives}
\begin{equation}
\bob{\bar\gamma_{\hh}\leq O(\bar\rho_{\hh})=O\!\left(\hh|\|D_1\|\|\bar u_0\|t\right)\leq 1.}
\end{equation}
\noindent \bob{We can now use Lemma~2 to estimate the truncation error. If $0<\gamma_{\hh}<1$, then}

\begin{equation}
\bob{\epsilon_\gamma\leq \frac{\gamma_{\hh}^{M+1}}{1-\gamma_{\hh}}\|\bar u_0\|.}
\end{equation}

\bob{Thus, a prescribed truncation accuracy $\epsilon_\gamma$ is obtained by choosing}

\begin{equation}
\bob{M\geq \left\lceil\frac{\log\left(\epsilon_\gamma(1-\gamma_{\hh})/\|\bar u_0\|\right)}{\log(\gamma_{\hh})}-1\right\rceil.}
\end{equation}

\bob{Similarly, if $0<\bar\gamma_{\hh}<1$, the finite-difference truncation error satisfies}

\begin{equation}
\bob{\epsilon_{\bar\gamma}\leq \frac{\bar\gamma_{\hh}^{M+1}}{1-\bar\gamma_{\hh}}\|\bar u_0\|.}
\end{equation}

\bob{The required number of deformation terms in the finite-difference computation is therefore}

\begin{equation}
\bob{\bar M\geq \left\lceil\frac{\log\left(\epsilon_{\bar\gamma}(1-\bar\gamma_{\hh})/\|\bar u_0\|\right)}{\log(\bar\gamma_{\hh})}-1\right\rceil.}
\end{equation}

\begin{figure*}[htpb!]
        \centering
        \includegraphics[trim={0cm 0cm 0cm 0cm},clip=true,scale=0.5]{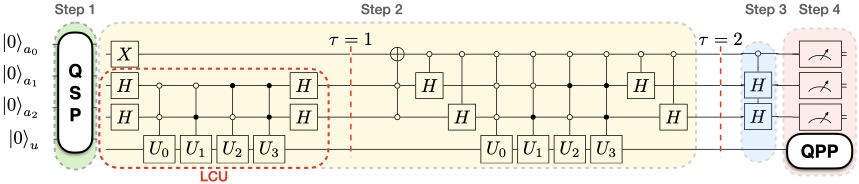}
        \caption{\justifying The figure shows an example of a 4 qubit LCU circuit to implement an iterative explicit-time simulation for $\tau=2$ time steps. Step 1 initializes all the qubits using an efficient quantum state preparation circuit. The first register $q(a_{0})$ is the counter qubit \cite{bharadwaj2024compact}. These qubits are initially set to one (representing the maximum number of time steps, in a binary format) and via successive bit-flip operations, this is counted down after each application of the LCU. The ancilla qubits $q(a_{1})$ and $q(a_{2})$ are used to implement the linear combination of the four unitaries $U_{0}$ to $U_{4}$. The last register $q(u)$ stores the velocity field at every time step. Step 2 is the iterative application of LCU performed to march forward in time. Step 3 performs the summation of each term of the homotopy series as per eq.\ \ref{eq:deformations} and finally the result is measured or post-processed in situ using a Quantum Post Processing circuit \cite{bharadwaj2023hybrid}.}\label{fig:lcu circuit}
\end{figure*}
\section{ Time Marching Compact Quantum Circuits}\label{sec:TMCQC}
To perform time marching simulations, we employ a state-of-the-art, compact, quantum algorithm proposed in Ref.~\cite{bharadwaj2024compact}, known as Time Marching Compact Quantum Circuits (TMCQC). These circuits are based on the concept of Linear Combination of Unitaries (LCU) \cite{childs2017quantum,schlimgen2021quantum}, where a given general, non-hermitian and non-unitary operator can be expressed as a weighted summation of unitary operations, that are efficiently implementable as quantum circuit with one- and two-qubit gates. The algorithm chosen here is the \bob{one-shot TMCQC5/6 version of this solver \cite{bharadwaj2024compact}, that has the near-optimal gate
complexity defined in Sec.~I, while the postselection/query overhead is
controlled under the assumptions discussed in Ref.~\cite{bharadwaj2024compact}. Besides the complexity, it is} designed to be an end-to-end method, viable on near-term quantum devices, thus justifying its choice over other existing quantum algorithms. We now briefly outline the idea underlying this algorithm and its complexity.
\subsection{Linear Combination of Unitaries}
Following \cite{bharadwaj2024compact,schlimgen2021quantum}, given a non-unitary, non-hermitian matrix $J$, one can decompose it into symmetric and anti-symmetric matrices, $S$ and $A$, respectively, as
\begin{align}
    S &= \frac{1}{2}(J+J^{\dag}) ~\text{\&}\\
    A &= \frac{1}{2}(J-J^{\dag}),
\end{align}
where $J =  S+A$. A further, exact decomposition of these matrices can be written as
\begin{align}
    S &= \lim_{\epsilon\rightarrow 0}\frac{i}{2\epsilon}(e^{-i\epsilon S }-e^{i\epsilon S}),\label{eq:sym} \\ 
    A &= \lim_{\epsilon\rightarrow 0}\frac{1}{2\epsilon}(e^{\epsilon A}-e^{-\epsilon A}),
    \label{eq:antisym}
\end{align}
where $\epsilon$ is an expansion parameter. By construction, it is easy to see that $e^{\pm iS}$ and $e^{\pm A}$ are both unitary operators. We denote these operators as $U_{0} = ie^{-i\epsilon S}$, $U_{1} = -ie^{i\epsilon S}$, $U_{2} = e^{\epsilon A}$ and $U_{3} = -e^{-\epsilon A}$, and express $M$ exactly as a weighted sum of purely unitary operators given by

    $J = \sum_{c=0}^{c=3}\beta_{c}U_{c}=\lim_{\epsilon\rightarrow 0}\frac{1}{2\epsilon}(U_{0}+U_{1}+U_{2}+U_{3})$. However, in practice, we approximate $J$ by choosing a small enough $\epsilon$, thus resulting in a decomposition with just 4 unitaries given by
    \begin{equation}
        \tilde{M} = \frac{1}{2\epsilon}(U_{0}+U_{1}+U_{2}+U_{3}).
        \label{eq:lcu}
    \end{equation}

In view of capabilities of NISQ and near-term devices, we can go a step further to reduce this decomposition to only two unitaries, by trading in one extra qubit to simply dilate the matrix $J$ into a hermitian one given by
\begin{equation}
    \hat{J} = \begin{pmatrix}
        0 & J\\
        J^{\dag} & 0
    \end{pmatrix}.
\end{equation}
Since $J_{ij} \in \mathbb{R}$, the matrix $\hat{J}$ is symmetric with no anti-symmetric part. This would leave us with a symmetric, hermitian, Hamiltonian $\hat{J}$ that can be decomposed into just two unitaries as \begin{equation}
   \hat{J} =\lim_{\epsilon\rightarrow 0}\frac{i}{2\epsilon}(e^{-i\epsilon \hat{J} }-e^{i\epsilon \hat{J}}) = \frac{1}{2\epsilon}(\hat{U}_{0}+\hat{U}_{1}).
   \label{eq: lcu2}
\end{equation} 
Here, $e^{-i\epsilon \hat{J} } = \hat{U}_{0}$ and $-e^{i\epsilon \hat{J}} = \hat{U}_{1}$. The size of input vector would likewise need to be doubled (costing only one extra qubit) and would be given by $\hat{b}=[0,b]$, where one half of the dilated vector is padded with zeros. From here on, let $C$ represent the number of unitaries corresponding to either two or four. We now examine how to implement the decomposition as a quantum circuit.
The basic LCU circuit (represented by LCU snippet highlighted by red-dashed lines, in figure \ref{fig:lcu circuit}) requires two quantum registers, both set to 0 initially: (1) $|\Psi\rangle_{u}$ -- to store the state vector that is to be operated on, and (2) $|\Psi\rangle_{a}$ -- a register with a total of $n_{a}=\log_{2}(C)$ ancillary qubits (here, $n_{a}\in\{1,2\}$). $|\Psi\rangle_{u}$ is prepared using the operator $R$ which, in our case, is simply a NOT gate to prepare the initial delta function. $|\Psi\rangle_{a}$, on the other hand, is prepared by an operator $V$ into a superposition state proportional to
\begin{equation}
    |\Psi\rangle_{a} = V|0\rangle^{\otimes n_{a}} = \sqrt{1/\beta}\sum_{c}\sqrt{\beta_{c}}|c\rangle,
\end{equation}
where $\beta = \sum_{c}\beta_{c}$. Since all $\vert\beta_{c}\vert$ are equal, the ancillary register can be prepared as a uniform superposition state by simply applying Hadamard gates on each qubit of the register. This preparation step is also efficient, since it requires only an $\mathcal{O}(1)$ depth operation. Now, using register $|\Psi\rangle_{a}$ as the control qubits, we apply the LCU unitaries as a series of uniformly controlled operations $U_{c}$ on $|\Psi\rangle_{u}$ which is represented by the operator $W=\sum_{c}|c\rangle \langle c|\otimes U_{c}$. Then, $|\Psi\rangle_{a}$ is reset to 0 by applying $V^{\dag}$. Finally the ancillary register is measured in the computational basis, yielding a state proportional to 
\begin{align}
    \nonumber 
    |\Psi\rangle &= R|0\rangle^{\otimes n_{a}}\otimes V|0\rangle^{\otimes n_{g}} \mapsto (V^{\dag}\otimes \mathbb{I})\otimes W(|\Psi\rangle_{a}\otimes |\Psi\rangle_{u}) \\&
    = |0\rangle^{n_{a}}\Big(\sum_{c}\beta_{c}U_{c}\Big)|\Psi\rangle_{u} + |\Psi\rangle_{\perp} \nonumber \\
    &= \frac{1}{\sqrt{\beta}}|0\rangle^{n_{a}}J|\Psi\rangle_{u} + |\Psi\rangle_{\perp}.
    \label{eq: LCU final state}
\end{align}
For brevity, we drop the required normalization constants and $|\Psi\rangle_{\perp}$ corresponds to an orthogonal subspace that stores the unwanted remainder from the above operation of eq.~(\ref{eq:lcu}). The post-selected solution subspace is then re-scaled classically by $\mathcal{O}(\|~|\Psi\rangle_{u}\|/\epsilon) = \mathcal{O}(1/\epsilon)$ to obtain the actual solution. However, this procedure also shows that, the circuit for linearly combining the unitaries applies the operator $M$ only probabilistically, and the solution subspace of the quantum state is prepared with a small but finite success probability $p_{\textrm{succ}}$. Since we are interested in applying such a decomposition for $\tau$ time steps, $p_{\textrm{succ}}$ would decay as function of $f(2^{-C},\epsilon^{-1},\tau,\||\Psi\rangle_{\tau}\|^{-1})$. The smaller the $p_{\textrm{succ}}$, the larger is the number of repeated circuit simulations (query complexity or shots) required to measure or sample the solution subspace accurately. \bob{
The compact time-marching construction is designed so that the LCU
postselection overhead can be controlled for the flow-type systems considered
in Ref.~\cite{bharadwaj2024compact}. In particular, matrix rescaling and
Richardson extrapolation can keep the relevant postselection factor from
becoming unfavorably large.
}

Proceeding further, one can use either implicit or explicit time integration circuits as shown in ref.~\cite{bharadwaj2024compact}. An example circuit of the explicit-in-time, iterative LCU approach (TMCQC2) is shown in figure \ref{fig:lcu circuit}. Alternatively, as shown in Appendix B, one could construct a single history-state approach \cite{liu2021efficient,bharadwaj2024compact}, where both explicit and implicit methods can be used to solve for all time steps in one go (TMCQC5/6), which also tends to exhibit an overall, near-optimal complexity. The explicit circuit construction in terms of basic one- and two- qubit gates may be done using different approaches. For instance, QISKIT transpilation is a functionality that allows one to construct optimized circuits with custom qubit topologies, on real IBM backend devices. However, the typical gate complexity scaling with this approach is still non-optimal in problem size. To ameliorate this one would have to employ the near-term strategies and LCU parallelization techniques proposed in \cite{bharadwaj2024compact}. 

For larger problem sizes of practical interest, given the availability of an adequate number of qubits and quantum volume on near-term devices, the asymptotic gate complexity scaling of the TMCQCs (in terms of 1- and 2-qubit gates) is attained using optimal Hamiltonian simulation algorithms \cite{bharadwaj2024compact}. In our case, this translates to implementing a Hamiltonian simulation of unitaries of the form $e^{\pm i\epsilon \hat{S}}$ (for $C=2$) and its powers. For an $s-$sparse matrix $\hat{J}$, of size $N\times N$, scaled such that $\|\cdot\|\leq 1$, the gate complexity to implement controlled unitaries of the form $e^{\pm i\epsilon \hat{S}}$, up to an admissible error of $\varepsilon_{U}$, would require 
\begin{equation}
    G_{U} = \mathcal{O}\big((s\epsilon+1) (\log N + \log^{2.5}(\epsilon/\varepsilon_{U}))\log(\epsilon/\varepsilon_{U})\big)\label{eq: GU}
\end{equation} one- and two-qubit gates \cite{berry2015hamiltonian}. The corresponding complexities of TMCQC2 and TMCQC5/6 mentioned above would be given, respectively, by \cite{bharadwaj2024compact}, $\mathcal{O}\Big(s\epsilon,\log(N),\textrm{polylog}(\epsilon/\varepsilon_{U}),\tau\Big)$ and $\mathcal{O}\Big(s\epsilon,\log(N\tau),\textrm{polylog}(\epsilon/\varepsilon_{U},1/\Gamma\varepsilon_{N}),\log^{-3}((\kappa-1)^{-1})\Big)$.
For the problem considered here, this corresponds to matrix $A$ given by eq.~(\ref{eq:linearized ODE}). In the above complexities $N$ is the size of the matrix, $\tau$ is the number of time steps, $s$ is the sparsity, $\kappa$ is the condition number, $\varepsilon_{U}$, $\varepsilon_{N}$ are the accuracies of Hamiltonian simulation and truncated Neumann series expansion of a matrix inverse (see Appendix \ref{sec: Trunc Neumann Series}), and $\Gamma^{-1}$ is a bound on a certain matrix norm, which is again defined in Appendix \ref{sec: Trunc Neumann Series}. \bob{
As one can notice, the one-shot TMCQC5/6 gate complexity is near-optimal in
the sense defined in Sec.~I: it is logarithmic or polylogarithmic
in the time-marching system size and accuracy parameters, while $s$ and
$\epsilon$ remain outside the logarithm. The corresponding total runtime also
depends on the postselection/query factor.
} 

Next, we discuss relevant measures of nonlinearity, followed by an analysis of the error and complexity estimates of the overall quantum algorithm. 

\section{\label{sec:length scales}Nonlinearity estimates using Kolmogorov scales}
\bob{The degree of nonlinearity that one can capture using a linearization approach such as the one presented here depends on the order of truncation $M$, the desired accuracy $\varepsilon$, and the physical length scales that must be resolved by the discretization. The standard physical measure of nonlinearity in fluid dynamics is the Reynolds number $Re=\bar U\bar D/\nu$, where $\bar U$ and $\bar D$ are characteristic velocity and length scales of the flow, respectively. This number measures the ratio of inertial to viscous effects: $Re\lesssim 1$ corresponds to a viscous-dominated regime, whereas $Re\gg 1$ corresponds to an inertia dominated regime in which nonlinear effects become increasingly important.}

\bob{In previous quantum algorithms for nonlinear dissipative equations, sufficient conditions for efficient truncation are often stated in terms of a norm-based nonlinearity parameter $R$ or $K$ \cite{liu2021efficient,xue2021quantum}. In a discretized Burgers-type problem, such parameters involve ratios of nonlinear derivative scales and dissipative derivative scales. Schematically, one may write a raw grid-dependent quantity as}
\begin{equation}
\bob{R_{N_g}:=K_{N_g}:=\frac{\|u_0^{(N_g)}\|\|\dd_1^{(N_g)}\|}{\|\dd_2^{(N_g)}\|}.}
\end{equation}
\bob{In formulations where the viscosity is included in the dissipative operator, the denominator is replaced by the corresponding norm or spectral scale of $\nu\dd_2^{(N_g)}$. The important point is that such raw quantities are not independent of the grid. For a grid spacing $\Delta x=L/N_g$, the finite-difference matrices satisfy}
\bob{\begin{align}
\|\dd_1^{(N_g)}\|&=O(\Delta x^{-1})=O(N_g/L),\\\|\dd_2^{(N_g)}\|&=O(\Delta x^{-2})=O(N_g^2/L^2).
\end{align}}
\bob{Thus quantities such as $\|\dd_2^{(N_g)}\|/\|\dd_1^{(N_g)}\|$ or $\|\dd_1^{(N_g)}\|/\|\dd_2^{(N_g)}\|$ do depend on $N_g$. The physical rescaling used here does not claim otherwise. Rather, following the physical-resolution argument of Ref.~\cite{gonzalez2024quantum}, the arbitrary choice of $N_g$ is removed by tying the grid to the smallest physical length scale that must be resolved.}

\bob{A turbulent or shock-forming flow contains a range of physical length scales generated by nonlinear interactions. The grid spacing must be sufficiently small to resolve the smallest dynamically relevant scale, denoted here by $\ell_{\rm phys}$. Therefore}
\begin{equation}
\bob{\Delta x\lesssim \ell_{\rm phys},\qquad N_{\rm phys}\sim \frac{L}{\ell_{\rm phys}}.}
\end{equation}
\bob{For the standard one-dimensional Burgers flow, the relevant small scale is the viscous shock width, which plays the role of a one-dimensional Kolmogorov scale. This gives $\ell_{\rm phys}\sim \eta\sim L/Re$, and hence}
\begin{equation}
\bob{N_{\eta}\sim \frac{L}{\eta}\sim Re.}
\end{equation}
\bob{For comparison, the corresponding resolution estimates in two- and three-dimensional turbulence are $N_{2D}\gtrsim Re^{1/2}$ and $N_{3D}\gtrsim Re^{3/4}$, respectively. If an additional physical feature, such as a forcing scale, is smaller than $\eta$, then $\ell_{\rm phys}$ must be replaced by that smaller scale.} \bob{We now define the homotopy nonlinearity parameter using the derivative-operator scales that appear in the PDE. For a fixed grid, the raw operator ratio entering the homotopy Reynolds parameter is
\begin{equation}
\frac{L}{\bar U}
\frac{\|\bar u_0^{(N_g)}\|_\infty\|\dd_2^{(N_g)}\|_\infty}{\|\dd_1^{(N_g)}\|_\infty}.
\end{equation}
This quantity is still grid dependent. To obtain the physically resolved homotopy Reynolds parameter we build in the appropriate rescaling using the grid spacing needed the smallest physical scale that must be resolved. We 
\begin{equation}
Re_{\rm H}
:=
\frac{\|\bar u_0\|_\infty}{\bar U}
\left.
\frac{L\|\dd_2\|_\infty}{\|\dd_1\|_\infty}
\right|_{\Delta x\sim \ell_{\rm phys}} .
\label{eq:ReH_definition}
\end{equation}
Since \(\|\dd_1\|_\infty=O(\Delta x^{-1})\) and \(\|\dd_2\|_\infty=O(\Delta x^{-2})\), this gives
\begin{equation}
Re_{\rm H}
=
O\!\left(
\frac{\|\bar u_0\|_\infty}{\bar U}
\frac{L}{\ell_{\rm phys}}
\right).
\label{eq:ReH_ellphys}
\end{equation}}

\bob{For the one-dimensional Burgers problem with \(\ell_{\rm phys}\sim L/Re\), this reduces to
\begin{equation}
Re_{\rm H}
=
O\!\left(
\frac{\|\bar u_0\|_\infty}{\bar U}Re
\right).
\label{eq:ReH_Burgers_scaling}
\end{equation}
For the normalized Burgers profiles used here, \(\|\bar u_0\|_\infty=O(\bar U)\), and hence \(Re_{\rm H}=O(Re)\). Thus the raw operator ratio depends on the numerical grid if \(N_g\) is varied freely, while \(Re_{\rm H}\) has no arbitrary grid dependence once \(\Delta x\) is fixed by the physically required resolution.}

\bob{We next connect this physical parameter to the homotopy truncation estimate. From Theorem~1, the finite-difference homotopy ratio satisfies}
\bob{\begin{equation}
\bar\rho_{\hat h}^{(N_g)}=\frac{|\hat h|}{2}\|\dd_1^{(N_g)}\|_\infty\|\bar u_0^{(N_g)}\|_\infty t.
\end{equation}}

\bob{Introducing the non-dimensional time \(t^*:=\bar U t/L\), and evaluating the grid at the physical resolution, this becomes, up to finite-difference constants,
\begin{equation}
\bar\rho_{\hat h}
=
O\!\left(
\frac{|\hat h|}{2}t^*Re_{\rm H}
\right).
\label{eq:rho_ReH}
\end{equation}}

\bob{The homotopy convergence condition is not $Re_{\rm H}<1$. Instead, the finite-time deformation accumulated over the integration window must remain smaller than one. Equivalently,}
\bob{\begin{equation}
Re_{\rm H}<\frac{2}{|\hat h|t^*}.
\end{equation}}
\bob{This form makes explicit the distinction from a Carleman-type all-time sufficient condition such as $R_{\rm C}<1$. The homotopy parameter $Re_{\rm H}$ itself may be larger than one; the admissible value increases as the integration window decreases.}

\bob{Using Lemma~2 and Theorem~1, if $0<\bar\rho_{\hat h}<1$, the numerical homotopy truncation error satisfies}
\bob{\begin{equation}
\varepsilon_{\bar\gamma}\leq \frac{\bar\rho_{\hat h}^{M+1}}{1-\bar\rho_{\hat h}}\|\bar u_0\|_{\infty}.
\end{equation}}
\bob{Thus, for fixed $t^*$ and $\hat h$, the error can be reduced geometrically by increasing $M$ as long as the finite-time condition above is satisfied.}
\bob{The corresponding nonlinear time window is}
\bob{\begin{equation}
t^*<\frac{2}{|\hat h|Re_{\rm H}}=:t^*_{\rm NS},\qquad \text{or equivalently}\qquad Re_{\rm H}<\frac{2}{|\hat h|t^*}.
\label{eq:tNS}
\end{equation}}
\bob{In dimensional variables, this gives
\begin{equation}
t_{\rm NS}:=\frac{L}{\bar U}\frac{2}{|\hat h|Re_{\rm H}}.
\end{equation}} 
\bob{For the one-dimensional Burgers resolution scaling used here, $\ell_{\rm phys}\sim L/Re$ and hence $Re_{\rm H}=O((\|\bar u_0\|_\infty/\bar U)Re)$. For the normalized base profile, $\|\bar u_0\|_\infty/\bar U=O(1)$, so $Re_{\rm H}=O(Re)$ and}
\bob{\begin{equation}
t_{\rm NS}=O\!\left(\frac{L}{\bar U|\hat h|}Re^{-1}\right).
\end{equation}}
\bob{Thus increasing the physical Reynolds number decreases the nonlinear time window over which the homotopy truncation is guaranteed to remain in the geometrically convergent regime. This is the sense in which the homotopy construction can operate with $Re_{\rm H}>1$: the restriction is not a direct smallness condition on $Re_{\rm H}$, but a finite-time admissibility condition whose time window scales like $Re_{\rm H}^{-1}$.}

\section{\label{sec:Complexity Analysis and Error Analysis}Error and Complexity Analysis}

\textit{Error Analysis} --- The accuracy of the final solution is determined by the errors introduced from the following three sources.
\begin{enumerate}[wide, labelwidth=!, labelindent=0pt]
 \item \textit{Homotopy Truncation Error} -- \bob{From Theorem~1 and Sec.~VI, the finite-difference homotopy truncation error is controlled by the finite-time condition}
\bob{\begin{equation}
Re_{\rm H}<\frac{2}{|\hat h|t^*}.
\end{equation}}

\bob{Equivalently, using the finite-time homotopy ratio \(\bar\rho_{\hat h}=O(|\hat h|t^*Re_{\rm H}/2)\), the truncation error satisfies}

\bob{\begin{equation}
\varepsilon_{\bar\gamma}:=\|\bar u(t)-\bar u^{(M)}(t)\|_\infty\leq \frac{\bar\rho_{\hat h}^{M+1}}{1-\bar\rho_{\hat h}}\|\bar u_0\|_\infty,\qquad 0<\bar\rho_{\hat h}<1.
\end{equation}}
\bob{Thus $Re_{\rm H}$ itself need not be smaller than one. The corresponding nonlinear integration window is $t_{\rm NS}=(L/\bar U)\,2/(|\hat h|Re_{\rm H})$, which decreases as $Re_{\rm H}$ increases.}
    \item \textit{Finite Difference Error} -- The numerical discretization in space and time of the linearized system of eqs.\ \ref{eq:lcu}, with step sizes $\Delta x$ and $\Delta t$, introduces additional errors that depend on the order of the finite difference approximation. Here, we employ a second-order, central finite difference scheme to approximate ODEs governing each of the $\mathcal{O}(M)$ terms in the truncated homotopy series. For a total of $\tau$ time steps, this yields an overall error (with respect to the \textit{exact} truncated solution) given by
    \begin{equation}
    \bob{\varepsilon_{\rm fd}=\max\left\{O(M\tau\Delta x^2),O(M\tau\Delta t^2)\right\}.}
    \end{equation}
    Furthermore, while implicit schemes are unconditionally stable for any choice of $\Delta t$, explicit schemes have stability restrictions, which require $\Delta t$ to be bounded by the von Neumann criteria. 
    In general, we require \begin{equation}
        \Delta t<\min\Bigg(\frac{-\textrm{Re}(\lambda_{k})}{|\lambda_{k}|^{2}}\Bigg) \approx \frac{1}{\|\mathbf{A}\|_{\infty}}.\label{eq:delta t bound}
    \end{equation} 
\bob{The explicit stability restriction is controlled by $\|A^{(M)}\|_\infty$. Let $U_0:=\|\bar u_0\|_\infty$, and let $\ell_0$ denote the physical variation length of the smooth base profile $\bar u_0$, so that $\|\partial_x\bar u_0\|_\infty=O(U_0/\ell_0)$ and $\|\partial_{xx}\bar u_0\|_\infty=O(U_0/\ell_0^2)$. With $d_1:=\|\dd_1\|_\infty$ and $d_2:=\|\dd_2\|_\infty$, define}
\begin{equation}
\bob{\Theta_M(\hh):=|\hh|\sum_{r=0}^{M-1}|1+\hh|^r.}
\end{equation}
\bob{For the order-$M$ lifted matrix,}
\begin{equation}
\bob{\|A^{(M)}\|_\infty\leq O\!\left(\Lambda_M(\hh)\right),}
\label{eq:A_infty_norm}
\end{equation}
\bob{where}
\bob{
\begin{align}
\Lambda_M(\hh):=\nu d_2+\Theta_M(\hh)\Bigg[\nu d_2&+M\left(1+U_0+\frac{U_0^2}{\ell_0^2}\right)d_1+\nonumber \\ +&M^2\left(\nu+\frac{U_0}{\ell_0}+\frac{U_0^2}{\ell_0}+\frac{U_0^2}{\ell_0^2}\right)\Bigg].
\end{align}}
\bob{The derivation of this bound is given in Appendix~\ref{app:AM_norm}. At the cancellation value $\hh=-1$, $\Theta_M(-1)=1$. For the normalized Burgers base profile used here, $L=1$, $U_0=O(1)$, and $\ell_0=O(1)$. Using the one-dimensional resolution scaling from Sec.~VI, $N_g\sim Re$, $\nu\sim Re^{-1}$, and $\Delta x\sim Re^{-1}$, this gives}
\begin{equation}
\bob{\Lambda_M(-1)=O\!\left[Re+MRe+M^2\right]=O\!\left[\max\{MRe,M^2\}\right].}
\end{equation}
\bob{Therefore the explicit timestep is chosen according to}
\begin{equation}
\bob{\Delta t\lesssim \Lambda_M(\hh)^{-1}.}
\end{equation}
\bob{The finite-difference error for the truncated homotopy system is}
\begin{equation}
\bob{\varepsilon_{\rm fd}=\max\left\{O(M\tau\Delta x^2),O(M\tau\Delta t^2)\right\}.}
\end{equation}
\bob{Using $\tau=t/\Delta t$ and choosing $\Delta t=\Theta(\Lambda_M^{-1})$,}
\begin{equation}
\bob{\varepsilon_{\rm fd}\leq \max\left\{O(Mt\Lambda_M\Delta x^2),O\!\left(\frac{Mt}{\Lambda_M}\right)\right\}.}
\end{equation}
\bob{Under the one-dimensional resolution scaling above,}
\bob{\begin{align}
\varepsilon_{\rm fd}\leq &\max\Bigg\{O\!\left(\frac{Mt\max\{MRe,M^2\}}{Re^2}\right),\nonumber \\ &~~~~~~~~~~~~~~~~~~~~~O\!\left(\frac{Mt}{\max\{MRe,M^2\}}\right)\Bigg\}.
\end{align}}
\bob{In the regime $M \ll Re$ , this simplifies to}
\begin{equation}
\bob{\varepsilon_{\rm fd}=O\!\left(\frac{M^2t}{Re}\right).}
\end{equation}

    \item \textit{Quantum Algorithm (TMCQC) Error} -- The third source of error stems from the quantum algorithm itself, which implements the time marching quantum circuits (TMCQCs). The error in the quantum solution using this method, given by $\varepsilon_{Q} = \|\bar{u}_{\textrm{fd}}(t)-u_{q}\|$, is bounded by $\mathcal{O}(\textrm{polylog}(\epsilon/\varepsilon_{Q}))$ as shown in ref.\ \cite{bharadwaj2024compact}, \bob{consistent with the compact-solver gate complexity stated above}, where $\epsilon\lessapprox 1$ is the factor used in the Richardson extrapolation \cite{bharadwaj2024compact}. The error estimates corresponding to the Richardson extrapolation are restated in Lemma 2 of Appendix \ref{sec: Trunc Neumann Series} for completeness.   
    \bob{Combining the three contributions by the triangle inequality gives the total error between the exact solution $u(t)$ and the quantum output $u_q(t)$ as}
\bob{\begin{align}
\varepsilon_{\rm total}(t)&:=\|u(t)-u_q(t)\|_\infty \\
&\leq \|u(t)-\bar u^{(M)}(t)\|_\infty+\|\bar u^{(M)}(t)-\bar u_{\rm fd}^{(M)}(t)\|_\infty+\nonumber \\&~~~~~~~~~~~~~~~~~~~~~~~~~~~~~~~~~~~~~+\|\bar u_{\rm fd}^{(M)}(t)-u_q(t)\|_\infty \\
&=\varepsilon_{\bar\gamma}(t)+\varepsilon_{\rm fd}(t)+\varepsilon_Q(t).
\label{eq:total_error_decomposition}
\end{align}}
\bob{Using the bounds above, this gives}
\bob{\begin{align}
\varepsilon_{\rm total}(t)
&\leq
\frac{\bar\rho_{\hat h}^{M+1}}{1-\bar\rho_{\hat h}}\|\bar u_0\|_\infty
+\nonumber \\&~~~+
\max\left\{
O(Mt\Lambda_M\Delta x^2),
O\!\left(\frac{Mt}{\Lambda_M}\right)
\right\}
+\varepsilon_Q(t),
\label{eq:total_error_bound}
\end{align}}

\bob{valid under the finite-time admissibility condition $Re_{\rm H}<2/(|\hat h|t^*)$. Under the one-dimensional Burgers resolution scaling and in the regime $M\ll Re$, the finite-difference contribution reduces to $\varepsilon_{\rm fd}=O(M^2t/Re)$, and therefore}
\begin{equation}
\bob{\varepsilon_{\rm total}(t)
\leq
\frac{\bar\rho_{\hat h}^{M+1}}{1-\bar\rho_{\hat h}}
\|\bar u_0\|_\infty+
O\!\left(\frac{M^2t}{Re}\right)+\varepsilon_Q(t).}
\label{eq:total_error_Re_bound}
\end{equation}

\end{enumerate}

\textit{Algorithm Complexity} --- This is primarily dictated by the TMCQC algorithms used here \cite{bharadwaj2024compact} to solve the truncated linear system of ODEs represented by eq.~(\ref{eq:linearized ODE}). For purposes of the present discussion, it suffices to consider the complexities of the TMCQC2 and TMCQC5/6 algorithms \cite{bharadwaj2024compact} that correspond to iterative explicit approach and explicit and implicit one-shot approaches respectively. For an overall quantum accuracy of $\varepsilon_{Q}$ and $N$, which is the size of the overall linearized homotopy system of equations, as mentioned earlier in Section \ref{sec:TMCQC}, the TMCQC circuits have gate complexities of $\mathcal{O}\Big(s\epsilon,\log(N_{g}M),\textrm{polylog}(\epsilon/\varepsilon_{Q}),\tau\Big)$ and $\mathcal{O}\Big(s\epsilon,\log(N_{g}M\tau),\textrm{polylog}(\epsilon/\Gamma\varepsilon_{Q}),\log^{-3}((\kappa-1)^{-1})\Big)$, respectively. For the problem considered here, $|\Gamma|=1$ for both explicit and implicit schemes. Note that, for simplicity, we have set here the accuracy requirements from the Hamiltonian simulation circuit and the truncated Neumann series approximation circuit for matrix inverses, to be equal to an overall quantum solution accuracy of $\varepsilon_{U} = \varepsilon_{N}=\varepsilon_{Q}$. Although these contributions may be strictly different in practice, the overall scaling still remains fundamentally unaltered. \bob{In essence, the algorithmic complexity is again near-optimal in the same sense as defined earlier. The quoted complexity scaling  is therefore conditional on the
following restrictions, each discussed earlier in the manuscript. First, the
homotopy series must be in its convergence regime: the ratio
$\gamma_{\hat h}$ in Lemma~1 must satisfy $0<\gamma_{\hat h}<1$, and the
truncation order $M$ must be chosen so that the homotopy truncation error is below the
prescribed tolerance. In the physically scaled estimate of Sec.~VI, this is equivalently restricted to the nonlinear time window in Eq.~\eqref{eq:tNS}, namely $t^*<2/(|\hat h|Re_{\rm H})$, or $t<t_{\rm NS}=(L/\bar U)\,2/(|\hat h|Re_{\rm H})$ in dimensional variables. For the one-dimensional Burgers scaling $Re_{\rm H}=O(Re)$, this gives $t_{\rm NS}=O(Re^{-1})$, so the guaranteed integration window decreases as the physical Reynolds number increases. Second, the finite-difference
time discretization must remain stable and accurate: for the explicit
scheme, $\Delta t$ must satisfy the von Neumann condition and in particular the estimate
of $\Delta t\leq \|A\|^{-1}_{\infty}$, given by eq.~\eqref{eq:A_infty_norm}; for one-shot or implicit formulations,
the corresponding scaled timestep/history matrix must satisfy the
Neumann-convergence condition stated in Appendix~\ref{sec: Appendix-Exp and Imp Circuits App},
namely $\rho(J)<1$ (or the stronger norm condition $\|J\|<1$ used in the
truncation bound). Third, the total error must satisfy the decomposition
in eq.~\eqref{eq:total_error_bound}, including homotopy truncation,
finite-difference, and quantum-solver errors. Finally, the compact
time-marching solver assumes efficient state preparation and sparse access
to the embedded matrix, and the output must be either the amplitude-encoded
solution or efficiently measurable observables rather than full tomography.
Besides these, it is important to also bound the post-selection or query complexity overhead which we discuss next. }

\textit{Query Complexity and Success Probability} --- The overall complexity of the algorithm is the product of gate complexity and query complexity. The latter depends on the probability of success $p_{\textrm{succ}}$ in measuring the solution subspace in the final quantum state. First, without loss of generality, the given homotopy embedded matrix can be scaled by a factor $\delta\lesssim\mathcal{O}(1/\epsilon)$, the effect of which can be rescaled post-selection (see Proposition 1 in the Appendix \ref{sec: Trunc Neumann Series}). The probability of a successful post-selection would thus be given by 
\begin{equation}
   p_{\textrm{succ}} \sim \Bigg(\frac{2\epsilon\delta\vert\vert  A\vert\Psi\rangle\vert\vert}{\sqrt{\beta}}\Bigg)^{2}.
   \label{eq: p_succ simple LCU}
\end{equation}
Further, we note that successive applications of the LCU for $\tau$ time steps lowers the success probability. This is further amplified by the simultaneous decay in the solution amplitudes, as is natural for dissipative PDEs, due to loss of energy. Therefore the effective probability of success would be given as
\begin{align}
   p_{\textrm{succ}} &= \Bigg(\frac{2\epsilon\delta\vert\vert A\vert\vert}{2}\Bigg)^{2\tau}\prod_{k=1}^{\tau}\frac{\vert\vert~\vert\Psi\rangle_{k}~\vert\vert^{2}}{\vert\vert~\vert\Psi\rangle_{k-1~}\vert\vert^{2}} \nonumber\\ &=  (\epsilon\delta\vert\vert A\vert\vert)^{2\tau}\frac{\vert\vert~\vert\Psi\rangle_{\tau}~\vert\vert^{2}}{\vert\vert~\vert\Psi\rangle_{0}~\vert\vert^{2}} .
\end{align} Therefore the number of shots required scales as
\begin{equation}
    N_{s} = \Bigg(\frac{1}{\epsilon\delta\vert\vert A\vert\vert}\Bigg)^{2\tau}\frac{\vert\vert~\vert\Psi\rangle_{0}~\vert\vert^{2}}{\vert\vert~\vert\Psi\rangle_{\tau}~\vert\vert^{2}} \approx \mathcal{O}\Bigg(\frac{\vert\vert~\vert\Psi\rangle_{0}~\vert\vert^{2}}{\vert\vert~\vert\Psi\rangle_{\tau}~\vert\vert^{2}}\Bigg),
    \label{eq: eta query complexity}
\end{equation} 
although, at the worst, the overall complexity would be scaled linearly by the ratio  $\frac{\vert\vert~\vert\Psi\rangle_{0}~\vert\vert}{\vert\vert~\vert\Psi\rangle_{\tau}~\vert\vert}$. However, the specific ways described in ref.\ \cite{bharadwaj2024compact} can be used to ameliorate this challenge. Together with Richardson extrapolation methods, one can readily choose an $\epsilon$ and a $\delta$ such that $\epsilon\delta\vert\vert A\vert\vert \gtrsim 1$ and, therefore, the prefactor $\eta = (1/\epsilon\delta\vert\vert A\vert\vert)$ can be maintained such that $\eta \lesssim 1$ and the overall query complexity $N_{s}$ can be made into a constant prefactor. For the problems considered here, the challenge of quantum state discrimination, described in ref.\ \cite{lewis2024limitations} does not directly apply; however, for chaotic systems, a similar analysis using the proposed algorithm remains to be explored, forming an integral part of ongoing efforts.

\bob{In short, the overall algorithm exhibits near-optimal scaling in the time-marching system parameters under the assumptions stated above. The parameters $s$ and $\epsilon$ can be further kept near unity using the methods outlined in Ref.~\cite{bharadwaj2024compact}. To maintain this efficiency while achieving a prescribed accuracy, the homotopy truncation must remain in the finite-time accumulation regime}

\bob{\begin{equation}
\bar\rho_{\hh}
=
O\!\left(
|\hh|t^{*}Re_{\rm H}
\right)
<1.
\end{equation}}

\bob{Equivalently, the admissible physically resolved nonlinearity and nonlinear integration window satisfy}
\bob{\begin{align}
&Re_{\rm H}<\frac{2}{|\hh|t^{*}},~~ t^{*}<\frac{2}{|\hh|Re_{\rm H}},~~\rm{therefore}, \\ & t_{\rm NS}=\frac{L}{\bar U}\frac{2}{|\hh|Re_{\rm H}},~~ \tau_{\rm NS}=\frac{t_{\rm NS}}{\Delta t}=\frac{L}{\bar U\Delta t}\frac{2}{|\hh|Re_{\rm H}}.
\end{align}}
\bob{We clarify that the present algorithm does not claim to circumvent a universal lower bound for Carleman linearization. Conditions such as $R_{\rm C}<1$ are sufficient conditions for efficient truncation of the Carleman tensor hierarchy, not universal restrictions on all nonlinear embeddings. After spatial discretization, a quadratic nonlinear system may be written as Eq.~\eqref{eq:quadratic PDE}, and Carleman linearization embeds it into the tensor hierarchy} $
\bob{z_C=(u,u^{\otimes 2},u^{\otimes 3},\ldots,u^{\otimes C}).}$ 
\bob{The present method uses a different approximation space. Instead of truncating a tensor-power hierarchy, we expand the solution along a homotopy path as defined in Eq.~\eqref{eq:deformations}, where $\bar u_0$ is a physically chosen solution of an auxiliary linear problem. The relevant convergence quantity is therefore the decay of successive deformation corrections}
\begin{equation}
\bob{\gamma_p=\frac{|\bar u_{p+1}|}{|\bar u_p|},\qquad \sup_p\gamma_p<1,}
\end{equation}
\bob{rather than the decay of Carleman tensor levels. The auxiliary operator and the convergence-control parameter $\hh$ provide a freedom in choosing this deformation path. Hence regimes for which the Carleman sufficient condition $R_{\rm C}<1$ is not satisfied may still be accessible to the homotopy algorithm over finite time intervals satisfying $Re_{\rm H}<2/(|\hat h|t^*)$. This distinction is important: the homotopy condition does not require $Re_{\rm H}<1$; rather, it allows larger $Re_{\rm H}$ over shorter nonlinear integration windows.}
\bob{This difference has three consequences. First, a physically chosen base solution $\bar u_0$ can absorb part of the dissipative dynamics, boundary conditions, and large-scale flow structure before nonlinear corrections are computed. Second, the raw discrete norm ratios are interpreted after tying the grid to the smallest physically relevant scale, leading to the homotopy nonlinearity parameter $Re_{\rm H}$. Third, the lifted space is not the full tensor algebra} $
\bob{\mathcal C_C=\mathrm{span}\{u^{\otimes k}:1\leq k\leq C\},}
$ \bob{but a pruned, problem-adapted closure generated by the homotopy recursion, schematically}
\begin{equation}
\bob{\mathcal H_M=\mathrm{span}\{\bar u_p,;\bar u_i\bar u_j,;\bar u_i\partial_x\bar u_j,;(\partial_x\bar u_i)(\partial_x\bar u_j),\ldots\}_{i+j\leq M},}
\end{equation}
\bob{with coefficients depending on the known base solution $\bar u_0$ and its derivatives.}
\bob{Finally, the resulting linear time-marching system is solved using the compact time-marching quantum ODE solver of Ref.~\cite{bharadwaj2024compact}. Thus the improvement over Carleman-based approaches comes from a different nonlinear truncation condition, a physically scaled nonlinearity estimate, and a pruned lifted space, rather than from a violation of known Carleman bounds. To concretely understand this, in Appendix \ref{app:carleman-comparison}, we have made a detailed comparative study with respect to a Carleman linearization baseline.}

\begin{figure*}[htpb!]
        \centering
        \subfloat[]{\hspace{-0.2cm}\includegraphics[trim={0.3cm 0.2cm 1.2cm 0.4cm},clip=true,scale=0.62]{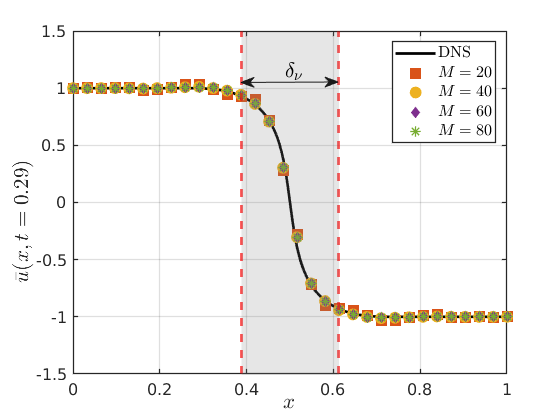}}
        \subfloat[]{\includegraphics[trim={0.2cm 0.1cm 1.0cm 0.4cm},clip=true,scale=0.61]{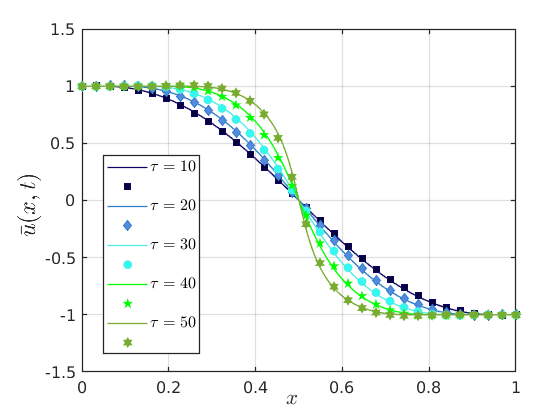}}
        
        \subfloat[]{\includegraphics[trim={0.4cm 0.2cm 1.0cm 0.4cm},clip=true,scale=0.6]{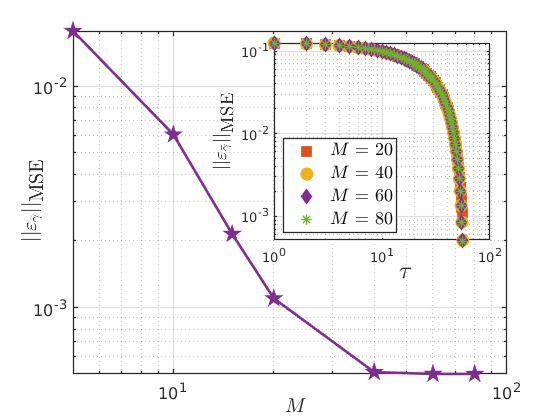}}
        \subfloat[]{\includegraphics[trim={0.1cm 0.1cm 0.1cm 0.1cm},clip=true,scale=0.6]{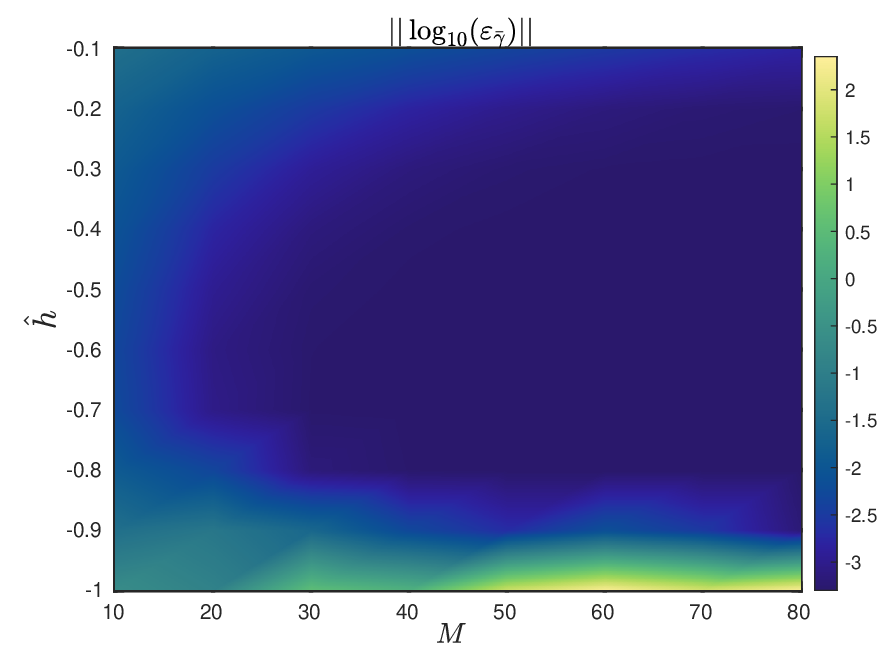}}
                \caption{\justifying (a) Shows the homotopy solutions computed for different orders of truncation (solid symbols) $M=10,20,40,60,80$, for $\hh=-0.5$, $\nu=0.001$, $\Delta t = 5\times10^{-3}$ and $N_{g}=32$. The fully classical DNS solution (black solid line) is also plotted for reference.   (b) Shows the homotopy solutions (solid symbols) of the time evolving flow field with respect to the classical DNS results (solid lines). (c) We plot here, the mean-squared-error of the homotopy solutions as a function of truncation order $M$, which is seen to decay exponentially. Further, the \textit{inset} shows the exponential decay of error, also as a function of time-steps $\tau$, plotted for increasing $M$. (d) Shows the contour plot of the MSE as a function of both $\hh\in [-1,-0.1]$ and $M\in\{10,20,40,60,80\}$, revealing that $\hh=\bob{-}0.5$ tends to yield an overall convergence behavior. However, it is also clear that for higher orders of truncation $M$, smaller $\hh$ yields a better accuracy of the solution. 
                }\label{fig:homtopy time evolution}
\end{figure*}
\section{Numerical Results}
To illustrate the working of the proposed algorithm and analyze the performance, we conduct numerical simulations of the Burgers flow, subject to the constraints defined in \textbf{Problem 1}. The domain is discretized into $N_{g}=32$ grid points using a second-order central finite difference scheme and integrated up to $\tau = \mathcal{O}(100)$ time steps. \bob{The simulations in this section are numerical (classical) direct homotopy-analysis computations, not gate-level simulations of the quantum algorithm. We recursively compute the deformation terms $\{\bar u_p\}_{p=0}^{M}$ on a classical grid and reconstruct the truncated solution $\bar u^{(M)}=\bar u_0+\sum_{p=1}^{M}\bar u_p$. These results are intended to verify the convergence and accuracy of the homotopy expansion that underlies the proposed, lifted quantum embedding.} We simulate the flow at a relatively lower viscosity (higher nonlinearity) of $\nu=0.001$ (one order of magnitude smaller compared to existing works \cite{liu2021efficient,xue2021quantum}). In order to assess the accuracy of the homotopy solution, we compute as reference a well resolved and accurate direct numerical simulation (DNS) of the Burgers flow problem. The results are summarized in figure \ref{fig:homtopy time evolution}, where we observe that homotopy results demonstrate excellent, quantitative agreement with the DNS results. This feature is readily observed in figure \ref{fig:homtopy time evolution}(a), where the homotopy solutions follow the DNS results more and more closely with increasing orders of homotopy truncation $M\in\{20,40,60,80\}$; see figure 3(b). In particular, the solution captures the smallest scale ${\eta}$, represented by the shock of width $\delta_{\nu}$. 

\bob{Using Eq.~\eqref{eq:ReH_definition}, the present simulation has an effective physically resolved \(Re_{\rm H}=O(10^2)\), which is larger than one and corresponds to a regime in which raw Carleman-style sufficient parameters such as \(R\) or \(K\) may exceed the small-nonlinearity bounds used in earlier analyses~\cite{liu2021efficient,xue2021quantum}.} \bob{This does not contradict the homotopy convergence estimate, because the sufficient restriction is finite-time: $Re_{\rm H}<2/(|\hat h|t^*)$. Thus larger physically resolved nonlinearity can be admitted when the integration window is correspondingly shorter.}

To quantify the last statement, we compute the mean-squared-error in the solution with respect to the DNS result, and show them as a function of increasing order of truncation $M$ and time, in figure \ref{fig:homtopy time evolution} (c) and its inset, respectively. The trends indicate that the error converges exponentially with truncation order $M$, and also decays as expected with time $t$, signifying the overall convergence behavior of the flow solutions as well as the exponential efficiency of the algorithm. Another important factor that controls the convergence of the solutions is $\hh$. 
\bob{From the contour plots of Fig.~\ref{fig:homtopy time evolution}(d), $\hh=-0.5$ gives the smallest overall error for the finite values of $M$ considered here. This should be interpreted as an empirical finite-resolution optimum, not as a contradiction of the analytical cancellation argument. In Eq.~\eqref{eq:Mth deformation Burgers}, the two prefactors are $(1+\hh)$ multiplying the inherited term $\partial_t\bar u_{M-1}$ and $\hh$ multiplying the newly generated nonlinear and diffusive correction. At $\hh=-1$, the inherited term is cancelled exactly, which gives a clean sufficient convergence estimate, but the new correction enters with full strength. In the present finite-$M$ Burgers computation this can over-correct the base diffusion solution and increase the discretization-level constants. By contrast, $\hh=-0.5$ gives coefficients of equal magnitude, $(1+\hh)=0.5$ and $\hh=-0.5$, thereby under-relaxing the update: it retains part of the previous deformation while reducing the strength of the newly generated correction. This balance produces the lower observed error for the problem and parameter range considered here.}

\bob{In fact, we go a step further to provide a broader set of PDE benchmarks by extending the above numerical analysis and executing the same algorithmic machinery on three additional nonlinear PDE problems, as outlined in Appendix \ref{app:additional_pde_benchmarks}. These are the Fisher--KPP reaction-- equation, the damped Kuramoto--Sivashinsky equation, and the real Ginzburg--Landau/Allen--Cahn equation. Together, these examples encompass local logistic reaction, cubic nonlinearity and saturation, quadratic transport, and fourth-order dissipative stiffness, thereby probing nonlinear mechanisms that are structurally distinct from those of the Burgers equation. For each case, we employ the same homotopy workflow: a physically motivated base solution is chosen, deformation corrections are retained up to order $M$, and the truncated homotopy solution is compared against an independently computed direct finite-difference reference solution. The reported diagnostics include final-time profile agreement, time-evolution agreement, and the mean-squared error as a function of truncation order. Across all three examples, the error decreases systematically with increasing $M$, and the proposed quantum homotopy algorithm accurately captures the underlying transient physics over the tested finite-time windows. While these benchmarks alone do not settle questions of higher-dimensional scalability or long-time chaotic dynamics, they demonstrate that the proposed embedding is not a Burgers-specific construction and highlight the robustness and generalizability of the approach across a broader class of nonlinear and stiff PDEs.}

In summary, numerical results reveal that the proposed homotopy embedding produces accurate results with a potential of preserving the overall efficiency of the algorithm. A full fledged quantum simulation, involving practical barriers such as noise and decoherence, forms a critical part of an upcoming manuscript, where details of the end-to-end quantum circuit implementation will be discussed.

\section{\label{sec:Conclusions}Summary and Outlook}

We propose a quantum algorithm, based on the concept of homotopy analysis method, to solve dissipative and time-dependent nonlinear flow problems. The proposed approach offers a robust and flexible embedding strategy to handle relatively higher levels of nonlinearity in the PDEs of the kind discussed here, compared to existing methods, \bob{relaxing some existing sufficient bounds in the
tested regime}. These additional benchmarks suggest broader applicability, while arbitrary nonlinearities remain outside the present analytical guarantees. This method is coupled with a near-optimal and end-to-end, time marching QLSA. This yields an overall efficient nonlinear, quantum solver that can perform both iterative and one-shot integration, using implicit and explicit time integration schemes. We take the example of a one-dimensional Burgers equation to discuss the details of the proposed method and analyze its performance. We first outline the embedding strategy and derive the general system of linearized equations. \bob{The embedding avoids an additional Carleman-type
truncation beyond the homotopy-series truncation itself.} We provide rigorous bounds for the truncation error, required orders of truncation, as well as stability and convergence criteria. We also outline the numerical setup and finite difference methods used for integration. We further outline the essentials of the time marching QLSA circuits used here, and followed it by a detailed error and complexity analysis of the algorithm. These results are further bolstered by performing numerical simulations of the problem discussed, revealing an excellent accuracy and performance, both qualitatively and quantitatively in comparison with a classical DNS simulation. \bob{We also define a physically resolved homotopy nonlinearity parameter $Re_{\rm H}$, tied to velocity scales and the smallest resolved physical length scale such as the Kolmogorov or viscous scale $\eta$. The sufficient homotopy condition is finite-time: $Re_{\rm H}<2/(|\hat h|t^*)$. Thus $Re_{\rm H}$ need not be smaller than one; larger physically resolved nonlinearity can be admitted over shorter integration windows. The corresponding nonlinear integration window is $t_{\rm NS}=(L/\bar U)\,2/(|\hat h|Re_{\rm H})$. For the one-dimensional Burgers scaling, $Re_{\rm H}=O(Re)$, so $t_{\rm NS}=O(Re^{-1})$: increasing the physical Reynolds number reduces the guaranteed time interval over which the homotopy truncation remains geometrically convergent.} Finally, given the end-to-end and near-term applicability of the underlying QLSA, the overall algorithm therefore has the potential for simulations of near-term devices. In essence, we propose a near-term quantum algorithm, which is not only an improvement over existing methods, but also provides a practical and near-optimal means for the simulation of nonlinear phenomena on near-term and fault-tolerant quantum devices. 

Certain interesting open questions remain. For instance, (a) computing an upper bound to the level of nonlinearity that one can achieve using the proposed method remains to be computed; (b) a gate-level simulation performance study with noise and decoherence, on real device backend; and (c) exploring the possibility of whether data-driven methods can be used to choose the truncation mechanism to simulate higher nonlinearities, which is an important enterprise to undertake. Besides these, there exists several variations of Homotopy Analysis Methods, that are worth translating into a quantum implementation. For instance, besides iterative homotopy methods, one could consider Boundary Element Methods using homotopy analysis\cite{liao1997boundary,liao2002direct}, where the PDEs are solved on the domain boundaries using a Boundary Integral Equation, thus reducing the dimensionality of the problem. This is analogous to a global, spectral method with inherent advantages in accuracy and resource requirements. These questions are central to ongoing and future investigations.

It is worth remarking that a preliminary version of the results, and methods proposed here, were initially contained in Ref.\cite{bharadwaj2024thesis}. A little later, that work was followed up by another work along similar lines in Ref.\cite{xue2024quantum}. We were made aware of this work during the preparation of this manuscript. Although the work addresses similar problems and is based on the homotopy technique, the underlying algorithmic formulations, efficiency and related results are quite different. In contrast to \cite{xue2024quantum}, the present work: (1) proposes an alternative linearization strategy already discussed in \cite{bharadwaj2024thesis,bharadwaj2023quantum}; this avoids the need for any auxiliary spaces;(2) uses a different and more recent compact QLSA algorithm \cite{bharadwaj2024compact} which is end-to-end and has a near-optimal complexities \bob{(as defined above)} designed for near-term and noisy quantum devices; (3) provides a general equation and methodology to construct arbitrary orders of homotopy equations with provable bounds on $\gamma_{\hh}$ and truncation errors $\varepsilon_{\gamma}$; (4) estimates allowable degree of nonlinearity as rescaled by physical, fundamental scales of fluid flows \cite{gonzalez2024efficient}.\\

\vspace{0.15cm}

\noindent\textbf{Acknowledgements} We thank Jörg Schumacher, Dhawal Buaria, Dylan Lewis, Javier Gonzalez-Conde, Mikel Sanz, Akash Rodhiya and Wael Itani for helpful discussions. We wish to acknowledge New York University's Greene supercomputing facility on which part of these simulations were performed. 
The work was supported partly by the NSF award 24-600, Number 2534977, under the title, TRAILBLAZER Quantum Computing and Machine Learning for Fluid Dynamics Research.

\noindent \textbf{Data Availability} All the data is included in the manuscript.\\

\noindent \textbf{Code Availability} The code will be made available following any reasonable request made to the authors.
\appendix
\section{Solving the one-dimensional diffusion problem}\label{app:analytical solution diffusion}

Let us consider the one-dimensional, diffusion equation given by,
\begin{equation}
    \frac{\partial u}{\partial \bob{t}} = \nu \frac{\partial^{2}u}{\partial x^{2}},
\end{equation}
in the domain $x\in[0,L]$, subject to non-homogeneous Dirichlet boundary conditions given by $u(0,t)=1$ \& $u(L,t)=-1$ and an initial condition $u(x,t)=\cos(\pi x/L)$. The solution proceeds by first decomposing the solution into steady and transient solutions as,
\begin{equation}
    u(x,t) = u_{s}(x,t=\infty) + u_{tr}, \label{eq: steady and trans decomp}
\end{equation}

where $u_{s}$ corresponds to the asymptotic steady state of the problem and $u_{tr}$, is the dynamic transient solution that varies as a function of time. We now solve for each term separately. First we note that $u_{s}$ satisfies,
\begin{equation}
     \frac{\partial u_{s}}{\partial t} = 0~
     \implies \frac{\partial^{2}u_{s}}{\partial x^{2}} = 0. \label{eq: u steady diffusion equation}
\end{equation}
The solution to the eq. (\ref{eq: u steady diffusion equation}) after substituting the boundary conditions, we obtain,
\begin{equation}
    u_{s} = 1-\frac{2x}{L}. \label{eq: u steady}
\end{equation}

Now to estimate the transient solution, we first note that $u_{tr}(x,t)$ satisfies,
\begin{equation}
    \frac{\partial u_{tr}}{\partial t} - \nu \frac{\partial^{2}u_{tr}}{\partial x^{2}} = 0.
\end{equation}
Consequently, this is subject to a homogeneous boundary conditions, $u_{tr}(0,t)=0$ \& $u_{tr}(L,t)=0$, with an initial condition $u_{tr}(x,0)=\cos(\pi x /L)-u_{s}$. We solve this system using the method of separation of variables. This proceeds by splitting $u_{tr}$ into space and time components given as $u_{tr}(x,t) = \bar{X}(x)\cdot\bar{T}(t)$. Now plugging this into the above equation and for some $\lambda\in \mathbb{R}$ we obtain two separate ODEs given by,
\begin{align}
\partial_{t}\bar{T} + \lambda^{2}\nu\bar{T} &= 0, \label{eq: T(t) eq}\\  \partial_{xx}\bar{X} + \lambda^{2}\bar{X} &= 0.\label{eq: X(x) eq}
\end{align}
Integrating the above equations we obtain,
\begin{equation}
\bar{T}(t) = e^{-\frac{4n^{2}\pi^{2}}{L^{2}}\nu t} ~~\textrm{~and~}~~ \bar{X}(x) = b_{n}\sin(2n\pi x/L) \label{eq: T(t) and X(x) solution} 
\end{equation}
where,
\begin{align}
    b_{n} &= \frac{2}{L}\int_{0}^{L}\sin\Bigg(\frac{n\pi x}{L}\Bigg)\Bigg\{ \cos\Bigg(\frac{\pi x}{L}\Bigg) - \Bigg(1-\frac{2x}{L}\Bigg)\Bigg\}\Bigg)dx \\&= \frac{2}{\pi n(4n^{2}-1)}, ~~n\geq1.
\end{align}
 Therefore the transient solution is written as a linear combination over $b_{n} \forall n$ and is given by,
\begin{equation}
    u_{tr}(x,t) = \sum_{n} b_{n}\sin(2n\pi x/L) \label{eq: utr solution}
\end{equation}
We now assemble these partial solutions into eq.(\ref{eq: steady and trans decomp}) to obtain the full solution given below, which thus completes the proof.
\begin{align}
    \bar{u}_{0}(x,t) = \Bigg(1-\frac{2x}{L}\Bigg) &+  \sum_{n=1}^{\infty} \Bigg\{\frac{2}{\pi n(4n^{2}-1)} \times \nonumber \\ &\times\exp\Bigg(-\frac{4\pi^{2}n^{2}\nu}{L^{2}}t\Bigg) \sin\Big(\frac{2n\pi x}{L}\Big)\Bigg\}
\end{align}

\section{Explicit and Implicit Methods}\label{sec: Appendix-Exp and Imp Circuits App}
\subsection{Explicit method - TMCQCs}
\label{subsec: Explicit TMCQC}
(a) \textit{Explicit expansion method}: The explicit forward time integration up to $\tau$ time steps translates to the application of the operator $(A_{E})^{\tau}$. A given matrix is first decomposed into a linear combination of unitaries, and the sum is raised to the power $\tau$ and expanded, yielding a new sum of unitaries with appropriate coefficients.

(b) \textit{Explicit iterative method}: To advance by $\tau$ time steps using the iterative method requires one to iteratively invoke the unitaries circuit to apply the original decomposition 
$\tau$ times. To extend the single step operation outlined {in Section \ref{sec:TMCQC} of the main text}, to perform $\tau$ iterations we introduce an additional ancillary register $q_{a0}$ with $n_{t}=\log(\tau)$ \textit{countdown} qubits. Let us consider a general non-unitary, non-hermitian matrix $J$, and consider for clarity the case for $\tau=2$ shown in figure \ref{fig:lcu circuit}. 
Including the additional ancillary qubit ($n_{t}=1$), with the first application of the LCU operators as before (marked by red-dashed line ($\tau=1$ on figure \ref{fig:lcu circuit}, we are left with the state proportional to 
$\vert1\rangle\frac{1}{\sqrt{\beta}}|0\rangle^{n_{a}}J|\Psi\rangle_{u} + \vert0\rangle|\Psi\rangle_{\perp}$, where
\begin{align}
    &|\Psi\rangle_{\perp} = \frac{1}{\sqrt{\beta}}\Big( \overbrace{(U_{0}+U_{1}+U_{2}+U_{3})}^{\text{J}}|00\rangle +\nonumber\\\nonumber &\underbrace{(U_{0}-U_{1}+U_{2}-U_{3})}_{\text{A}}|01\rangle +\underbrace{(U_{0}+U_{1}-U_{2}-U_{3})}_{\text{B}}|10\rangle \\  &+\underbrace{(U_{0}-U_{1}-U_{2}+U_{3})}_{\text{C}}|11\rangle \Big).
\end{align} The countdown qubits are all set to unity initially as the state $|\tau-1\rangle$ which, after every time step, are reduced by one in binary, using bit flips to finally reach $|0\rangle$, thus having counted $\tau$ time steps. Now we use the ancillary qubit $q_{a0}$ to tag the subspace of the wave function, thus associating it with a time step counter and apply a controlled NOT gate with $q_{a1}$ and $q_{a2}$ (requiring both to be 0), which yields the state
    $\vert1\rangle|0\rangle^{n_{a}}J|\Psi\rangle_{u} + \vert0\rangle|\Psi\rangle_{\perp}$. 
We then reapply the unitaries circuit similar to the first iteration, but this time, all the operations will be controlled additionally by $q_{a0}$ (requiring to be set to 1, marking the solution subspace) as shown in figure \ref{fig:lcu circuit} by the red-dashed line ($\tau=2$). This gives \begin{align}
    \vert \Psi \rangle = &\vert1\rangle\vert00\rangle\frac{1}{\sqrt{\beta'}}J^{2}\vert b\rangle +  \vert1\rangle\overbrace{\frac{1}{\sqrt{\beta'}}\Big(|01\rangle A+|10\rangle B +|11\rangle C \Big)J\vert b\rangle}^{\vert\Psi\rangle_{\perp_{1}}} \nonumber \\ &+  \vert0\rangle\underbrace{\frac{1}{\sqrt{\beta}}\Big(|01\rangle A+|10\rangle B +|11\rangle C \Big)\vert b\rangle}_{\vert\Psi\rangle_{\perp_{2}}}.
\end{align}
Finally we apply a NOT gate on $q_{a0}$ giving the state
    $\vert \Psi \rangle = \vert0\rangle\vert00\rangle\frac{1}{\sqrt{\beta'}}J^{2}\vert b\rangle + \vert1\rangle\vert\Psi\rangle_{\perp_{1}}  + \vert1\rangle\vert\Psi\rangle_{\perp_{2}}$,
and then measure the first three qubits in the computational basis. When one lands an output of (for the first three qubits), $\vert000\rangle$ upon measurement, we are left with a state proportional to $J^{2}\vert b\rangle$.

(c) \textit{Explicit one-shot method}: Alternatively, a matrix inversion problem of the form
\begin{equation}
    \mathbf{A}_{EO}\tilde{u} = \mathbf{b}_{EO}
\end{equation}
can be setup to solve for the velocity field at all time steps together. $\mathbf{A}_{EO}$ has a double-banded structure as written in eq.\ref{eq:FE_OBM}, $(A_{EO})_{ij} = \mathbf{I}~ \forall ~ i=j $ (see below). However, $\forall~ i\leq \tau$, $ (A_{EO})_{ij} = -(\mathbf{I} - \mathbf{A}_{e}) ~\text{for}~ j=i-1 $ and $\forall~ i\geq \tau$,  $ (A_{EO})_{ij} = -\mathbf{I} ~\text{for}~ j=i-1 $. Further, $(b_{EO})_{i} = \{u_{in} \forall ~ i=0 $; $= \mathbf{f}\Delta t 
~\forall~ 0<i\leq \tau$; and $= 0~ \forall ~ i>\tau\}$. Here, $\textbf{f}=0$ identically. Following this, the inverse of matrix $\mathbf{A}_{EO}$ is approximated either as a truncated Neumann series (see later sections, for details on error bounds) or as a truncated Fourier series approach as outlined in \cite{childs2017quantum}, which also provides error bounds on the truncation. Here, we consider the former approach. First, we rewrite the inverse as $(\textbf{I}-(\textbf{I}-\mathbf{A}_{EO}))^{-1}$. 
Then, under the constraint of $\|\textbf{I}-\mathbf{A}_{EO}\|<1$, the matrix inverse is given by
\begin{equation}
    \mathbf{A}^{-1}_{EO} \approx  \sum_{p=0}^{P}(\textbf{I}-\mathbf{A}_{EO})^{p} = \sum_{p=0}^{P}(\tilde{\mathbf{A}}_{EO})^{p}.
\end{equation} 
\begin{figure*}
\begin{equation}\begin{bmatrix}
\mathbf{I} & 0 & \cdots &  &  &  &0\\
-(\mathbf{I} - \mathbf{A}_{e}) & \mathbf{I} &  & &  & 
\\
0 & \ddots & \ddots    &\\
\vdots &  & -(\mathbf{I} - \mathbf{A}_{e}) & \mathbf{I} & &\\
 & & &-\mathbf{I} & \mathbf{I} & &\\
 &  &  &   & \ddots & \ddots\\
0 &  & \cdots & & 0 & -\mathbf{I} & \mathbf{I} \\

\end{bmatrix} \begin{bmatrix}
u^{0}\\u^{1}\\ \vdots\\ u^{\tau}\\ u^{\tau+1}\\ \vdots\\ u^{\tau+\bob{c}}\\ 
\end{bmatrix} = \begin{bmatrix}
u^{in}\\b_{0}\\ \vdots\\ b_{\tau-1}\\ 0\\ \vdots\\ 0\\ 
\end{bmatrix} 
\label{eq:FE_OBM}
\end{equation}
\begin{equation}
\begin{bmatrix}
\mathbf{I} & 0 & \cdots &  &  &  &0\\
-\mathbf{I} & (\mathbf{I} - \mathbf{A}_{i}) &  & &  & 
\\
0 & \ddots & \ddots    &\\
\vdots &  & -\mathbf{I}  & (\mathbf{I} - \mathbf{A}_{i}) & &\\
 & & &-\mathbf{I} & \mathbf{I} & &\\
 &  &  &   & \ddots & \ddots\\
 0&  & \cdots & &0  & -\mathbf{I} & \mathbf{I} \\

\end{bmatrix} \begin{bmatrix}
u^{0}\\u^{1}\\ \vdots\\ u^{\tau}\\ u^{\tau+1}\\ \vdots\\ u^{\tau+\bob{c}}\\ 
\end{bmatrix} = \begin{bmatrix}
u^{in}\\b_{0}\\ \vdots\\ b_{\tau-1}\\ 0\\ \vdots\\ 0\\ 
\end{bmatrix} 
\label{eq:BE_OBM}
\end{equation}
\end{figure*}
\subsection{Implicit method - TMCQCs}
\label{subsec: Implicit TMCQC}

(a) \textit{Implicit iterative method}: This method again, requires one to simply multiply the initial state by the matrix $\mathbf{A}_{NI}$, $\tau$ times iteratively or serially giving
\begin{equation}
    u^{\tau} = (\mathbf{A}_{NI})^{\tau}u^{0},
\end{equation} where the above matrix $\mathbf{A}_{NI}$ is a truncated Neumann series approximation to the inverse of $\mathbf{A}_{I}$, given by
\begin{equation}
    \mathbf{A}_{NI} = \mathbf{A}^{-1}_{I} \approx  \sum_{p=0}^{P}\mathbf{A}^{p} 
\end{equation} computed up to $P$ terms. This approximation is convergent only when $\zeta$ is chosen such that $\|\zeta\mathbf{A} \|<1$.

(b) \textit{Implicit expansion method}: This method proceeds exactly as the explicit expansion method, except that the matrix $\mathbf{A_{NI}}$ is used to compute the expansion.

(c) \textit{Implicit one-shot method}: One can also alternatively set up a matrix inversion problem of the form
\begin{equation}
    \mathbf{A}_{IO}\tilde{u} = \mathbf{b}_{IO},
\end{equation}
to solve for the velocity field at all time steps together. The matrix will be of the form shown in eq.\ref{eq:BE_OBM}, where $(A_{IO})_{ij} = (\mathbf{I} - \mathbf{A}_{i}) ~ \forall ~ i=j $ and $1<i<\tau$. However, $\forall~ i\leq \tau$, $ (A_{IO})_{ij} = -\mathbf{I} ~\text{for}~ j=i-1 $ and $\forall~ i\geq \tau$,  $ (A_{IO})_{ij} = -\mathbf{I} ~\text{for}~ j=i-1 $. Further, $(b_{IO})_{i} = \{u_{in} \forall ~ i=0 $; $= -\mathbf{f}\Delta t 
~\forall~ 0<i\leq \tau $; and $= 0~ \forall ~ i>\tau\}$. Here again, $\textbf{f}=0$ identically. Further the inverse of matrix $\mathbf{A}_{IO}$ is again given by the truncated Neumann approximation as,
\begin{equation}
    \mathbf{A}^{-1}_{IO} \approx  \sum_{p=0}^{P}(\textbf{I}-\mathbf{A}_{IO})^{p} = \tilde{\mathbf{A}}_{IO}
\end{equation} 

\section{Truncated Neumann Series}
\label{sec: Trunc Neumann Series}

The matrix inversion problem of the kind $(\mathbf{I}-\mathbf{J})^{-1}$, under the condition $\rho(\mathbf{J})<1$ (where $\rho$ is the spectral radius), can be rewritten using the Neumann power series as follows:
\begin{equation}
    (\mathbf{I}-\mathbf{J})^{-1} = \sum_{p=0}^{\infty}\mathbf{J}^{p}.
    \label{eq: Neumann series}
\end{equation}
To approximate the inverse using such a summation, we compute a truncated sum of eq.(\ref{eq: Neumann series}) up to $p=P_{min}-1$ terms. Then it can be shown that for Laplacian type operators outlined earlier, the error bound of truncation $\varepsilon_{N}$ is given by \textit{Theorem 1}.

\textbf{Lemma 1} \textit{Consider a central finite difference based, homotopy embedded matrix, $\mathbf{J}\in\mathbb{R}^{N\times N}$, corresponding to a flow field discretized into $N$ grid points, integrated up to time $T$ (with $\tau$ time steps), with an appropriately chosen $\zeta_{\textrm{CFL}}$ such that $\|\mathbf{J}\|<1$. If the inverse $(\mathbf{I}-\mathbf{J})^{-1}$, is approximated by the truncated Neumann series as $\sum_{p=0}^{P_{min}-1}\mathbf{J}^{p}$, the truncation error of $\varepsilon_{N} = \| (\mathbf{I}-\mathbf{J})^{-1} - \sum_{p=0}^{P_{min}-1}\mathbf{J}^{p}\|$ is bounded from above by}

\begin{equation}
    \varepsilon_{N} \leq \mathcal{O}( \| \mathbf{J}\|^{P_{min}}) = \mathcal{O}((\kappa-1)^{P_{min}}),
    \label{eq: eps_N bound}
\end{equation}
\textit{where $\kappa =  \| \mathbf{I}-\mathbf{J}\| \cdot \| (\mathbf{I}-\mathbf{J})^{-1}\| \leq \Gamma$ is the condition number and $P_{min}$ is the number of terms needed to admit a specified $\varepsilon_{N}$ is bounded below accordingly by}

\begin{equation}
    P_{min} = \mathcal{O}\Bigg(\Bigg\lceil\Bigg( \frac{\log(1/\varepsilon_{N})}{\log(1/(\kappa-1))} \Bigg)\Bigg\rceil\Bigg).
    \label{eq: pmin}
\end{equation}

\textbf{Proof -} Starting with the Neumann series representation as in eq.
\ref{eq: Neumann series}, consider a truncated series $\mathbf{R}_{P_{min}}$ computed up to $P_{min}$ terms
\begin{equation}
    \mathbf{R}_{P_{min}} = \mathbf{I}+\mathbf{J}+\mathbf{J}^{2}+\cdots +\mathbf{J}^{P_{min}-1}.
\end{equation}
The truncation error would therefore be given by
\begin{align}
    \varepsilon_{N} &= \| (\mathbf{I}-\mathbf{J})^{-1} - \mathbf{R}_{P_{min}}\| = \| \sum_{p=0}^{\infty}\mathbf{J}^{p} - \sum_{p=0}^{P_{min}-1}\mathbf{J}^{p}\| \nonumber \\ \nonumber
    &=  \| \mathbf{J}^{P_{min}}+\mathbf{J}^{P_{min}+1} + \cdots \| \\ 
    &=  \| \mathbf{J}^{P_{min}}\Big(\sum_{p=0}^{\infty}\mathbf{J}^{p}\Big) \|
    =  \| \mathbf{J}^{P_{min}}(\mathbf{I}-\mathbf{J})^{-1} \| \\
    &\leq  \| \mathbf{J}^{P_{min}}\| \cdot \| (\mathbf{I}-\mathbf{J})^{-1} \|
    \label{eq:product terms}
\end{align}
Consider the first factor in this eq.(\ref{eq:product terms}), given the submultiplicativity of the matrix norm we have
\begin{equation*}
= \| \mathbf{J}^{P_{min}}\|
    \leq \prod_{P_{min}}\| \mathbf{J}\| = \| \mathbf{J}\|^{P_{min}}
    \label{eq:factor1}.
\end{equation*}

For the second factor in eq.(\ref{eq:product terms}), again noting that $\| \mathbf{J}\| \leq \| \mathbf{J}\|_{\infty}$, and additionally by invoking the Varah bound \cite{varah1975lower}, we obtain 
\begin{align}
    &\| (\mathbf{I}-\mathbf{J})^{-1} \|_{\infty} \leq \frac{1}{\Gamma} \textrm{~where,} \nonumber \\ &\Gamma = \min_{i}\Big\{\vert (\mathbf{I}-\mathbf{J})_{i,i}\vert - \sum_{i\neq j}\vert (\mathbf{I}-\mathbf{J})_{i,j}\vert\Big\} \nonumber \\\implies &\| (\mathbf{I}-\mathbf{J})^{-1} \|_{\infty}
    \leq \frac{1}{\Gamma(\zeta_{\textrm{CFL}},\hh,\nu)}
    \label{eq:factor2}.
\end{align}
To compute the bound for $\| \mathbf{J}\|$, we proceed as follows. The resource requirement for a matrix inversion depends on the condition number $\kappa = \| \mathbf{I}-\mathbf{J}\| \cdot \| (\mathbf{I}-\mathbf{J})^{-1}\|$. So let us consider the inequality 
\begin{align}
    \| (\mathbf{I}-\mathbf{J}) \| &\leq \| (\mathbf{I}-\mathbf{J}) \|_{\infty} \nonumber\\
    \frac{\kappa}{\| (\mathbf{I}-\mathbf{J})^{-1}\|}  &\leq \| \mathbf{I} \|_{\infty} + \| \mathbf{J} \|_{\infty} \implies \kappa \leq \mathcal{O}\Bigg(\frac{2}{\Gamma}\Bigg).
    \label{eq:kappa leq 2}
\end{align}
From the above considerations we can see that
\begin{equation}
    1 \leq \kappa - \vert \vert \mathbf{J} \| \leq \frac{2}{\Gamma}-1 \implies \vert \vert \mathbf{J} \| \leq \kappa -1.
\end{equation}
Therefore we get,
\begin{align}
     \varepsilon_{N} &\leq \mathcal{O}\Bigg(\frac{(\kappa-1)^{P_{min}}}{\Gamma}\Bigg) \\\implies P_{min} &=\mathcal{O}\Bigg( \Bigg\lceil\frac{\log(1/\Gamma\varepsilon_{N})}{\log(1/ \vert \vert \mathbf{J}\|)}\Bigg\rceil\Bigg)\\ &= \mathcal{O}\Bigg(\Bigg\lceil\Bigg( \frac{\log(1/\Gamma\varepsilon_{N})}{\log(1/(\kappa-1))} \Bigg)\Bigg\rceil\Bigg).
\end{align}

\textbf{Proposition 1} \textit{A suitable combination of \bob{scaling factor $\zeta$ together with the coefficients of the finite difference operator}, exists such that for both explicit and implicit time schemes, the corresponding finite difference matrix operators $\mathbf{J}$ (or $\mathbf{(I-J)})$ can be re-scaled by a positive real number $\delta \lesssim \mathcal{O}(1/\epsilon)$ such that $Q = \epsilon\delta\| \mathbf{J}\| \gtrsim 1$, while still ensuring $\| \delta \mathbf{J}\| = \delta\| \mathbf{J}\|<1$} and $\zeta<\zeta_{\textrm{cfl}}$ for implicit and explicit schemes, respectively.

\textbf{Lemma 2} \textit{The time marching operators as approximated by a second order, Richardson extrapolated, the block encoding of linear combination of unitaries for (a) TMCQC1,2 (b) TMCQC3,4 and (c) TMCQC5,6 admit an error $\varepsilon = \vert\mathbf{X} - \mathbf{Y}\vert_{\textrm{max}}$,of at most (a) $\mathcal{O}(\tau\epsilon^{4})$, (b) $\mathcal{O}(\tau P^{3}_{min}\epsilon^{4})$ and (c) $\mathcal{O}(P^{2}_{min}\epsilon^{4})$.}

\textbf{Proposition 2} (see \textit{Proposition 9} of \cite{childs2017quantum})
\textit{Given a Hermitian operator $\mathbf{O}$ such that $\|\mathbf{O}^{-1}\| \leq 1$ and $\bar{\mathbf{O}}$ is an approximation of $\mathbf{O}$ such that $\|\mathbf{O}-\bar{\mathbf{O}}\| \leq \varepsilon < 0.5$, the corresponding solution states $\vert\mathbf{u}\rangle = \mathbf{O}\vert\mathbf{u}\rangle/\|\mathbf{O}\vert\mathbf{u}\rangle\|$ and $\vert\bar{\mathbf{u}}\rangle = \bar{\mathbf{O}}\vert\bar{\mathbf{u}}\rangle/\|\bar{\mathbf{O}}\vert\bar{\mathbf{u}}\rangle\|$ satisfy $\|~\|\mathbf{u}\rangle-\vert\bar{\mathbf{u}}\rangle~\| \leq 4\varepsilon$}.

\bob{\section{Bound on $\|A^{(M)}\|_\infty$}\label{app:AM_norm}}

\bob{We derive the bound on $\|A^{(M)}\|_\infty$ used in Sec.~VII. Let $A^{(M)}$ be the lifted matrix generated by the order-$M$ closed homotopy hierarchy. If $A^{(M)}=[A^{(M)}_{ij}]$ is written in block form, where each block has size $N_g\times N_g$, then}
\begin{equation}
\bob{\|A^{(M)}\|_\infty\leq \max_i\sum_j\|A^{(M)}_{ij}\|_\infty.}
\end{equation}
\bob{Let $U_0:=\|\bar u_0\|_\infty$, and let $\ell_0$ be the physical variation length of the prescribed smooth base profile $\bar u_0$. We assume}
\begin{equation}
\bob{\|\partial_x\bar u_0\|_\infty\leq C_1\frac{U_0}{\ell_0},\qquad \|\partial_{xx}\bar u_0\|_\infty\leq C_2\frac{U_0}{\ell_0^2}.}
\end{equation}
\bob{Using the coefficient definitions $\alpha_0=\bar u_0$, $\alpha_1=\partial_x\bar u_0$, $\alpha_3=\partial_x(\bar u_0^2)$, and $\alpha_4=\partial_{xx}(\bar u_0^2)$, this gives}
\begin{align}
\bob{\|\alpha_1\|_\infty}&\bob{\leq C_1\frac{U_0}{\ell_0},}\\
\bob{\|\alpha_3\|_\infty}&\bob{=\|\partial_x(\bar u_0^2)\|_\infty\leq 2C_1\frac{U_0^2}{\ell_0},}\\
\bob{\|\alpha_4\|_\infty}&\bob{=\|\partial_{xx}(\bar u_0^2)\|_\infty\leq 2(C_2+C_1^2)\frac{U_0^2}{\ell_0^2}.}
\end{align}
\bob{The grid dependence enters through the finite-difference matrices. Define}
\begin{equation}
\bob{d_1:=\|\dd_1\|_\infty,\qquad d_2:=\|\dd_2\|_\infty.}
\end{equation}
\bob{For second-order central differences on a grid with spacing $\Delta x=L/N_g$,}
\begin{equation}
\bob{d_1=O(\Delta x^{-1})=O(N_g/L),~~ d_2=O(\Delta x^{-2})=O(N_g^2/L^2).}
\end{equation}
\bob{The estimate for $\|A^{(M)}\|_\infty$ has three structural contributions. First, the diffusive part contributes at most one second-derivative block per lifted equation, giving $O(\nu d_2)$. Second, the $p$-th deformation equation contains at most $O(p)\leq O(M)$ first-derivative convolution blocks. Using the bounds above for the base-profile coefficients, these terms contribute}
\begin{equation}
\bob{O\!\left(M\left(1+U_0+\frac{U_0^2}{\ell_0^2}\right)d_1\right).}
\end{equation}
\bob{Third, differentiating lifted product variables generates zeroth-order gradient-product closure terms. A product containing at most $M$ factors produces at most $O(M^2)$ such pairwise gradient couplings. Their coefficients are controlled by $\nu$, $\|\alpha_1\|_\infty$, $\|\alpha_3\|_\infty$, and $\|\alpha_4\|_\infty$, giving}
\begin{equation}
\bob{O\!\left(M^2\left(\nu+\frac{U_0}{\ell_0}+\frac{U_0^2}{\ell_0}+\frac{U_0^2}{\ell_0^2}\right)\right).}
\end{equation}
\bob{Repeated substitution of lower-order deformation equations introduces inherited factors of $1+\hh$. Their absolute contribution is bounded by}
\begin{equation}
\bob{\Theta_M(\hh):=|\hh|\sum_{r=0}^{M-1}|1+\hh|^r.}
\end{equation}
\bob{Combining the diffusive, first-derivative convolution, product-closure, and inherited homotopy contributions gives}
\begin{equation}
\bob{\|A^{(M)}\|_\infty\leq O\!\left(\Lambda_M(\hh)\right),}
\end{equation}
\bob{where}
\bob{
\begin{align}
\Lambda_M(\hh):=\nu d_2&+\Theta_M(\hh)\Bigg[\nu d_2+M\left(1+U_0+\frac{U_0^2}{\ell_0^2}\right)d_1+\nonumber \\ &~~~~~~~+M^2\left(\nu+\frac{U_0}{\ell_0}+\frac{U_0^2}{\ell_0}+\frac{U_0^2}{\ell_0^2}\right)\Bigg].
\end{align}}
\bob{This proves the bound on $\|A^{(M)}\|_\infty$ used in the main text.}

\section{Richardson Extrapolation}
\label{app:Richardson Extrapolation}
The success probability with the algorithms outlined so far can be enhanced via Richardson extrapolation \cite{schlimgen2021quantum}, which offers an elegant way to reduce the required number of shots. Conversely, it could be used to improve the accuracy for a fixed number of total available shots of the sample the solution. This tool allows us to simulate the unitaries decomposition even for $\epsilon \sim \mathcal{O}(1)$, this being crucial to control the query complexity of the overall algorithm, as discussed in the previous section. The concept of this extrapolation is as follows:  Given an operator $U(\Gamma,\epsilon)$, estimating its output at $\lim{\epsilon\rightarrow 0}$, could be done through extrapolation as shown in \cite{richardson1927viii,schlimgen2021quantum}. From this we can write
\begin{equation}
    U(\Gamma,0) = \frac{U(\Gamma,\epsilon_{1}) - \gamma^{2}U(\Gamma,\epsilon_{2})}{(1-\gamma^{2})},
    \label{eq: richardson U's}
\end{equation}
where $\epsilon_{1}>\epsilon_{2}$ and $\gamma = \epsilon_{1}/\epsilon_{2}$ is the order of extrapolation, where the error of extrapolation scales as $\mathcal{O}(\epsilon^4)$. $\Gamma$ here represents, collectively, any arbitrary set of parameters on which the operator could depend. Higher powers of $\gamma$ lead to higher orders and more accurate extrapolations. As we shall demonstrate later through simulations, even for $\epsilon_{1} \textrm{~and~} \epsilon_{2}$ close to unity, the solution can be computed accurately through extrapolation. This method could also serve as an aid to possible amplitude amplification procedures \cite{berry2014exponential}. The extrapolation procedure can be viewed alternatively as follows. Given a fixed number of shots $N_{s}$, we can extrapolate the solutions that were obtained with $\epsilon_{1,2}$ and $N_{s}$ shots as 
\begin{equation}
     \vert \mathbf{u}\rangle_{0} = \frac{\vert \mathbf{u}\rangle_{\epsilon_{1}} - \gamma^{2}\vert \mathbf{u}\rangle_{\epsilon_{2}}}{(1-\gamma^{2})},
    \label{eq: richardson}
\end{equation} to produce an extrapolated solution with higher accuracy. The procedure can be repeated for higher orders as well, giving more accurate extrapolations. If we compute the gradient of the above expression with respect to $\gamma$, we note that the gradient is maximum at about 1. Therefore $\gamma$ is a number close to but greater than 1; that is, if $\epsilon_{1,2}$ are close to each other and $\epsilon_{1}>\epsilon_{2}$, then the effect of extrapolation is amplified.  Applying this tool in our TMCQC simulations has the benefits of lowering the required $N_{s}$ for a given accuracy, improving the query complexity, and allowing large $\epsilon$ simulations to help distinguish them from the noise on real hardware devices.

\bob{\section{Additional Applications and PDE benchmarks}\label{app:additional_pde_benchmarks}}

\bob{In the main text we have considered the one-dimensional viscous Burgers equation as the primary worked example because it contains the advective quadratic nonlinearity characteristic of flow problems and admits a direct connection between the homotopy nonlinearity parameter $Re_H$ and the physical Reynolds number. To test whether the numerical behavior is specific to Burgers or instead reflects the broader homotopy embedding strategy, we also consider three additional nonlinear PDE benchmarks. These examples are still moderate-scale, one-dimensional tests, but they probe different nonlinear mechanisms and PDE stiffness structures, including logistic reaction, cubic nonlinearity, and fourth-order stiff dissipative dynamics with quadratic transport. In all cases, the homotopy solution is compared against an classically-computed, high-resolution, direct numerical simulation (DNS) result, and the same diagnostics are reported. This includes the final-time solution profile comparison, time-evolution agreement, and mean-squared error as a function of homotopy truncation order $M$ and time $t$.}

\bob{\subsection{Example 1: Fisher--KPP reaction--diffusion equation}}
\label{app:fisher_kpp}
\begin{figure*}[htpb!]
        \subfloat[]{\hspace{-0.2cm}
        \includegraphics[trim={4cm 1cm 1cm 2cm},clip=true,scale=0.17]{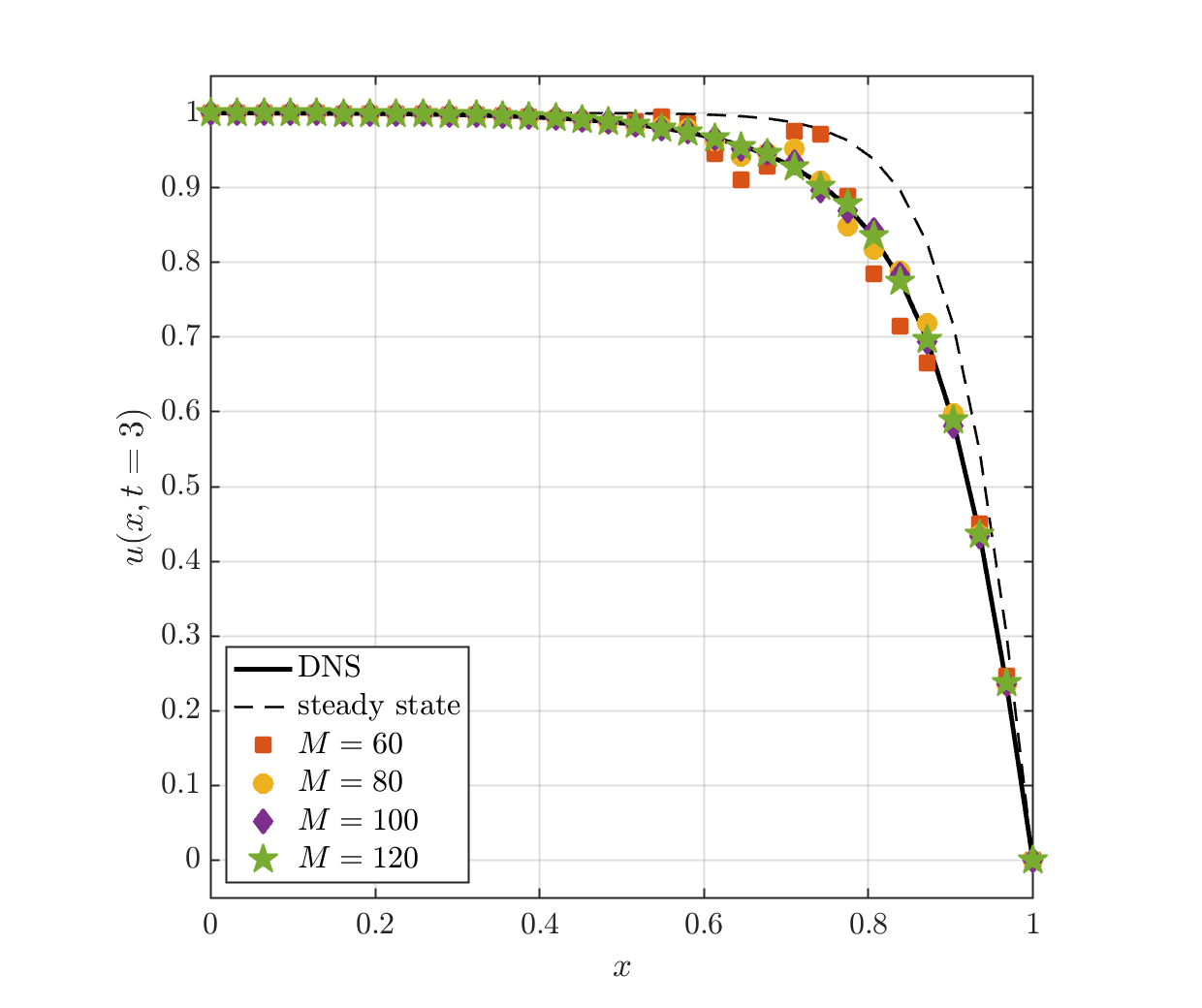}}
        \subfloat[]{\hspace{-0.7cm}\includegraphics[trim={4cm 1cm 1cm 2cm},clip=true,scale=0.17]{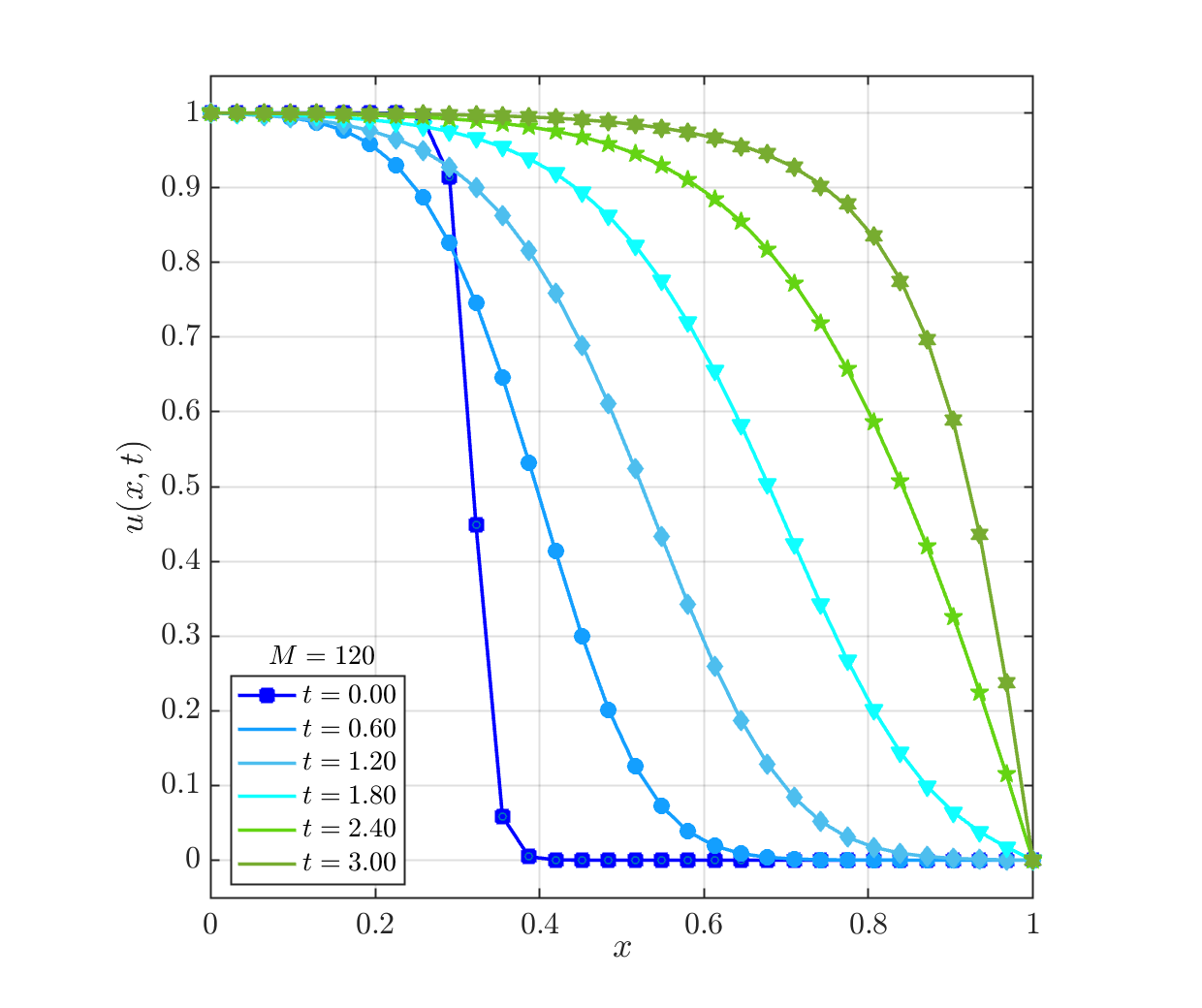}}
        \subfloat[]{\hspace{-0.7cm}\includegraphics[trim={2.4cm 1cm 1cm 2cm},clip=true,scale=0.17]{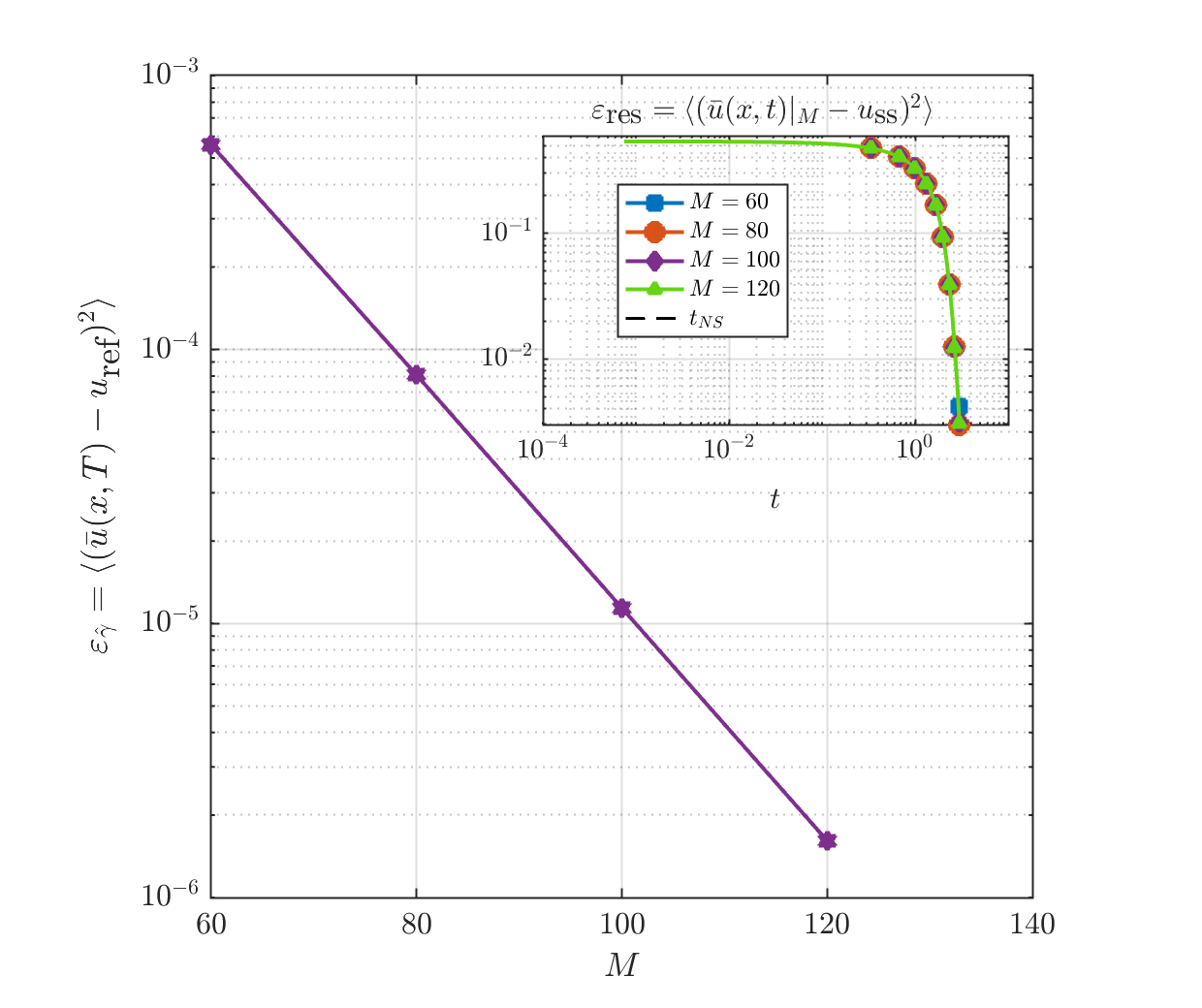}}
        \caption{\justifying \bob{Fisher--KPP reaction--diffusion benchmark is shown here. The equation is solved on $x\in[0,1]$ with Dirichlet data $u(0,t)=1$ and $u(1,t)=0$, using $N=32$, $T=3$, $\nu=10^{-2}$, $r=2.8$, and $\hat h=-0.25$. (a) Final-time profile $u(x,t=T)$ plotted for increasing homotopy truncation orders $M$, compared with the direct nonlinear finite-difference reference solution at t=T. The steady state profile is also plotted (black dashed-line) for reference. The homotopy approximation converges systematically toward the reference as $M$ is increased. (b) The time evolution of homotopy solution profile for $M=120$ is shown. The propagation and relaxation of the reaction--diffusion front over the full time interval $t=[0,T]$. (c) Shows the exponential convergence of the mean-squared error (computed with respect to the reference DNS solution) $\epsilon_H=\langle(\bar u(x,T)-u_{\rm ref}(x,T))^2\rangle$, as a function of increasing truncation orders $M$. The inset on the other hand, shows the residual distance $\epsilon_{\rm res}=\langle(\bar u(x,t)M-u{\rm_{ss}})^2\rangle$ from the steady state over time is shown. }}\label{fig:fisher example}
\end{figure*}
\bob{The first additional benchmark case is the Fisher--Kolmogorov--Petrovsky--Piskunov equation,
\begin{equation}
    \frac{\partial u}{\partial t}=\nu \frac{\partial^2 u}{\partial x^2}+r u(1-u), \qquad x\in[0,1].
\end{equation}
Here $u(x,t)$ is a scalar (NOTE: it is not velocity) and is a \textit{dimensionless}, normalized, concentration or population density, $x$ is the spatial coordinate, $t$ is time, $\nu>0$ is the diffusion coefficient, and $r>0$ is the local reaction or growth rate. The term $\nu u_{xx}$ represents spatial spreading and smoothing by diffusion, while $r u(1-u)$ is the logistic reaction term. This equation is a standard model for reaction--diffusion fronts and appears in population dynamics, combustion-type front propagation, and autocatalytic reaction systems \cite{fisher1937wave,kolmogorov1937study,aronson1975nonlinear}. It is useful here because its nonlinearity is not advective, as in Burgers, but instead arises from a logistic reaction term. The problem therefore tests whether the homotopy embedding remains accurate when the nonlinear term is local in space while the stiffness comes from the competition between diffusion and reaction-driven front motion.}

For the numerical benchmark we impose Dirichlet boundary conditions
\begin{equation}
    u(0,t)=1,\qquad u(1,t)=0,
\end{equation}
and use a front-like initial condition of the form
\begin{equation}
    u(x,0)=\frac{1}{2}\left[1-\tanh\left(\frac{x-x_0}{\delta}\right)\right],
\end{equation}
with the endpoint values reset to satisfy the boundary conditions exactly and $x_{0}=0.32$ and $\delta=0.025$. The results from the numerical simulation of this problem using the homotopy algorithm is presented in figure \ref{fig:fisher example}. 

The reference solution is obtained by a high-resolution, direct numerical simulation (DNS), classical PDE solver of the nonlinear Fisher--KPP equation. The steady profile used in the relaxation diagnostic is obtained by solving
\begin{equation}
    0=\nu u_{ss,xx}+r u_{ss}(1-u_{ss}),\qquad u_{ss}(0)=1,\quad u_{ss}(1)=0.
\end{equation}
\bob{For Fisher--KPP, the natural physical time scale is the reaction time
\begin{equation}
    t_H=\frac{1}{r}.
\end{equation}
The corresponding reaction--diffusion front length scale is
\begin{equation}
    \ell_H=\sqrt{\frac{\nu}{r}}.
\end{equation}
Here \(U_H\) denotes the characteristic dimensionless amplitude of the normalized Fisher--KPP front. Since \(u\) is a normalized concentration or population density, \(U_H\) is not a velocity scale and carries no physical units; for the present front connecting \(u=1\) to \(u=0\), we have \(U_H\approx 1\). Unlike Burgers, the nonlinearity here is local reaction rather than advective transport. Thus the relevant nonlinear scale is \(rU_H^2\), while the diffusive scale across a front of width \(\ell_H\) is \(\nu U_H/\ell_H^2\). Both scales have dimensions of inverse time, so their ratio is dimensionless. We therefore define the Fisher--KPP homotopy Reynolds parameter as
\begin{equation}
    Re^{\rm FKPP}_{\rm H}
    :=
    \frac{rU_H^2}{\nu U_H/\ell_H^2}
    =
    U_H\frac{r\ell_H^2}{\nu}.
\end{equation}
Using \(\ell_H^2=\nu/r\), this gives
\begin{equation}
    Re^{\rm FKPP}_{\rm H}\sim U_H.
\end{equation}
For the normalized front amplitude \(U_H\approx 1\), we obtain
\begin{equation}
    Re^{\rm FKPP}_{\rm H}=O(1).
\end{equation}
The finite-time homotopy condition has the same structure as in the Burgers analysis, but with the reaction time \(t_H\) replacing the advective time scale:
\begin{equation}
    \frac{|\hat h|}{2}Re^{\rm FKPP}_{\rm H}\frac{t}{t_H}<1.
\end{equation}
Hence
\begin{equation}
    t_{\rm NS}
    =
    t_H\frac{2}{|\hat h|Re^{\rm FKPP}_{\rm H}}
    \approx
    \frac{1}{r}\frac{2}{|\hat h|}.
\end{equation}
With the parameters used here, \(r=2.8\) and \(\hat h=-0.25\), this gives \(t_{\rm NS}\approx 2.86\), which lies close to the final simulation time \(T=3\). Thus this benchmark tests the homotopy approximation near the finite-time nonlinear admissibility boundary, but for a reaction--diffusion nonlinearity rather than an advective Burgers nonlinearity.}

\bob{The numerical results shown in figure \ref{fig:fisher example}, depicts a good convergence and accuracy performance of the homotopy solution. In panel figure \ref{fig:fisher example}(a) , the final-time homotopy profiles can be seen to capture accurately, the direct reference solution with increasing $M$. Similarly, figure \ref{fig:fisher example}(b), shows that for the largest chosen homotopy order $M=120$, the homotopy solution accurately tracks the time-evolving front as computed by DNS, over the full simulation time window. To quantify the above observations more carefully, in figure \ref{fig:fisher example}(c), we plot the final-time mean-squared error
\begin{equation}
    \mathrm{MSE}(M;T)=\frac{1}{N_g}\sum_{j=1}^{N_g}\left[\bar{u}^{(M)}(x_j,T)-u_{\mathrm{ref}}(x_j,T)\right]^2,
\end{equation}
which is seen to decay approximately exponentially with $M$. The accompanying time-resolved relaxation diagnostic shown in the inset,
\begin{equation}
    \mathrm{MSE}_{ss}^{(M)}(t)=\frac{1}{N_g}\sum_{j=1}^{N_g}\left[\bar{u}^{(M)}(x_j,t)-u_{ss}(x_j)\right]^2,
\end{equation}
clearly quantifies the expected approach toward the steady front, with the nonlinear time scale $t_{NS}$ marked. This benchmark therefore confirms that the homotopy construction is not limited to advective quadratic nonlinearities and remains effective for reaction--diffusion fronts.}\\

\begin{figure*}[htpb!]
        \subfloat[]{\hspace{-0.2cm}
        \includegraphics[trim={4cm 1cm 1cm 2cm},clip=true,scale=0.17]{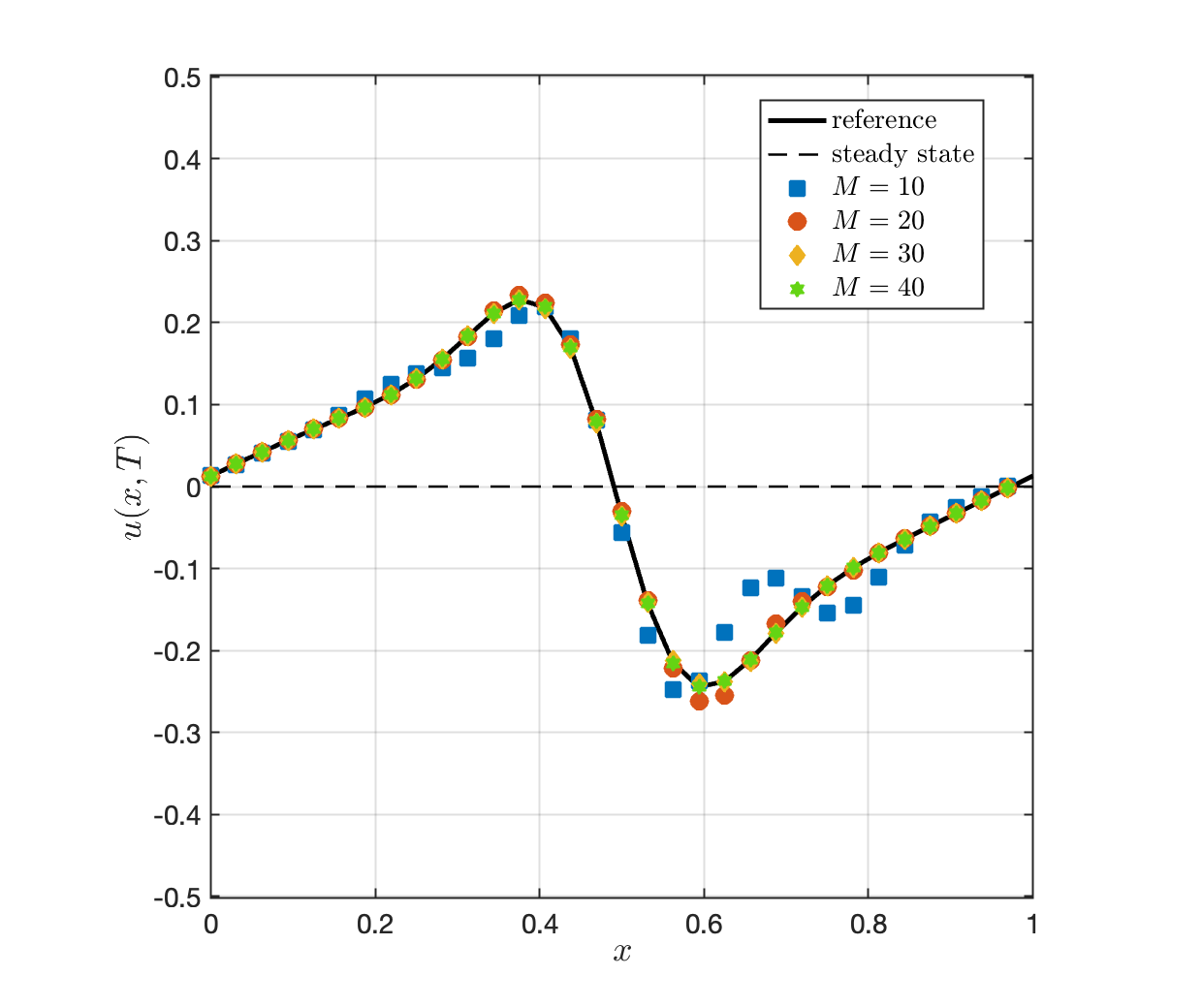}}
        \subfloat[]{\hspace{-0.7cm}\includegraphics[trim={4cm 1cm 1cm 2cm},clip=true,scale=0.17]{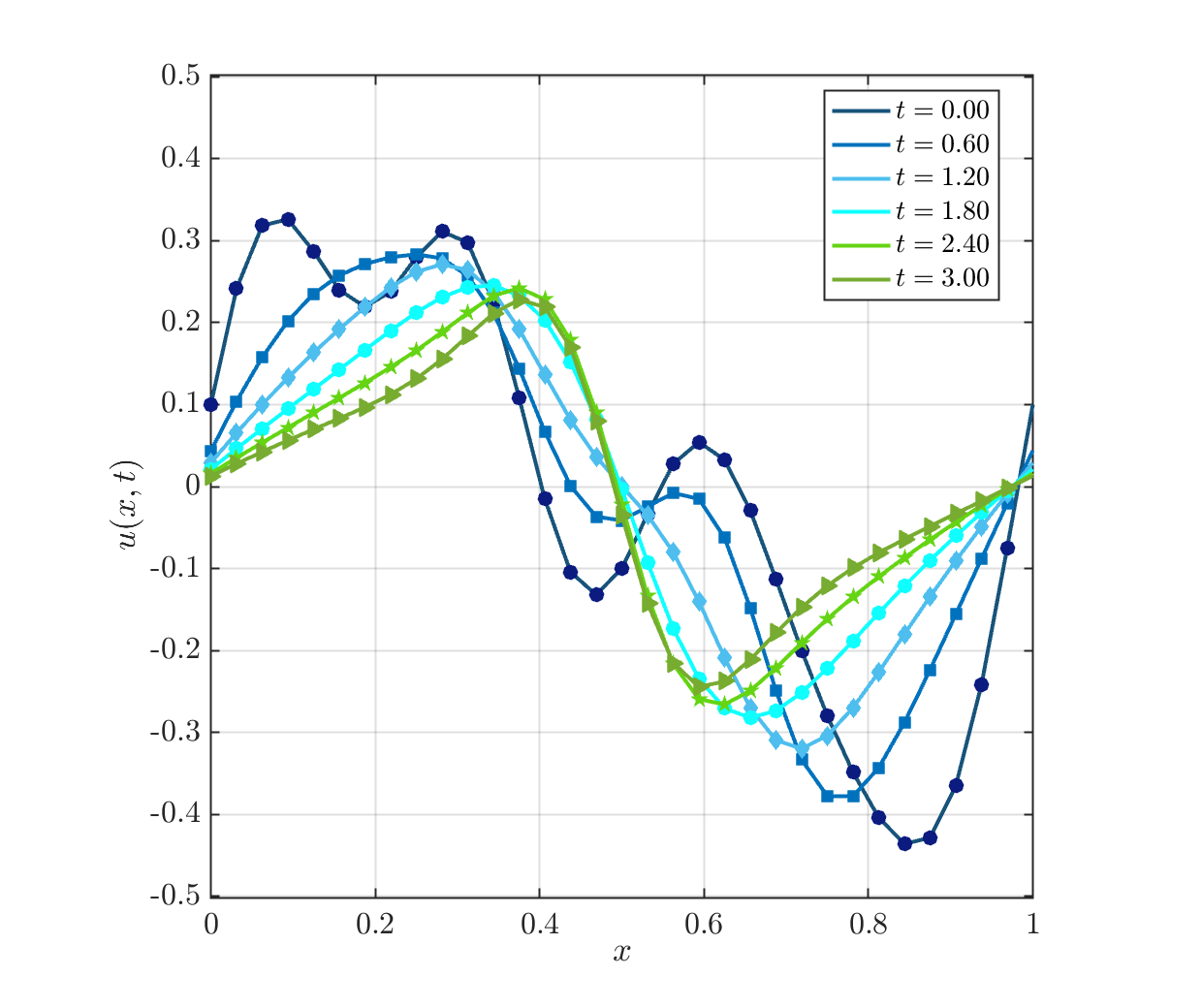}}
        \subfloat[]{\hspace{-0.7cm}\includegraphics[trim={2.4cm 1cm 1cm 2cm},clip=true,scale=0.17]{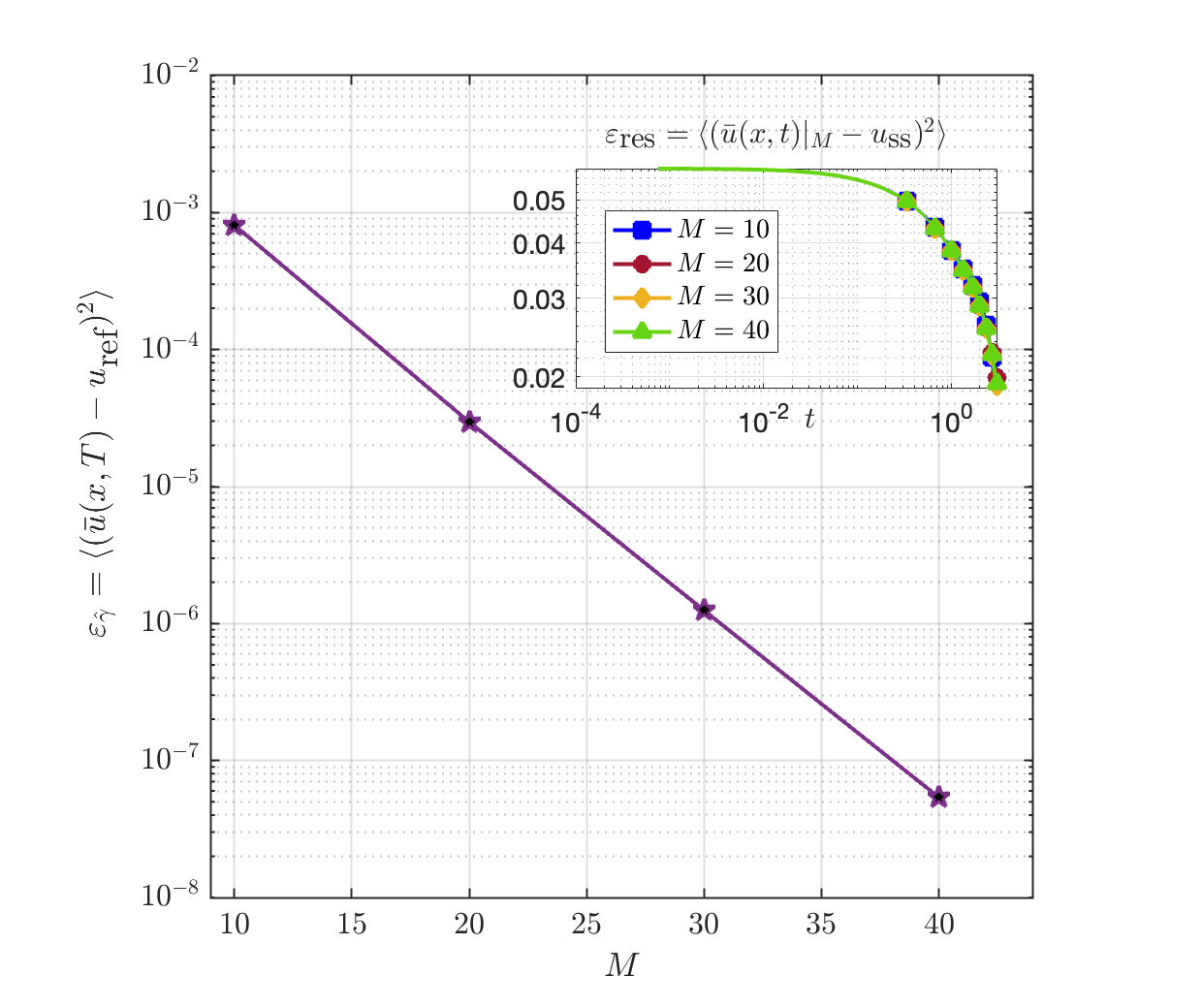}}
                       \caption{\justifying \bob{Damped Kuramoto--Sivashinsky PDE benchmark is shown here. The equation is solved on \(x\in[0,1]\) with periodic boundary conditions, using \(N_g=32\), \(T=3\), \(\tau=4000\), \(a=10^{-3}\), \(b=10^{-5}\), \(\mu=0.15\), and \(\hat h=-0.25\). (a) Final-time profile \(u(x,t=T)\) plotted for increasing homotopy truncation orders \(M\), compared with the direct nonlinear reference solution (black solid-line) at \(t=T\) . The steady state profile \(u_{ss}=0\) is also plotted for reference (black dashed-line). The homotopy approximation converges systematically toward the reference as \(M\) is increased. (b) The time evolution of the homotopy solution profile for the largest retained truncation order \(M=40\) is shown. The transient deformation features and relaxation of the nonlinear fourth-order system is captured over the full time interval \(t=[0,T]\). (c) Shows the exponential convergence of the mean-squared error, computed with respect to the reference DNS solution, \(\epsilon_H=\langle(\bar u(x,T)-u_{\rm ref}(x,T))^2\rangle\), as a function of increasing truncation order \(M\). The inset on the other hand, shows the residual distance \(\epsilon_{\rm res}=\langle(\bar u^{(M)}(x,t)-u_{ss})^2\rangle\) from the steady state over time.}}\label{fig:kuramoto example}
\end{figure*}

\subsection{\bob{Example 2: Damped Kuramoto--Sivashinsky equation}}
\label{app:ks}

\bob{The second additional benchmark case is the damped Kuramoto--Sivashinsky equation,
\begin{equation}
    \frac{\partial u}{\partial t}
    =
    -c_{\rm nl}u\frac{\partial u}{\partial x}
    -
    a\frac{\partial^2u}{\partial x^2}
    -
    b\frac{\partial^4u}{\partial x^4}
    -
    \mu u,
    \qquad x\in[0,L].
\end{equation}
Here \(u(x,t)\) is a scalar field, \(x\) is the spatial coordinate, \(t\) is time, \(c_{\rm nl}\) controls the strength of the quadratic transport nonlinearity, \(a\) controls the destabilizing second-derivative term, \(b>0\) controls the stabilizing fourth-order dissipation, and \(\mu>0\) is a linear damping rate. This equation is useful as a benchmark because it combines a Burgers-type quadratic transport nonlinearity with a stiff fourth-order dissipative term. The usual Kuramoto--Sivashinsky equation is well known for pattern formation, interfacial instabilities, and spatiotemporal dynamics \cite{kuramoto1976persistent,sivashinsky1977nonlinear,sivashinsky1980irregular}. The additional damping term \(-\mu u\) makes the solution relax toward the steady state \(u_{ss}=0\), and therefore makes the time-resolved steady-state diagnostic meaningful.}

\bob{For the numerical benchmark we impose periodic boundary conditions on \(x\in[0,1]\), and use the periodic initial condition
\begin{align}
    u(x,0)
    &=
    0.28\sin(2\pi x/L)
    +
    0.16\sin(4\pi x/L)
    \nonumber\\
    &\quad+
    0.10\cos(6\pi x/L)
    +
    0.06\sin(8\pi x/L).
\end{align}
The results from the numerical simulation of this problem using the homotopy algorithm are presented in figure~\ref{fig:kuramoto example}. The reference solution is obtained by direct time marching of the nonlinear damped Kuramoto--Sivashinsky equation, independent of the homotopy recursion. The steady profile used in the relaxation diagnostic is
\begin{equation}
    u_{ss}(x)=0.
\end{equation}}

\bob{For the damped Kuramoto--Sivashinsky equation, the relevant physical time scale is set by the stabilizing linear rate at the dominant active wavenumber. Writing the linear part as
\begin{equation}
    Lu=-a u_{xx}-b u_{xxxx}-\mu u,
\end{equation}
and considering a dominant wavenumber $k_H$, corresponding to a Fourier mode $\hat{u}_{k_H}(t)=e^{ik_Hx}, $ substituting this in the above linear part, we obtain
the stabilizing linear rate at this scale is
\begin{equation}
    \gamma_H=\mu+bk_H^4-ak_H^2,
\end{equation}
provided \(\gamma_H>0\). Then the corresponding natural ``diffusion" time scale would be
\begin{equation}
    t_H=\frac{1}{\gamma_H}.
\end{equation}
The corresponding active length scale is
\begin{equation}
    \ell_H=\frac{1}{k_H}.
\end{equation}
Unlike Fisher--KPP, the nonlinearity here is not local reaction but quadratic transport. Thus the nonlinear scale is \(c_{\rm nl}U_H^2/\ell_H=c_{\rm nl}U_H^2k_H\), while the stabilizing linear scale is \(\gamma_HU_H\). We therefore define the damped Kuramoto--Sivashinsky homotopy Reynolds parameter as
\begin{equation}
    Re^{\rm KS}_{\rm H}
    :=
    \frac{c_{\rm nl}U_H^2k_H}{\gamma_HU_H}
    =
    \frac{c_{\rm nl}U_Hk_H}{\gamma_H}.
\end{equation}}

\bob{The finite-time homotopy condition has the same structure as in the Burgers and Fisher--KPP analyses, but with the linear relaxation time \(t_H\) replacing the advective or reaction time scale:
\begin{equation}
    \frac{|\hat h|}{2}Re^{\rm KS}_{\rm H}\frac{t}{t_H}<1.
\end{equation}
Hence
\begin{equation}
    t_{\rm NS}
    =
    t_H\frac{2}{|\hat h|Re^{\rm KS}_{\rm H}}
    =
    \frac{2}{|\hat h|c_{\rm nl}U_Hk_H}.
\end{equation}
Thus the nonlinear time window is controlled by the actual quadratic transport rate at the active scale, rather than by a Burgers-type viscosity Reynolds number. In the reported benchmark, the nonlinear coefficient is chosen so that the accumulated nonlinear deformation remains close to the finite-time admissibility boundary while the stiff fourth-order dissipative dynamics remains resolved.}

\bob{The numerical results shown in figure~\ref{fig:kuramoto example}, depicts a good convergence and accuracy performance of the homotopy solution for this stiff fourth-order PDE. In panel figure~\ref{fig:kuramoto example}(a), the final-time homotopy profiles can be seen to capture accurately, the direct reference solution with increasing \(M\). Similarly, figure~\ref{fig:kuramoto example}(b), shows that for the largest chosen homotopy order \(M=40\), the homotopy solution accurately tracks the time-evolving periodic profile as computed by DNS, over the full simulation time window. To quantify the above observations more carefully, in figure~\ref{fig:kuramoto example}(c), we plot the final-time mean-squared error
\begin{equation}
    \mathrm{MSE}(M;T)
    =
    \frac{1}{N_g}
    \sum_{j=1}^{N_g}
    \left[
    \bar u^{(M)}(x_j,T)
    -
    u_{\rm ref}(x_j,T)
    \right]^2,
\end{equation}
which is seen to decay approximately exponentially with \(M\). The accompanying time-resolved relaxation diagnostic shown in the inset,
\begin{equation}
    \mathrm{MSE}^{(M)}_{ss}(t)
    =
    \frac{1}{N_g}
    \sum_{j=1}^{N_g}
    \left[
    \bar u^{(M)}(x_j,t)
    -
    u_{ss}(x_j)
    \right]^2,
\end{equation}
clearly quantifies the expected relaxation toward the zero steady state, with the nonlinear time scale \(t_{\rm NS}\) marked. This benchmark therefore confirms that the homotopy construction remains effective for stiff fourth-order dissipative dynamics coupled to quadratic transport.}

\subsection{\bob{Example 3: Real Ginzburg--Landau / Allen--Cahn equation}}
\label{app:rgl}
\begin{figure*}[htpb!]
        \subfloat[]{\hspace{-0.2cm}
        \includegraphics[trim={4cm 1cm 1cm 2cm},clip=true,scale=0.17]{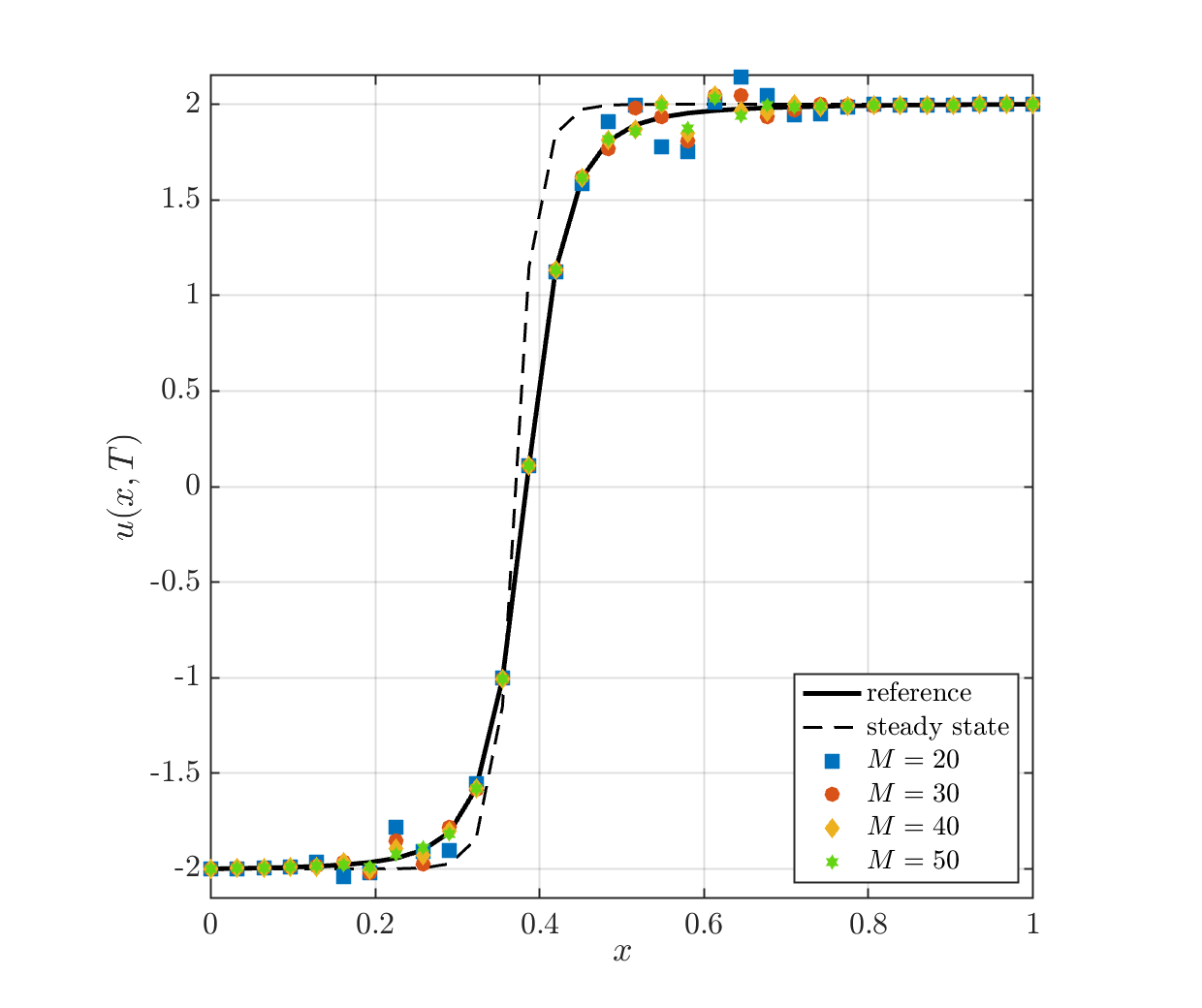}}
        \subfloat[]{\hspace{-0.7cm}\includegraphics[trim={4cm 1cm 1cm 2cm},clip=true,scale=0.17]{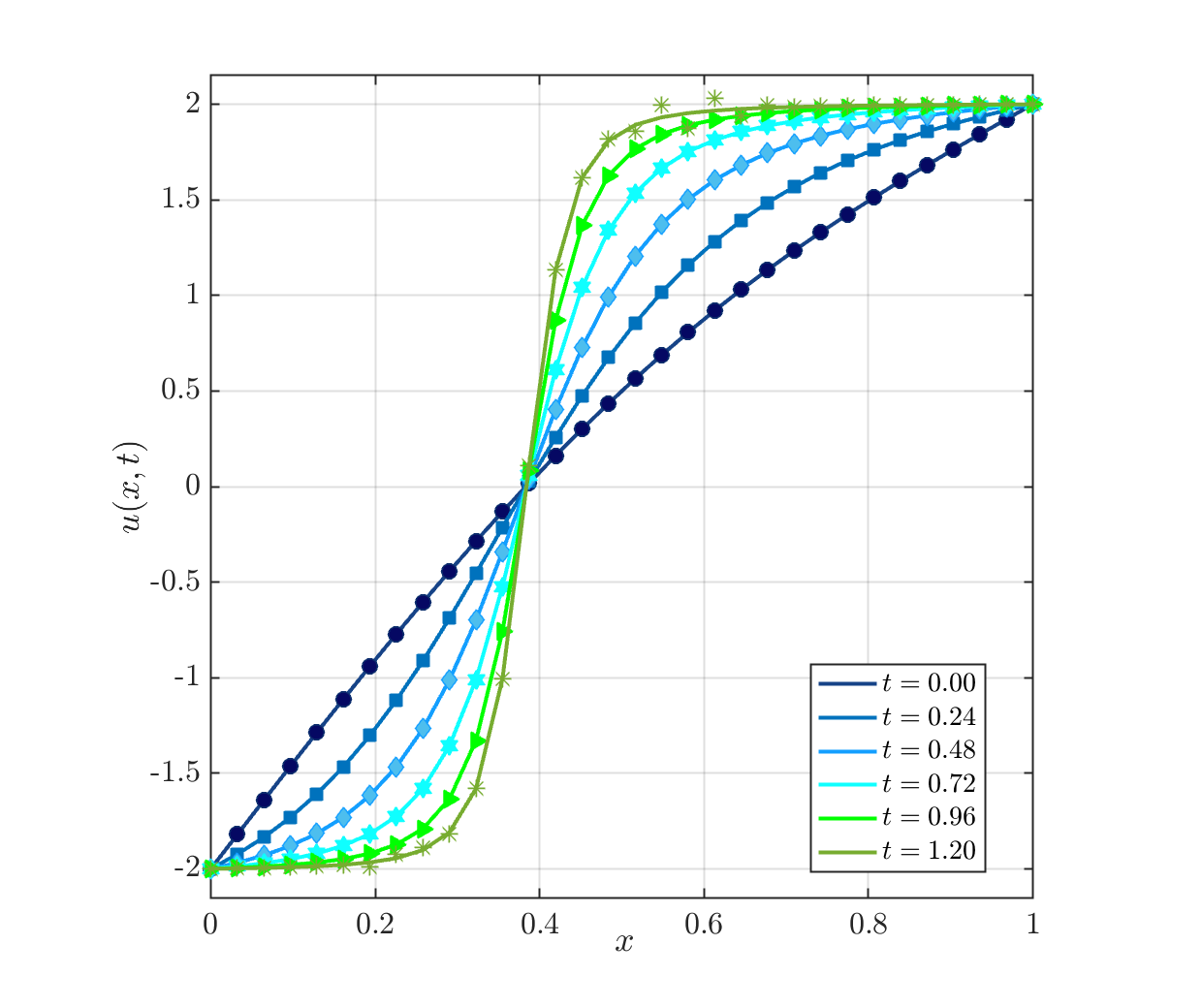}}
        \subfloat[]{\hspace{-0.7cm}\includegraphics[trim={2.4cm 1cm 1cm 2cm},clip=true,scale=0.17]{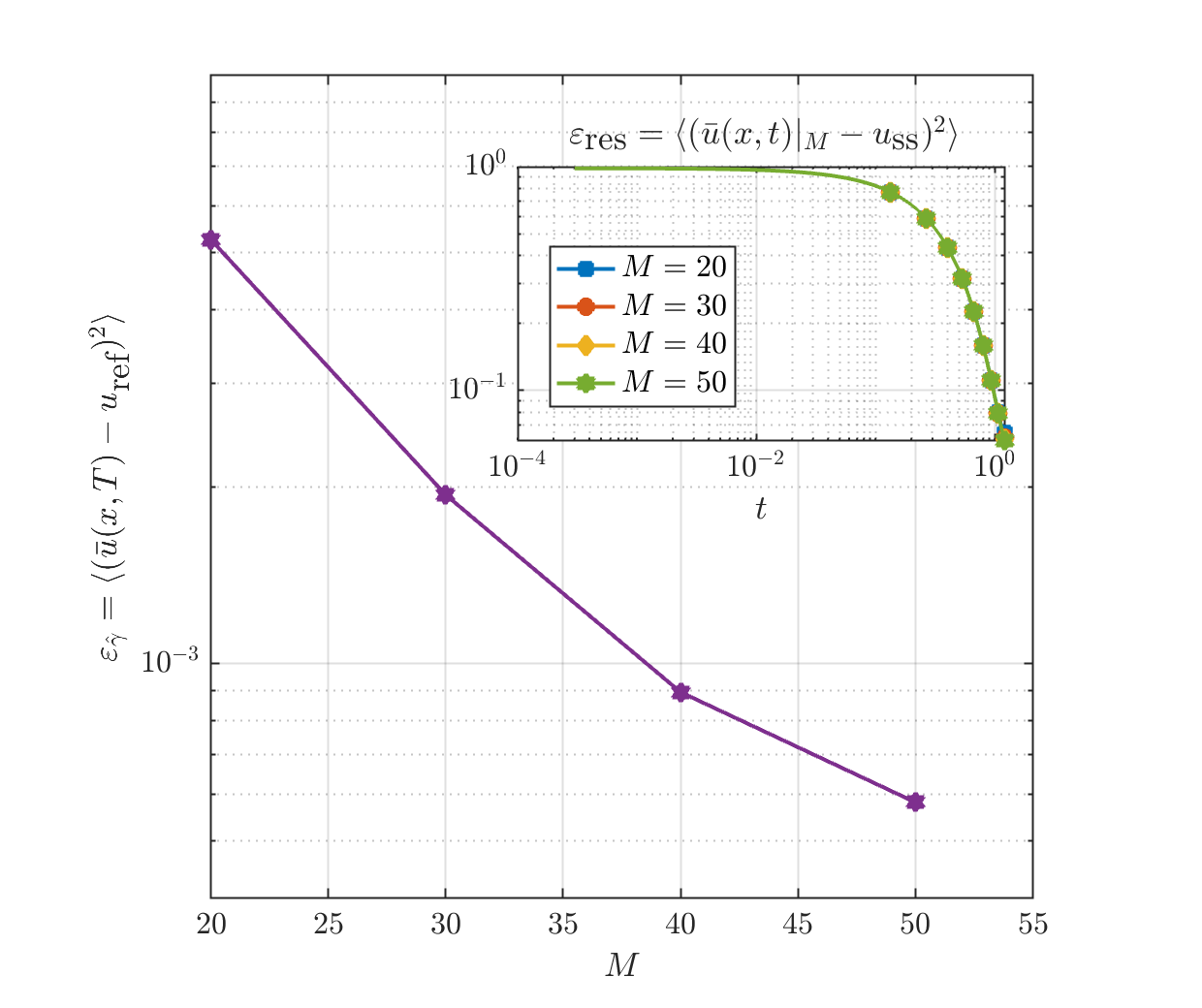}}
        \caption{\justifying \bob{Real Ginzburg--Landau/Allen--Cahn benchmark is shown here. The equation is solved on \(x\in[0,1]\) with Dirichlet data \(u(0,t)=-u_\ast\) and \(u(1,t)=u_\ast\), where \(u_\ast=\sqrt{\alpha/\beta}\), using \(N_g=32\), \(T=1.2\), \(\nu=10^{-3}\), \(\alpha=2\), \(\beta=0.5\), and \(\hat h=-0.25\). (a) Final-time profile \(u(x,t=T)\) plotted for increasing homotopy truncation orders \(M\), compared with the direct nonlinear reference solution at \(t=T\). The steady state profile is also plotted for reference. The homotopy approximation converges systematically toward the reference as \(M\) is increased. (b) The time evolution of the homotopy solution profile for the largest retained truncation order \(M=50\) is shown. The relaxation from the initial condition toward the nonlinear saturated states \(\pm u_\ast\) is captured over the full time interval \(t=[0,T]\). (c) Shows the exponential convergence of the mean-squared error, computed with respect to the reference DNS solution, \(\epsilon_H=\langle(\bar u(x,T)-u_{\rm ref}(x,T))^2\rangle\), as a function of increasing truncation order \(M\). The inset on the other hand, shows the residual distance \(\epsilon_{\rm res}=\langle(\bar u^{(M)}(x,t)-u_{ss})^2\rangle\) from the steady state over time.}}\label{fig:rgl example}
\end{figure*}
\bob{The third benchmark application is that of the real Ginzburg--Landau equation, also commonly written in Allen--Cahn form,
\begin{equation}
    \frac{\partial u}{\partial t}
    =
    \nu\frac{\partial^2u}{\partial x^2}
    +
    \alpha u
    -
    \beta u^3,
    \qquad x\in[0,1].
\end{equation}
Here \(u(x,t)\) is a scalar order parameter, \(x\) is the spatial coordinate, \(t\) is time, \(\nu>0\) is the diffusion coefficient, \(\alpha>0\) is the linear growth rate, and \(\beta>0\) controls the cubic saturation. This equation is a canonical model for bistable relaxation, phase ordering, and domain-wall dynamics \cite{allen1979microscopic,bray1994theory,cross1993pattern}. It is useful here because its nonlinearity is cubic and saturating, and is therefore structurally distinct from the quadratic advective nonlinearity in Burgers, the logistic reaction nonlinearity in Fisher--KPP, and the stiff fourth-order transport problem above. Although, our discussion and calcualtions in the main text pertain to quadratic and dissipative nonlinear PDEs, we go a step further to explore (at least numerically and in the absence of theoretical guarantees) whether or not, the proposed algorithms performs well. One way to extract a suitable scale in this case is by considering the steady and spatially uniform state ($u_t=0; u_{xx}=0)$ equilibrium state, which yeilds
\begin{equation}
    u_\ast=\sqrt{\frac{\alpha}{\beta}},
\end{equation}
and the equation supports a sharp, shock-like profiles connecting boundary values that may be set to \(-u_\ast\) and \(u_\ast\).}
\bob{Therefore, for the numerical benchmark we impose Dirichlet boundary conditions
\begin{equation}
    u(0,t)=-u_\ast,\qquad u(1,t)=u_\ast.
\end{equation}
To avoid an artificially easy test, the initial condition is chosen away from a steady shock-like profile. A representative choice is the broad ramp-like profile
\begin{equation}
    u(x,0)=u_L+(u_R-u_L)x+A\sin(\pi x),
\end{equation}
with the endpoint values reset to \(u_L=-u_\ast\) and \(u_R=u_\ast\). The results from the numerical simulation of this problem using the homotopy algorithm are presented in figure~\ref{fig:rgl example}. The reference solution is obtained by direct time marching of the nonlinear real Ginzburg--Landau equation, independent of the homotopy recursion. The steady profile used in the relaxation diagnostic is obtained by solving
\begin{equation}
    0
    =
    \nu u_{ss,xx}
    +
    \alpha u_{ss}
    -
    \beta u_{ss}^3,
    \qquad
    u_{ss}(0)=-u_\ast,\quad
    u_{ss}(1)=u_\ast.
\end{equation}} 
\bob{For the real Ginzburg--Landau/Allen--Cahn equation, the natural physical time scale is the reaction time
\begin{equation}
    t_H=\frac{1}{\alpha}.
\end{equation}
The corresponding reaction--diffusion interfacial length scale is
\begin{equation}
    \ell_H=\sqrt{\frac{\nu}{\alpha}}.
\end{equation}
Unlike Burgers, the nonlinearity here is local cubic reaction rather than advective transport. Thus the relevant nonlinear scale is \(\beta U_H^3\), while the diffusive scale across an interface of width \(\ell_H\) is \(\nu U_H/\ell_H^2\). We therefore define the real Ginzburg--Landau homotopy Reynolds parameter as
\begin{equation}
    Re^{\rm RGL}_{\rm H}
    :=
    \frac{\beta U_H^3}{\nu U_H/\ell_H^2}
    =
    \frac{\beta U_H^2\ell_H^2}{\nu}.
\end{equation}
Using \(\ell_H^2=\nu/\alpha\), this gives
\begin{equation}
    Re^{\rm RGL}_{\rm H}
    =
    \frac{\beta U_H^2}{\alpha}.
\end{equation}
For the normalized saturated amplitude \(U_H=u_\ast=\sqrt{\alpha/\beta}\), we obtain
\begin{equation}
    Re^{\rm RGL}_{\rm H}=O(1).
\end{equation}}

\bob{The finite-time homotopy condition has the same structure as in the Burgers and Fisher--KPP analyses, but with the reaction time \(t_H\) replacing the advective time scale:
\begin{equation}
    \frac{|\hat h|}{2}Re^{\rm RGL}_{\rm H}\frac{t}{t_H}<1.
\end{equation}
Hence
\begin{equation}
    t_{\rm NS}
    =
    t_H\frac{2}{|\hat h|Re^{\rm RGL}_{\rm H}}
    \approx
    \frac{1}{\alpha}\frac{2}{|\hat h|}.
\end{equation}
With the parameters used here, \(\alpha=2\) and \(\hat h=-0.25\), this gives \(t_{\rm NS}\approx 4\), which lies well beyond the final simulation time \(T=1.2\). Thus this benchmark tests the homotopy approximation inside the finite-time nonlinear admissibility boundary, but for a cubic reaction--diffusion nonlinearity rather than logistic reaction or advective transport.}

\bob{The numerical results shown in figure~\ref{fig:rgl example}, depicts a good convergence and accuracy performance of the homotopy solution. In panel figure~\ref{fig:rgl example}(a), the final-time homotopy profiles can be seen to capture accurately, the direct reference solution with increasing \(M\). Similarly, figure~\ref{fig:rgl example}(b), shows that for the largest chosen homotopy order \(M=50\), the homotopy solution accurately tracks the time-evolving relaxation toward the saturated kink profile as computed by DNS, over the full simulation time window. To quantify the above observations more carefully, in figure~\ref{fig:rgl example}(c), we plot the final-time mean-squared error
\begin{equation}
    \mathrm{MSE}(M;T)
    =
    \frac{1}{N_g}
    \sum_{j=1}^{N_g}
    \left[
    \bar u^{(M)}(x_j,T)
    -
    u_{\rm ref}(x_j,T)
    \right]^2,
\end{equation}
which is seen to decay approximately exponentially with \(M\). The accompanying time-resolved relaxation diagnostic shown in the inset,
\begin{equation}
    \mathrm{MSE}^{(M)}_{ss}(t)
    =
    \frac{1}{N_g}
    \sum_{j=1}^{N_g}
    \left[
    \bar u^{(M)}(x_j,t)
    -
    u_{ss}(x_j)
    \right]^2,
\end{equation}
clearly quantifies the expected approach toward the steady kink profile, with the nonlinear time scale \(t_{\rm NS}\) marked. This benchmark therefore confirms that the homotopy construction is not limited to quadratic or logistic nonlinearities and remains effective for cubic reaction--diffusion saturation.}

\subsection{\bob{Summary of the additional PDE benchmarks}}
\label{app:benchmark_summary}

\bob{Together, these additional benchmarks broaden the numerical evidence beyond the Burgers equation and demonstrates the wider applicability of the proposed algorithm. Fisher--KPP tests a logistic reaction--diffusion front, the damped Kuramoto--Sivashinsky equation tests quadratic transport coupled to fourth-order stiffness, and the real Ginzburg--Landau equation tests cubic saturation and kink relaxation. In each case, the same homotopy workflow is used -- a physically meaningful base solution is chosen, deformation corrections are computed up to order $M$, and the truncated solution is compared against an independently computed, direct numerical simulation solution.}

\bob{The results show consistent convergence with truncation order across different nonlinear structures, accurately capturing the physics of the problem. The final-time MSE decreases with $M$ in all three examples, while the time-evolution plots show that the method captures the transient dynamics rather than merely matching the final profile. The time-resolved steady-state diagnostics provide an additional physical relaxation check and place the observed behavior in relation to the nonlinear time scale $t_{NS}$. These tests do not by themselves settle all questions of higher-dimensional scalability or chaotic long-time dynamics, but they demonstrate that the proposed homotopy embedding is not a Burgers-specific construction. Instead, the same finite-time admissibility idea and truncation diagnostics remain effective across reaction, cubic saturation, and stiff fourth-order nonlinear PDE benchmarks.}

\bob{\section{Carleman linearization baseline comparison for Burgers equation}\label{app:carleman-comparison}}

\begin{figure*}[htpb!]
        \subfloat[]{
        \includegraphics[trim={1cm 0.5cm 1cm 2cm},clip=true,scale=0.165]{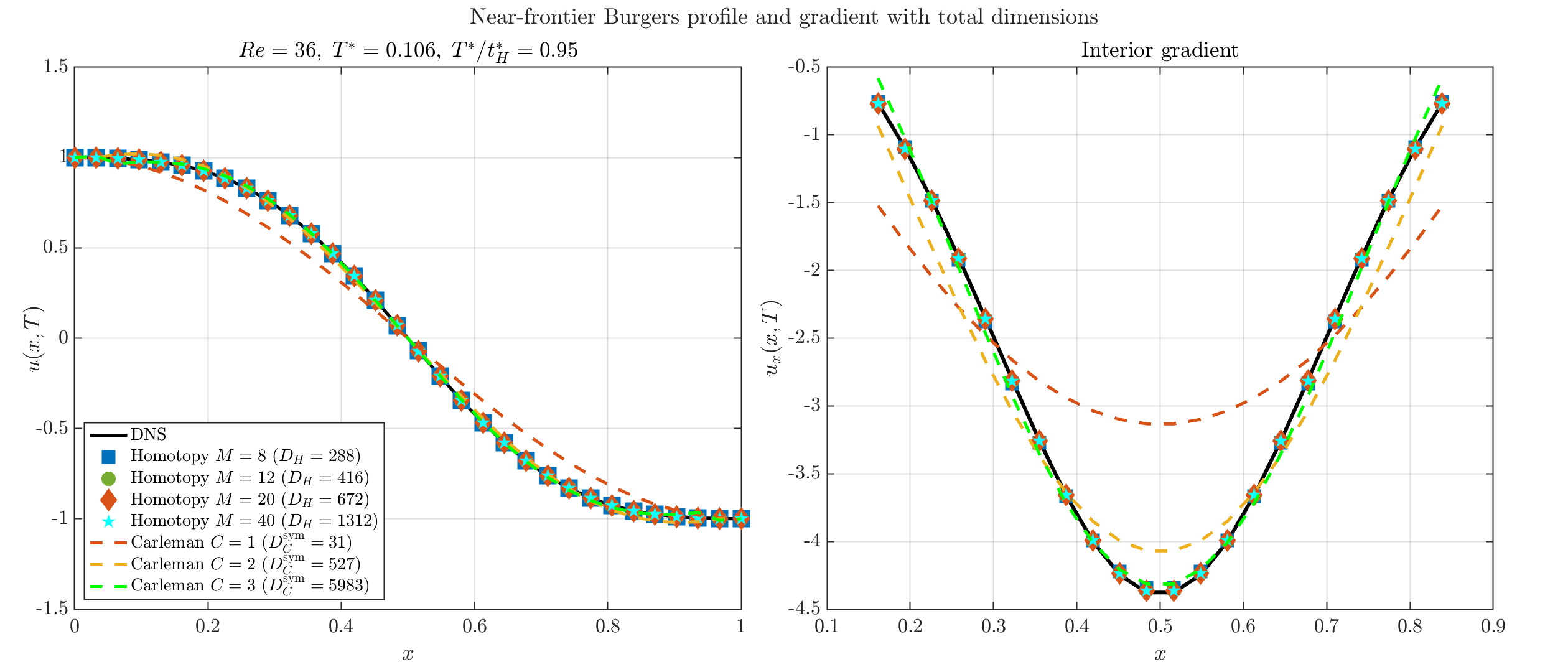}}
        
        \subfloat[]{\includegraphics[trim={0.2cm 0.2cm 1cm 2cm},clip=true,scale=0.185]{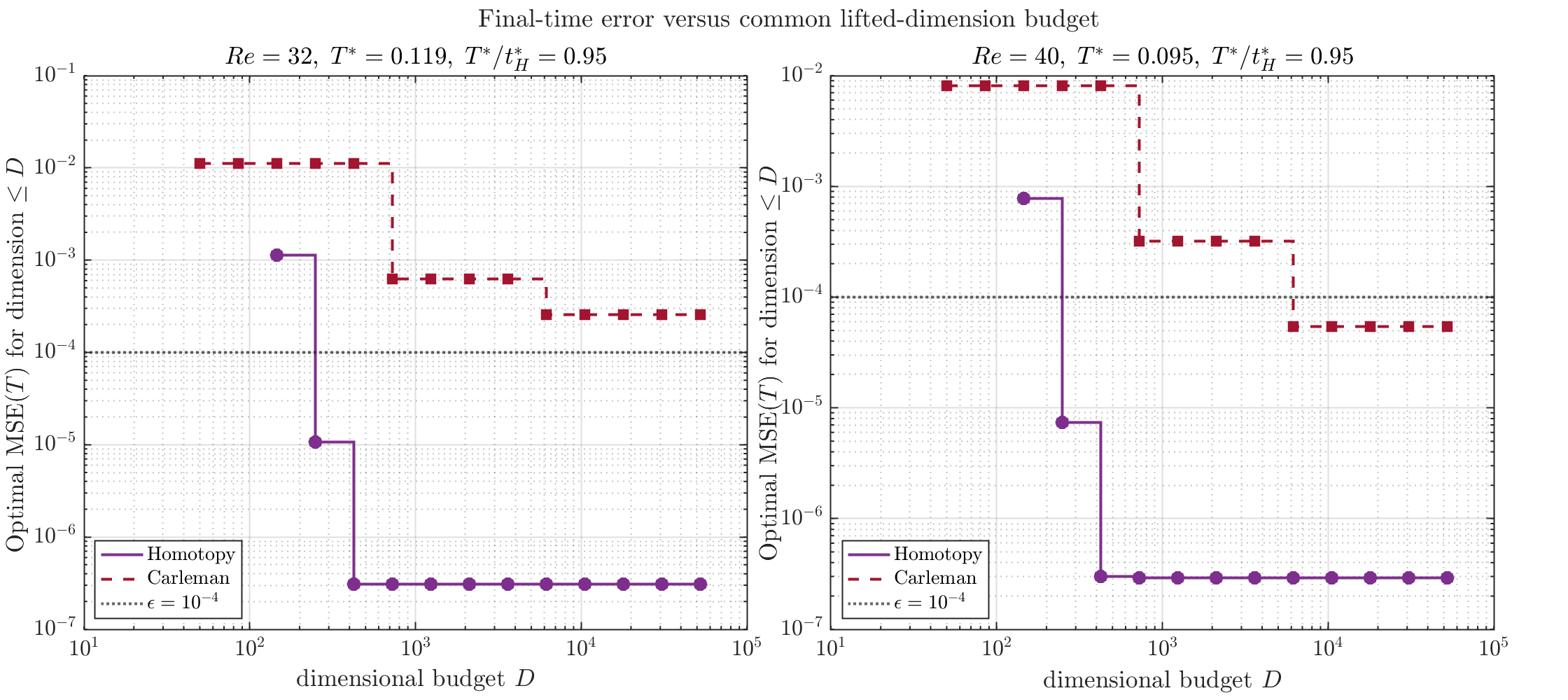}}
        
        \subfloat[]{\hspace{-0.15cm}\includegraphics[trim={1.3cm 0.5cm 3.2cm 0.8cm},clip=true,scale=0.19]{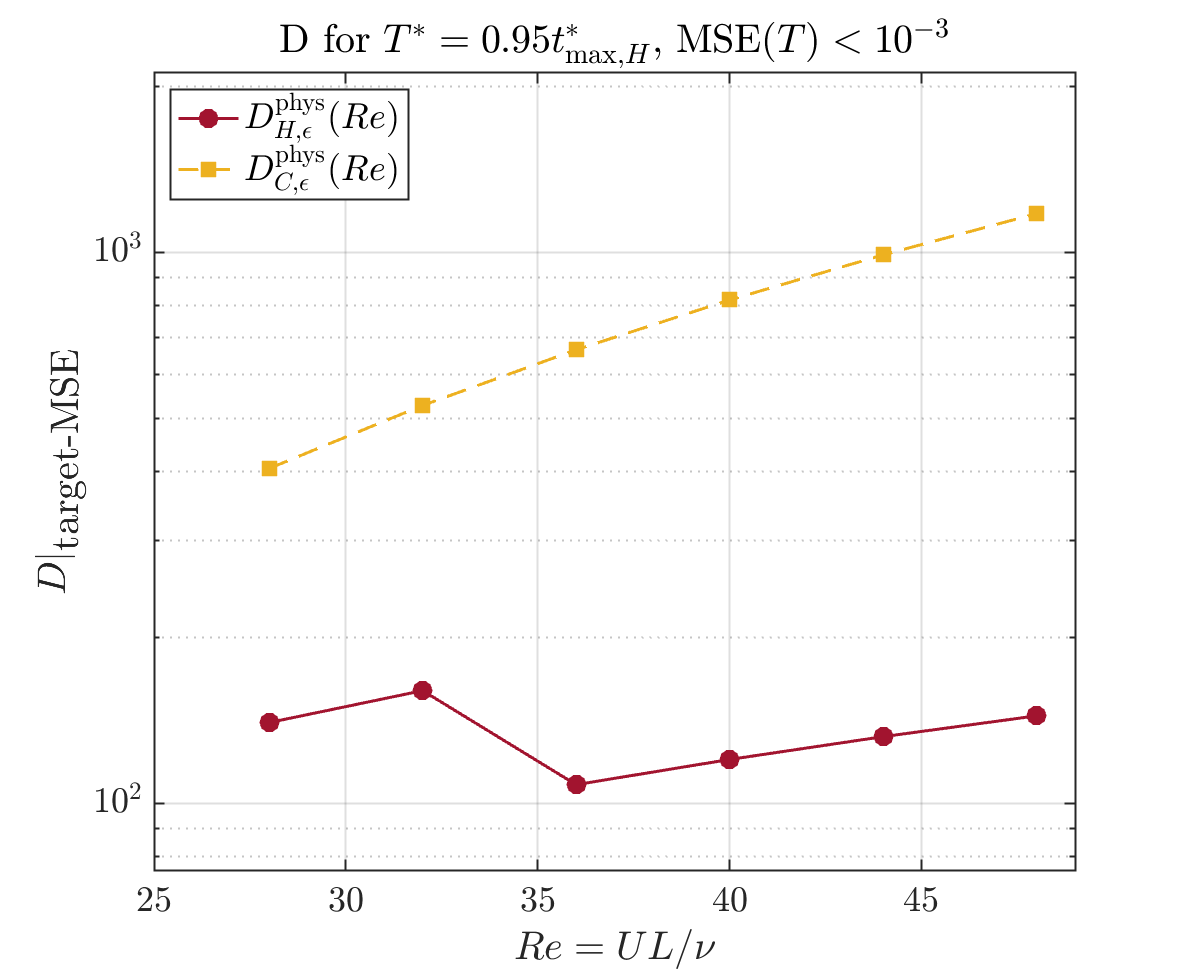}}
        \subfloat[]{\includegraphics[trim={0.6cm 0.5cm 3.2cm 0.8cm},clip=true,scale=0.19]{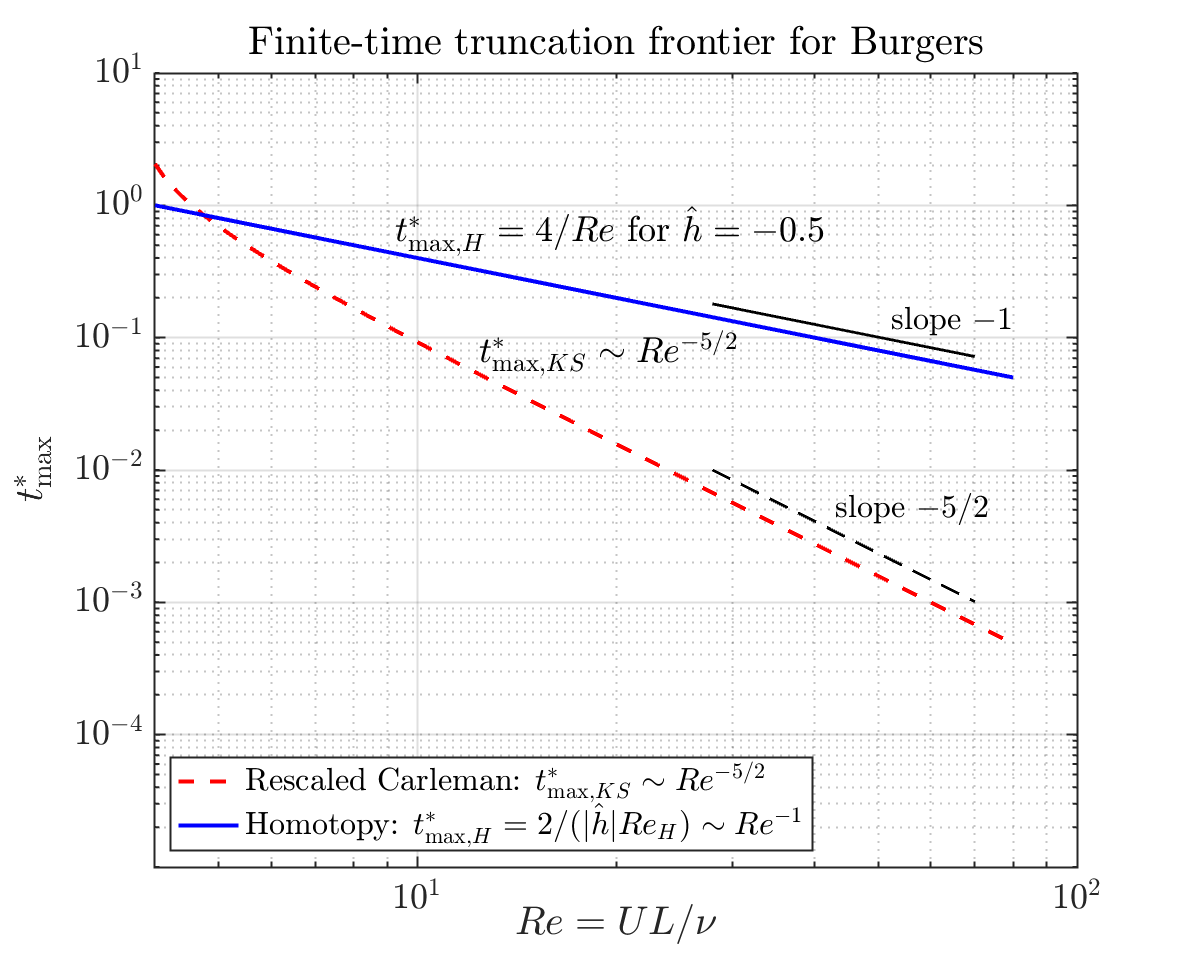}}
        \caption{\justifying \bob{Comparison between the present homotopy embedding and Carleman linearization for the viscous Burgers equation.
    (a) Near-frontier profile and interior-gradient comparison at $Re=36$ and $T^\ast=0.95\,t^\ast_{\max,H}$.  The right panel compares the interior derivative $u_x$ to avoid boundary-stencil artifacts and to test the steep Burgers transition layer.
    (b) Best final-time MSE attainable under a common lifted-dimension budget $D$ for representative Reynolds numbers. 
    (c) Performance-controlled lifted dimension required to reach ${\rm MSE}(T)<10^{-3}$ at $T^\ast=0.95\,t^\ast_{\max,H}(Re)$.  The plotted dimensions include the physical Burgers resolution scaling $N_{\rm phys}\sim Re$.
    (d) Finite-time admissible windows.  The physically rescaled Carleman window scales as $t^\ast_{\max,KS}\sim Re^{-5/2}$, while the homotopy window scales as $t^\ast_{\max,H}\sim Re^{-1}$ for the normalized Burgers setting.}}\label{fig:carleman-comparison}
\end{figure*}

\bob{In this discussion, we turn to a controlled comparison between the present homotopy embedding and a Carleman linearization. For this, we keep as our baseline, the one-dimensional viscous Burgers equation considered in Problem 1 of the main text. Let $u_{\rm ref}(x,t)$ denote the reference solution, obtained from a Direct Numerical Simulation (DNS), that is a fully-nonlinear, classical finite-difference solver. All measures of error and performance of each method, are computed with respect to this solution. This comparative analysis can be organized around three key aspects: (1) Error metrics, (2) Resource metrics for the same accuracy and (3) Finite-time integration and convergence frontiers.}\\

\subsection{\bob{Performance metrics}}
\label{app:carleman-performance-metrics}
\bob{First, we compute the solution using homotopy, Carleman and classical DNS. In Panel (a) of Fig.~\ref{fig:carleman-comparison} both the solution profile and the interior gradient computed around the shock profile. The homotopy approximation shows excellent agreement with the reference profile and with the interior derivative even at moderate homotopy orders. Low-order Carleman approximations capture only the broad profile but fails to accurately resolve the interior gradient at lower orders; although higher Carleman orders improve the solution profile and the gradient profile approximations, but it only does so at a larger lifted dimension cost compared to the homotopy case. To quantify these visual observations and to understand the performance of these methods concretely, we make a few more detailed calculations described below.}

\bob{For a given linearly-embedded, approximation $u_{\rm Q}$ (homotopy or Carleman method), the primary error metric used here, is the final-time mean-square error
\begin{equation}
    {\rm MSE}(T)
    =
    \frac{1}{N}
    \sum_{j=1}^{N}
    \left[
    u_{\rm Q}(x_j,T)
    -
    u_{\rm ref}(x_j,T)
    \right]^2,
    \label{eq:profile-mse-carl-comp}
\end{equation}
that are shown in both panels (b) and (c) of Fig.~\ref{fig:carleman-comparison}.  It gives a common scalar measure for both embeddings at the same Reynolds number and the same final time.}

For Burgers-like problems however, the profile error alone can be incomplete when the solution profile has detailed or intricate features. A PDE solver may approximate the point-wise profile reasonably well, while still inaccurately resolving the transition-layer thickness or shape. Therefore we also include an interior-gradient comparison.  Let $D_x$ denote the finite-difference first-derivative matrix.  We compare
$
    (D_xu_{\rm Q}(T))_j
    \qquad\text{with}\qquad
    (D_xu_{\rm ref}(T))_j
$
on an interior index set
$
    I_{\rm int}
    =
    \{j: x_j\in[x_{\rm min},x_{\rm max}]\},
$
where $x_{\rm min}$ and $x_{\rm max}$ exclude the Dirichlet boundary region.  In the reported plots we use an interior window $0.15\le x\le 0.85$. 

 The boundary points are excluded because finite-difference derivative estimates near imposed Dirichlet data can contain boundary-stencil artifacts that are not representative of the interior nonlinear steepening.

\subsection{\bob{Comparing algorithmic costs at matched accuracy}}
\label{app:carleman-resource-comparison}

\bob{The Carleman and homotopy truncation parameters are not directly comparable.  A Carleman truncation order $C$ and a homotopy order $M$ represent different lifted spaces.  We therefore compare the methods through the total lifted/evolved dimension needed to reach a fixed error target.}

\bob{A Carleman linearization lifting, at order $C$ contains all tensor powers,
\begin{equation}
    y_C
    =
    \left(
    u,\;
    u^{\otimes 2},\;
    \ldots,\;
    u^{\otimes C}
    \right).
\end{equation}
If the augmented state dimension is $d$, the full tensor dimension scales as
\begin{equation}
    D_C^{\rm full}(C)
    =
    \sum_{k=1}^{C} d^k .
\end{equation}
In the comparison we report the more favorable symmetric-monomial count
\begin{equation}
    D_C^{\rm sym}(C)
    =
    \sum_{k=1}^{C}
    \binom{d+k-1}{k}.
    \label{eq:carleman-sym-dim}
\end{equation}
This gives the Carleman baseline the benefit of removing redundant tensor permutations. The homotopy embedding instead evolves a system of deformation correction terms.  For $M$ retained homotopy orders and $N$ spatial degrees of freedom, the evolved dimension is counted as
\begin{equation}
    D_H(M) \simeq N(M+1).
    \label{eq:ham-dim-count}
\end{equation}
Thus, throughout the comparison, we compare $D_H$ with $D_C^{\rm sym}$ rather than comparing $M$ with $C$.}

\bob{For a fixed lifted-dimension budget $D$, define the best attainable final-time errors
\begin{equation}
    \mathcal E_H(D;Re)
    =
    \min_{M:\,D_H(M)\le D}
    {\rm MSE}_H(M,Re;T),
    \label{eq:best-error-ham}
\end{equation}
and
\begin{equation}
    \mathcal E_C(D;Re)
    =
    \min_{C:\,D_C^{\rm sym}(C)\le D}
    {\rm MSE}_C(C,Re;T).
    \label{eq:best-error-carl}
\end{equation}
Panel (b) of Fig.~\ref{fig:carleman-comparison} reports these quantities for representative, increasing Reynolds numbers, where the solutions are tested close to the finite time window $t^*$.  This panel answers the finite-size question -- for the same lifted-dimension budget $D$, which embedding reaches a smaller error?  In the tested near-frontier cases, the homotopy embedding reaches a lower final-time MSE at a substantially smaller dimension.}

\bob{We also define a performance-controlled dimension.  For a target error $\varepsilon$, let
\begin{equation}
    M_\varepsilon(Re)
    =
    \min
    \left\{
    M:
    {\rm MSE}_H(M,Re;T)<\varepsilon
    \right\},
    \label{eq:M-eps-def}
\end{equation}
and
\begin{equation}
    C_\varepsilon(Re)
    =
    \min
    \left\{
    C:
    {\rm MSE}_C(C,Re;T)<\varepsilon
    \right\}.
    \label{eq:C-eps-def}
\end{equation}
The corresponding performance-controlled dimensions are
\begin{equation}
    D_{H,\epsilon}(Re)
    =
    D_H\!\left(M_\epsilon(Re)\right),
    \qquad
    D_{C,\epsilon}(Re)
    =
    D_C^{\rm sym}\!\left(C_\epsilon(Re)\right).
    \label{eq:target-dim-fixed-grid}
\end{equation}}

\bob{To connect this finite-size comparison to the physical Burgers scaling, we also report the dimension at the physically resolved grid size.  The Burgers transition-layer thickness scales as $\delta \sim L/Re,$
so resolving this layer requires $N_{\rm phys}(Re)\sim Re.$
After the truncation order has been selected by the measured error, we estimate the corresponding physical-resolution dimensions as
\begin{equation}
    D^{\rm phys}_{H,\varepsilon}(Re)
    =
    N_{\rm phys}(Re)\,[M_\varepsilon(Re)+1],
    \label{eq:phys-dim-ham-eps}
\end{equation}
and
\begin{equation}
    D^{\rm phys}_{C,\varepsilon}(Re)
    =
    \sum_{k=1}^{C_\varepsilon(Re)}
    \binom{d_{\rm phys}(Re)+k-1}{k},
    \label{eq:phys-dim-carl-eps}
\end{equation}
where $d_{\rm phys}$ is the augmented physical state dimension.  This resource measure includes both effects: the spatial resolution required by the PDE and the truncation order required to reach the chosen accuracy.}

\bob{Panel (c) of Fig.~\ref{fig:carleman-comparison} plots $D^{\rm phys}_{H,\varepsilon}(Re)$ and $D^{\rm phys}_{C,\varepsilon}(Re)$ for $\varepsilon=10^{-3}$, using a near-frontier final time
\begin{equation}
T^\ast(Re)=0.95\,t^\ast_{\max,H}(Re).
    \label{eq:near-frontier-time-carl}
\end{equation}
The same final time is used for both homotopy and Carleman at each Reynolds number.  Thus panel (c) compares the resources required to solve the same physical problem at each $Re$, close to the homotopy single-window frontier.  The homotopy dimension remains substantially below the Carleman dimension across the tested range.  Small non-monotonic variations can occur because $M_\epsilon$ and $C_\epsilon$ are selected from discrete truncation-order sets; these local deviations are due to the drops in $t^*$ computed for the specific $Re$ values. These should not be interpreted as asymptotic scaling laws.}

\subsection{\bob{Finite-time simulation window and convergence}}
\label{app:carleman-time-window}

\bob{The third part of the comparison concerns the time interval over which truncation convergence is expected.  This is distinct from the error-versus-dimension comparison above.  Typical linear-embedding-based quantum algorithms, such the ones that use Carleman methods \cite{liu2021efficient,gonzalez2024quantum} often state an all-time convergence condition (exponential error convergence with truncation order, for all times $t$) of the form
\begin{equation}
    R<1,
    \label{eq:carleman-all-time}
\end{equation}
where $R$ is a dimensionless measure of nonlinear strength relative to linear dissipation.  This is a sufficient condition for convergence over arbitrarily long time intervals.  However, it is generally conservative for finite-time simulations.  A finite-time Carleman condition can instead be expressed in the form
\begin{equation}
    R\left(1-e^{-t^\ast}\right)<1.
    \label{eq:carleman-finite-time-condition}
\end{equation}
When $R>1$, this gives a finite admissible time
\begin{equation}
    t^\ast_{\max,C}
    =
    -\log\left(1-\frac{1}{R}\right).
    \label{eq:carleman-finite-time-window}
\end{equation}
Thus $R<1$ is an all-time sufficient condition, whereas Eq.~\eqref{eq:carleman-finite-time-condition} is a finite-time admissibility condition.}

\bob{For Burgers, it is important to evaluate $R$ using the physical resolution required by the PDE.  Since $N_{\rm phys}\sim Re$, the physically rescaled Carleman parameter behaves as \cite{gonzalez2024quantum}
\begin{equation}
    R_{KS}(Re)\sim Re^{5/2},
    \label{eq:RKS-scaling}
\end{equation}
which leads to
\begin{equation}
    t^\ast_{\max,KS}(Re)
    =
    -\log\left(
    1-\frac{1}{R_{KS}(Re)}
    \right),
    \qquad R_{KS}>1.
    \label{eq:tmax-KS}
\end{equation}
For large Reynolds number,
\begin{equation}
    t^\ast_{\max,KS}(Re)\sim Re^{-5/2}.
    \label{eq:tmax-KS-scaling}
\end{equation}}

The homotopy embedding has a different finite-time control parameter. Using the physically resolved homotopy Reynolds parameter
\begin{equation}
Re_{\rm H}
=
\frac{\|\bar u_0\|_\infty}{\bar U}
\left.
\frac{L\|\dd_2\|_\infty}{\|\dd_1\|_\infty}
\right|_{\Delta x\sim\ell_{\rm phys}},
\end{equation}
the accumulated homotopy nonlinear load is
\begin{equation}
    \rho_H(t)
    =
    \frac{|\hat h|}{2}Re_H t^\ast .
    \label{eq:rhoH-def}
\end{equation}

Therefore
\begin{equation}
    t^\ast_{\max,H}(Re)
    =
    \frac{2}{|\hat h|Re_H}.
    \label{eq:tmax-H}
\end{equation}
For the normalized Burgers setting used here, \(\ell_{\rm phys}\sim L/Re\) and \(\|\bar u_0\|_\infty=O(\bar U)\), so \(Re_{\rm H}=O(Re)\). With \(\hat h=-0.5\),
\begin{equation}
    t^\ast_{\max,H}(Re)
    =
    \frac{4}{Re},
    \qquad
    t^\ast_{\max,H}(Re)\sim Re^{-1}.
    \label{eq:tmax-H-scaling}
\end{equation}

\bob{Panel (d) of Fig.~\ref{fig:carleman-comparison} compares these finite-time admissible windows.  The all-time condition $R<1$ is not the relevant criterion for a finite simulation; rather, the relevant question is how long the truncation remains in its finite-time convergence regime.  Under physical Burgers scaling, the rescaled Carleman finite-time window shrinks like $Re^{-5/2}$, while the homotopy finite-time window shrinks like $Re^{-1}$.  This slower decrease gives the homotopy embedding a larger single-window admissible time at higher Reynolds number.  The resource tests in panels (b) and (c) complement this finite-time picture by showing how much lifted dimension is needed to reach a fixed accuracy within that window.}

\subsection{\bob{Summary of the comparison}}

\bob{The comparison in Fig.~\ref{fig:carleman-comparison} supports the following conclusions.  First, the homotopy approximation captures both the final profile and the interior gradient near the finite-time nonlinear frontier, while low-order Carleman approximations under-resolve the interior gradient.  Second, when error is plotted against a common lifted-dimension budget, the homotopy embedding reaches lower error at smaller dimension in the tested cases.  Third, when the truncation order is selected to meet a target accuracy and the physical Burgers resolution is accounted for, the homotopy embedding requires a smaller lifted dimension than the Carleman lift over the tested range.  Finally, the finite-time window analysis explains why this advantage is expected to be most visible near nonlinear finite-time frontiers: the physically rescaled Carleman admissible time decreases more rapidly with $Re$, whereas the homotopy admissible time decreases only as $Re^{-1}$ in the normalized Burgers setting.}

\bob{This does not imply that Carleman linearization is ineffective in all regimes.  Carleman improves with increasing $C$ and can be competitive in weakly nonlinear or short-time cases. It is also likely that Carleman method has a better nonlinearity bound, for all-time convergence guarantees. The numerical evidence instead indicates that, for the tested Burgers near-frontier regimes, the present homotopy embedding reaches the same error targets with a smaller lifted dimension and a more favorable finite-time scaling. It is quite natural that any linearization based method could asymptote to a similar performance metrics at large times and at high nonlinearities. Thus our focus on finite-time convergence criteria is motivated by near-term hardware constraints.}

\bibliography{refs}

\begin{thebibliography}{104}%
\makeatletter
\providecommand \@ifxundefined [1]{%
 \@ifx{#1\undefined}
}%
\providecommand \@ifnum [1]{%
 \ifnum #1\expandafter \@firstoftwo
 \else \expandafter \@secondoftwo
 \fi
}%
\providecommand \@ifx [1]{%
 \ifx #1\expandafter \@firstoftwo
 \else \expandafter \@secondoftwo
 \fi
}%
\providecommand \natexlab [1]{#1}%
\providecommand \enquote  [1]{``#1''}%
\providecommand \bibnamefont  [1]{#1}%
\providecommand \bibfnamefont [1]{#1}%
\providecommand \citenamefont [1]{#1}%
\providecommand \href@noop [0]{\@secondoftwo}%
\providecommand \href [0]{\begingroup \@sanitize@url \@href}%
\providecommand \@href[1]{\@@startlink{#1}\@@href}%
\providecommand \@@href[1]{\endgroup#1\@@endlink}%
\providecommand \@sanitize@url [0]{\catcode `\\12\catcode `\$12\catcode `\&12\catcode `\#12\catcode `\^12\catcode `\_12\catcode `\%12\relax}%
\providecommand \@@startlink[1]{}%
\providecommand \@@endlink[0]{}%
\providecommand \url  [0]{\begingroup\@sanitize@url \@url }%
\providecommand \@url [1]{\endgroup\@href {#1}{\urlprefix }}%
\providecommand \urlprefix  [0]{URL }%
\providecommand \Eprint [0]{\href }%
\providecommand \doibase [0]{https://doi.org/}%
\providecommand \selectlanguage [0]{\@gobble}%
\providecommand \bibinfo  [0]{\@secondoftwo}%
\providecommand \bibfield  [0]{\@secondoftwo}%
\providecommand \translation [1]{[#1]}%
\providecommand \BibitemOpen [0]{}%
\providecommand \bibitemStop [0]{}%
\providecommand \bibitemNoStop [0]{.\EOS\space}%
\providecommand \EOS [0]{\spacefactor3000\relax}%
\providecommand \BibitemShut  [1]{\csname bibitem#1\endcsname}%
\let\auto@bib@innerbib\@empty
\bibitem [{\citenamefont {Bharadwaj}\ and\ \citenamefont {Sreenivasan}(2025{\natexlab{a}})}]{bharadwaj2025}%
  \BibitemOpen
  \bibfield  {author} {\bibinfo {author} {\bibfnamefont {S.~S.}\ \bibnamefont {Bharadwaj}}\ and\ \bibinfo {author} {\bibfnamefont {K.~R.}\ \bibnamefont {Sreenivasan}},\ }\bibfield  {title} {\bibinfo {title} {Towards simulating fluid flows with quantum computing},\ }\href@noop {} {\bibfield  {journal} {\bibinfo  {journal} {Sādhanā}\ }\textbf {\bibinfo {volume} {50}},\ \bibinfo {pages} {57} (\bibinfo {year} {2025}{\natexlab{a}})}\BibitemShut {NoStop}%
\bibitem [{\citenamefont {Bharadwaj}\ and\ \citenamefont {Sreenivasan}(2025{\natexlab{b}})}]{bharadwaj2024compact}%
  \BibitemOpen
  \bibfield  {author} {\bibinfo {author} {\bibfnamefont {S.~S.}\ \bibnamefont {Bharadwaj}}\ and\ \bibinfo {author} {\bibfnamefont {K.~R.}\ \bibnamefont {Sreenivasan}},\ }\bibfield  {title} {\bibinfo {title} {Compact quantum algorithms for time-dependent differential equations},\ }\href@noop {} {\bibfield  {journal} {\bibinfo  {journal} {Phys. Rev. Res.}\ }\textbf {\bibinfo {volume} {7}},\ \bibinfo {pages} {023262} (\bibinfo {year} {2025}{\natexlab{b}})}\BibitemShut {NoStop}%
\bibitem [{\citenamefont {Harrow}\ \emph {et~al.}(2009)\citenamefont {Harrow}, \citenamefont {Hassidim},\ and\ \citenamefont {Lloyd}}]{harrow2009quantum}%
  \BibitemOpen
  \bibfield  {author} {\bibinfo {author} {\bibfnamefont {A.~W.}\ \bibnamefont {Harrow}}, \bibinfo {author} {\bibfnamefont {A.}~\bibnamefont {Hassidim}},\ and\ \bibinfo {author} {\bibfnamefont {S.}~\bibnamefont {Lloyd}},\ }\bibfield  {title} {\bibinfo {title} {Quantum algorithm for linear systems of equations},\ }\href@noop {} {\bibfield  {journal} {\bibinfo  {journal} {Phys. Rev. Lett.}\ }\textbf {\bibinfo {volume} {103}},\ \bibinfo {pages} {150502} (\bibinfo {year} {2009})}\BibitemShut {NoStop}%
\bibitem [{\citenamefont {Berry}(2014)}]{berry2014high}%
  \BibitemOpen
  \bibfield  {author} {\bibinfo {author} {\bibfnamefont {D.~W.}\ \bibnamefont {Berry}},\ }\bibfield  {title} {\bibinfo {title} {High-order quantum algorithm for solving linear differential equations},\ }\href@noop {} {\bibfield  {journal} {\bibinfo  {journal} {Journal of Physics A: Mathematical and Theoretical}\ }\textbf {\bibinfo {volume} {47}},\ \bibinfo {pages} {105301} (\bibinfo {year} {2014})}\BibitemShut {NoStop}%
\bibitem [{\citenamefont {Berry}\ \emph {et~al.}(2017)\citenamefont {Berry}, \citenamefont {Childs}, \citenamefont {Ostrander},\ and\ \citenamefont {Wang}}]{berry2017quantum}%
  \BibitemOpen
  \bibfield  {author} {\bibinfo {author} {\bibfnamefont {D.~W.}\ \bibnamefont {Berry}}, \bibinfo {author} {\bibfnamefont {A.~M.}\ \bibnamefont {Childs}}, \bibinfo {author} {\bibfnamefont {A.}~\bibnamefont {Ostrander}},\ and\ \bibinfo {author} {\bibfnamefont {G.}~\bibnamefont {Wang}},\ }\bibfield  {title} {\bibinfo {title} {Quantum algorithm for linear differential equations with exponentially improved dependence on precision},\ }\href@noop {} {\bibfield  {journal} {\bibinfo  {journal} {Commun. Math. Phys.}\ }\textbf {\bibinfo {volume} {356}},\ \bibinfo {pages} {1057} (\bibinfo {year} {2017})}\BibitemShut {NoStop}%
\bibitem [{\citenamefont {Subasi}\ \emph {et~al.}(2019)\citenamefont {Subasi}, \citenamefont {Somma},\ and\ \citenamefont {Orsucci}}]{subacsi2019quantum}%
  \BibitemOpen
  \bibfield  {author} {\bibinfo {author} {\bibfnamefont {Y.}~\bibnamefont {Subasi}}, \bibinfo {author} {\bibfnamefont {R.~D.}\ \bibnamefont {Somma}},\ and\ \bibinfo {author} {\bibfnamefont {D.}~\bibnamefont {Orsucci}},\ }\bibfield  {title} {\bibinfo {title} {Quantum algorithms for systems of linear equations inspired by adiabatic quantum computing},\ }\href@noop {} {\bibfield  {journal} {\bibinfo  {journal} {Phys. Rev. Lett.}\ }\textbf {\bibinfo {volume} {122}},\ \bibinfo {pages} {060504} (\bibinfo {year} {2019})}\BibitemShut {NoStop}%
\bibitem [{\citenamefont {Costa}\ \emph {et~al.}(2019)\citenamefont {Costa}, \citenamefont {Jordan},\ and\ \citenamefont {Ostrander}}]{costa2019quantum}%
  \BibitemOpen
  \bibfield  {author} {\bibinfo {author} {\bibfnamefont {P.~C.~S.}\ \bibnamefont {Costa}}, \bibinfo {author} {\bibfnamefont {S.}~\bibnamefont {Jordan}},\ and\ \bibinfo {author} {\bibfnamefont {A.}~\bibnamefont {Ostrander}},\ }\bibfield  {title} {\bibinfo {title} {Quantum algorithm for simulating the wave equation},\ }\href@noop {} {\bibfield  {journal} {\bibinfo  {journal} {Phys. Rev. A}\ }\textbf {\bibinfo {volume} {99}},\ \bibinfo {pages} {012323} (\bibinfo {year} {2019})}\BibitemShut {NoStop}%
\bibitem [{\citenamefont {Liu}\ \emph {et~al.}(2021)\citenamefont {Liu}, \citenamefont {Kolden}, \citenamefont {Krovi}, \citenamefont {Loureiro}, \citenamefont {Trivisa},\ and\ \citenamefont {Childs}}]{liu2021efficient}%
  \BibitemOpen
  \bibfield  {author} {\bibinfo {author} {\bibfnamefont {J.-P.}\ \bibnamefont {Liu}}, \bibinfo {author} {\bibfnamefont {H.~{\O}.}\ \bibnamefont {Kolden}}, \bibinfo {author} {\bibfnamefont {H.~K.}\ \bibnamefont {Krovi}}, \bibinfo {author} {\bibfnamefont {N.~F.}\ \bibnamefont {Loureiro}}, \bibinfo {author} {\bibfnamefont {K.}~\bibnamefont {Trivisa}},\ and\ \bibinfo {author} {\bibfnamefont {A.~M.}\ \bibnamefont {Childs}},\ }\bibfield  {title} {\bibinfo {title} {Efficient quantum algorithm for dissipative nonlinear differential equations},\ }\href@noop {} {\bibfield  {journal} {\bibinfo  {journal} {Proc. Nat. Acad. Sci.}\ }\textbf {\bibinfo {volume} {118}},\ \bibinfo {pages} {e2026805118} (\bibinfo {year} {2021})}\BibitemShut {NoStop}%
\bibitem [{\citenamefont {Childs}\ \emph {et~al.}(2021)\citenamefont {Childs}, \citenamefont {Liu},\ and\ \citenamefont {Ostrander}}]{childs2021high}%
  \BibitemOpen
  \bibfield  {author} {\bibinfo {author} {\bibfnamefont {A.~M.}\ \bibnamefont {Childs}}, \bibinfo {author} {\bibfnamefont {J.-P.}\ \bibnamefont {Liu}},\ and\ \bibinfo {author} {\bibfnamefont {A.}~\bibnamefont {Ostrander}},\ }\bibfield  {title} {\bibinfo {title} {High-precision quantum algorithms for partial differential equations},\ }\href@noop {} {\bibfield  {journal} {\bibinfo  {journal} {Quantum}\ }\textbf {\bibinfo {volume} {5}},\ \bibinfo {pages} {574} (\bibinfo {year} {2021})}\BibitemShut {NoStop}%
\bibitem [{\citenamefont {Costa}\ \emph {et~al.}(2022)\citenamefont {Costa}, \citenamefont {An}, \citenamefont {Sanders}, \citenamefont {Su}, \citenamefont {Babbush},\ and\ \citenamefont {Berry}}]{costa2022optimal}%
  \BibitemOpen
  \bibfield  {author} {\bibinfo {author} {\bibfnamefont {P.~C.~S.}\ \bibnamefont {Costa}}, \bibinfo {author} {\bibfnamefont {D.}~\bibnamefont {An}}, \bibinfo {author} {\bibfnamefont {Y.~R.}\ \bibnamefont {Sanders}}, \bibinfo {author} {\bibfnamefont {Y.}~\bibnamefont {Su}}, \bibinfo {author} {\bibfnamefont {R.}~\bibnamefont {Babbush}},\ and\ \bibinfo {author} {\bibfnamefont {D.~W.}\ \bibnamefont {Berry}},\ }\bibfield  {title} {\bibinfo {title} {Optimal scaling quantum linear-systems solver via discrete adiabatic theorem},\ }\href@noop {} {\bibfield  {journal} {\bibinfo  {journal} {PRX quantum}\ }\textbf {\bibinfo {volume} {3}},\ \bibinfo {pages} {040303} (\bibinfo {year} {2022})}\BibitemShut {NoStop}%
\bibitem [{\citenamefont {Costa}\ \emph {et~al.}(2025)\citenamefont {Costa}, \citenamefont {Schleich}, \citenamefont {Morales},\ and\ \citenamefont {Berry}}]{costa2023further}%
  \BibitemOpen
  \bibfield  {author} {\bibinfo {author} {\bibfnamefont {P.~C.~S.}\ \bibnamefont {Costa}}, \bibinfo {author} {\bibfnamefont {P.}~\bibnamefont {Schleich}}, \bibinfo {author} {\bibfnamefont {M.~E.~S.}\ \bibnamefont {Morales}},\ and\ \bibinfo {author} {\bibfnamefont {D.~W.}\ \bibnamefont {Berry}},\ }\bibfield  {title} {\bibinfo {title} {Further improving quantum algorithms for nonlinear differential equations via higher-order methods and rescaling},\ }\href {https://doi.org/10.1038/s41534-025-01084-z} {\bibfield  {journal} {\bibinfo  {journal} {npj Quantum Inf.}\ }\textbf {\bibinfo {volume} {11}},\ \bibinfo {pages} {141} (\bibinfo {year} {2025})}\BibitemShut {NoStop}%
\bibitem [{\citenamefont {Krovi}(2023)}]{krovi2023improved}%
  \BibitemOpen
  \bibfield  {author} {\bibinfo {author} {\bibfnamefont {H.}~\bibnamefont {Krovi}},\ }\bibfield  {title} {\bibinfo {title} {Improved quantum algorithms for linear and nonlinear differential equations},\ }\href@noop {} {\bibfield  {journal} {\bibinfo  {journal} {Quantum}\ }\textbf {\bibinfo {volume} {7}},\ \bibinfo {pages} {913} (\bibinfo {year} {2023})}\BibitemShut {NoStop}%
\bibitem [{\citenamefont {Liu}\ \emph {et~al.}(2023)\citenamefont {Liu}, \citenamefont {An}, \citenamefont {Fang}, \citenamefont {Wang}, \citenamefont {Low},\ and\ \citenamefont {Jordan}}]{liu2023efficient}%
  \BibitemOpen
  \bibfield  {author} {\bibinfo {author} {\bibfnamefont {J.-P.}\ \bibnamefont {Liu}}, \bibinfo {author} {\bibfnamefont {D.}~\bibnamefont {An}}, \bibinfo {author} {\bibfnamefont {D.}~\bibnamefont {Fang}}, \bibinfo {author} {\bibfnamefont {J.}~\bibnamefont {Wang}}, \bibinfo {author} {\bibfnamefont {G.~H.}\ \bibnamefont {Low}},\ and\ \bibinfo {author} {\bibfnamefont {S.}~\bibnamefont {Jordan}},\ }\bibfield  {title} {\bibinfo {title} {Efficient quantum algorithm for nonlinear reaction–diffusion equations and energy estimation},\ }\href@noop {} {\bibfield  {journal} {\bibinfo  {journal} {Commun. Math. Phys.}\ }\textbf {\bibinfo {volume} {404}},\ \bibinfo {pages} {963} (\bibinfo {year} {2023})}\BibitemShut {NoStop}%
\bibitem [{\citenamefont {An}\ \emph {et~al.}(2023{\natexlab{a}})\citenamefont {An}, \citenamefont {Liu},\ and\ \citenamefont {Lin}}]{an2023linear}%
  \BibitemOpen
  \bibfield  {author} {\bibinfo {author} {\bibfnamefont {D.}~\bibnamefont {An}}, \bibinfo {author} {\bibfnamefont {J.-P.}\ \bibnamefont {Liu}},\ and\ \bibinfo {author} {\bibfnamefont {L.}~\bibnamefont {Lin}},\ }\bibfield  {title} {\bibinfo {title} {Linear combination of hamiltonian simulation for nonunitary dynamics with optimal state preparation cost},\ }\href@noop {} {\bibfield  {journal} {\bibinfo  {journal} {Phys. Rev. Lett.}\ }\textbf {\bibinfo {volume} {131}},\ \bibinfo {pages} {150603} (\bibinfo {year} {2023}{\natexlab{a}})}\BibitemShut {NoStop}%
\bibitem [{\citenamefont {Fang}\ \emph {et~al.}(2023)\citenamefont {Fang}, \citenamefont {Lin},\ and\ \citenamefont {Tong}}]{fang2023time}%
  \BibitemOpen
  \bibfield  {author} {\bibinfo {author} {\bibfnamefont {D.}~\bibnamefont {Fang}}, \bibinfo {author} {\bibfnamefont {L.}~\bibnamefont {Lin}},\ and\ \bibinfo {author} {\bibfnamefont {Y.}~\bibnamefont {Tong}},\ }\bibfield  {title} {\bibinfo {title} {Time-marching based quantum solvers for time-dependent linear differential equations},\ }\href@noop {} {\bibfield  {journal} {\bibinfo  {journal} {Quantum}\ }\textbf {\bibinfo {volume} {7}},\ \bibinfo {pages} {955} (\bibinfo {year} {2023})}\BibitemShut {NoStop}%
\bibitem [{\citenamefont {Bharadwaj}\ and\ \citenamefont {Sreenivasan}(2023)}]{bharadwaj2023hybrid}%
  \BibitemOpen
  \bibfield  {author} {\bibinfo {author} {\bibfnamefont {S.~S.}\ \bibnamefont {Bharadwaj}}\ and\ \bibinfo {author} {\bibfnamefont {K.~R.}\ \bibnamefont {Sreenivasan}},\ }\bibfield  {title} {\bibinfo {title} {Hybrid quantum algorithms for flow problems},\ }\href@noop {} {\bibfield  {journal} {\bibinfo  {journal} {Proc. Nat. Acad. Sci.}\ }\textbf {\bibinfo {volume} {120}},\ \bibinfo {pages} {e2311014120} (\bibinfo {year} {2023})}\BibitemShut {NoStop}%
\bibitem [{\citenamefont {An}\ \emph {et~al.}(2023{\natexlab{b}})\citenamefont {An}, \citenamefont {Childs},\ and\ \citenamefont {Lin}}]{an2023quantum}%
  \BibitemOpen
  \bibfield  {author} {\bibinfo {author} {\bibfnamefont {D.}~\bibnamefont {An}}, \bibinfo {author} {\bibfnamefont {A.~M.}\ \bibnamefont {Childs}},\ and\ \bibinfo {author} {\bibfnamefont {L.}~\bibnamefont {Lin}},\ }\bibfield  {title} {\bibinfo {title} {Quantum algorithm for linear non-unitary dynamics with near-optimal dependence on all parameters},\ }\href@noop {} {\bibfield  {journal} {\bibinfo  {journal} {arXiv:2312.03916}\ } (\bibinfo {year} {2023}{\natexlab{b}})}\BibitemShut {NoStop}%
\bibitem [{\citenamefont {Gonzalez-Conde}\ \emph {et~al.}(2025)\citenamefont {Gonzalez-Conde}, \citenamefont {Lewis}, \citenamefont {Bharadwaj},\ and\ \citenamefont {Sanz}}]{gonzalez2024quantum}%
  \BibitemOpen
  \bibfield  {author} {\bibinfo {author} {\bibfnamefont {J.}~\bibnamefont {Gonzalez-Conde}}, \bibinfo {author} {\bibfnamefont {D.}~\bibnamefont {Lewis}}, \bibinfo {author} {\bibfnamefont {S.~S.}\ \bibnamefont {Bharadwaj}},\ and\ \bibinfo {author} {\bibfnamefont {M.}~\bibnamefont {Sanz}},\ }\bibfield  {title} {\bibinfo {title} {Quantum carleman linearisation efficiency in nonlinear fluid dynamics},\ }\href@noop {} {\bibfield  {journal} {\bibinfo  {journal} {Phys. Rev. Res.}\ }\textbf {\bibinfo {volume} {7}},\ \bibinfo {pages} {023254} (\bibinfo {year} {2025})}\BibitemShut {NoStop}%
\bibitem [{\citenamefont {Dalzell}(2024)}]{dalzell2024shortcut}%
  \BibitemOpen
  \bibfield  {author} {\bibinfo {author} {\bibfnamefont {A.~M.}\ \bibnamefont {Dalzell}},\ }\bibfield  {title} {\bibinfo {title} {A shortcut to an optimal quantum linear system solver},\ }\href@noop {} {\bibfield  {journal} {\bibinfo  {journal} {arXiv:2406.12086v1}\ } (\bibinfo {year} {2024})}\BibitemShut {NoStop}%
\bibitem [{\citenamefont {Lubasch}\ \emph {et~al.}(2025)\citenamefont {Lubasch}, \citenamefont {Kikuchi}, \citenamefont {Wright},\ and\ \citenamefont {Keever}}]{lubasch2025quantum}%
  \BibitemOpen
  \bibfield  {author} {\bibinfo {author} {\bibfnamefont {M.}~\bibnamefont {Lubasch}}, \bibinfo {author} {\bibfnamefont {Y.}~\bibnamefont {Kikuchi}}, \bibinfo {author} {\bibfnamefont {L.}~\bibnamefont {Wright}},\ and\ \bibinfo {author} {\bibfnamefont {C.~M.}\ \bibnamefont {Keever}},\ }\bibfield  {title} {\bibinfo {title} {Quantum circuits for partial differential equations in {Fourier} space},\ }\href {https://doi.org/10.1103/tbzc-w9x8} {\bibfield  {journal} {\bibinfo  {journal} {Phys. Rev. Research}\ }\textbf {\bibinfo {volume} {7}},\ \bibinfo {pages} {043326} (\bibinfo {year} {2025})}\BibitemShut {NoStop}%
\bibitem [{\citenamefont {Berry}\ and\ \citenamefont {Costa}(2024)}]{berry2024quantum}%
  \BibitemOpen
  \bibfield  {author} {\bibinfo {author} {\bibfnamefont {D.~W.}\ \bibnamefont {Berry}}\ and\ \bibinfo {author} {\bibfnamefont {P.~C.}\ \bibnamefont {Costa}},\ }\bibfield  {title} {\bibinfo {title} {Quantum algorithm for time-dependent differential equations using dyson series},\ }\href@noop {} {\bibfield  {journal} {\bibinfo  {journal} {Quantum}\ }\textbf {\bibinfo {volume} {8}},\ \bibinfo {pages} {1369} (\bibinfo {year} {2024})}\BibitemShut {NoStop}%
\bibitem [{\citenamefont {Brearley}\ and\ \citenamefont {Laizet}(2024)}]{brearley2024quantum}%
  \BibitemOpen
  \bibfield  {author} {\bibinfo {author} {\bibfnamefont {P.}~\bibnamefont {Brearley}}\ and\ \bibinfo {author} {\bibfnamefont {S.}~\bibnamefont {Laizet}},\ }\bibfield  {title} {\bibinfo {title} {Quantum algorithm for solving the advection equation using hamiltonian simulation},\ }\href@noop {} {\bibfield  {journal} {\bibinfo  {journal} {Phys. Rev. A}\ }\textbf {\bibinfo {volume} {110}},\ \bibinfo {pages} {012430} (\bibinfo {year} {2024})}\BibitemShut {NoStop}%
\bibitem [{\citenamefont {Leong}\ \emph {et~al.}(2022)\citenamefont {Leong}, \citenamefont {Ewe},\ and\ \citenamefont {Koh}}]{leong2022variational}%
  \BibitemOpen
  \bibfield  {author} {\bibinfo {author} {\bibfnamefont {F.~Y.}\ \bibnamefont {Leong}}, \bibinfo {author} {\bibfnamefont {W.-B.}\ \bibnamefont {Ewe}},\ and\ \bibinfo {author} {\bibfnamefont {D.~E.}\ \bibnamefont {Koh}},\ }\bibfield  {title} {\bibinfo {title} {Variational quantum evolution equation solver},\ }\href@noop {} {\bibfield  {journal} {\bibinfo  {journal} {Sci. Rep.}\ }\textbf {\bibinfo {volume} {12}},\ \bibinfo {pages} {10817} (\bibinfo {year} {2022})}\BibitemShut {NoStop}%
\bibitem [{\citenamefont {Lubasch}\ \emph {et~al.}(2020)\citenamefont {Lubasch}, \citenamefont {Joo}, \citenamefont {Moinier}, \citenamefont {Kiffner},\ and\ \citenamefont {Jaksch}}]{lubasch2020variational}%
  \BibitemOpen
  \bibfield  {author} {\bibinfo {author} {\bibfnamefont {M.}~\bibnamefont {Lubasch}}, \bibinfo {author} {\bibfnamefont {J.}~\bibnamefont {Joo}}, \bibinfo {author} {\bibfnamefont {P.}~\bibnamefont {Moinier}}, \bibinfo {author} {\bibfnamefont {M.}~\bibnamefont {Kiffner}},\ and\ \bibinfo {author} {\bibfnamefont {D.}~\bibnamefont {Jaksch}},\ }\bibfield  {title} {\bibinfo {title} {Variational quantum algorithms for nonlinear problems},\ }\href@noop {} {\bibfield  {journal} {\bibinfo  {journal} {Phys. Rev. A}\ }\textbf {\bibinfo {volume} {101}},\ \bibinfo {pages} {010301} (\bibinfo {year} {2020})}\BibitemShut {NoStop}%
\bibitem [{\citenamefont {Ingelmann}\ \emph {et~al.}(2024)\citenamefont {Ingelmann}, \citenamefont {Bharadwaj}, \citenamefont {Pfeffer}, \citenamefont {Sreenivasan},\ and\ \citenamefont {Schumacher}}]{ingelmann2024two}%
  \BibitemOpen
  \bibfield  {author} {\bibinfo {author} {\bibfnamefont {J.}~\bibnamefont {Ingelmann}}, \bibinfo {author} {\bibfnamefont {S.~S.}\ \bibnamefont {Bharadwaj}}, \bibinfo {author} {\bibfnamefont {P.}~\bibnamefont {Pfeffer}}, \bibinfo {author} {\bibfnamefont {K.~R.}\ \bibnamefont {Sreenivasan}},\ and\ \bibinfo {author} {\bibfnamefont {J.}~\bibnamefont {Schumacher}},\ }\bibfield  {title} {\bibinfo {title} {Two quantum algorithms for solving the one-dimensional advection-diffusion equation},\ }\href@noop {} {\bibfield  {journal} {\bibinfo  {journal} {Comput. Fluids}\ }\textbf {\bibinfo {volume} {281}},\ \bibinfo {pages} {106369} (\bibinfo {year} {2024})}\BibitemShut {NoStop}%
\bibitem [{\citenamefont {Wright}\ \emph {et~al.}(2024)\citenamefont {Wright}, \citenamefont {Keever}, \citenamefont {First}, \citenamefont {Johnston}, \citenamefont {Tillay}, \citenamefont {Chaney}, \citenamefont {Rosenkranz},\ and\ \citenamefont {Lubasch}}]{wright2024noisy}%
  \BibitemOpen
  \bibfield  {author} {\bibinfo {author} {\bibfnamefont {L.}~\bibnamefont {Wright}}, \bibinfo {author} {\bibfnamefont {C.~M.}\ \bibnamefont {Keever}}, \bibinfo {author} {\bibfnamefont {J.~T.}\ \bibnamefont {First}}, \bibinfo {author} {\bibfnamefont {R.}~\bibnamefont {Johnston}}, \bibinfo {author} {\bibfnamefont {J.}~\bibnamefont {Tillay}}, \bibinfo {author} {\bibfnamefont {S.}~\bibnamefont {Chaney}}, \bibinfo {author} {\bibfnamefont {M.}~\bibnamefont {Rosenkranz}},\ and\ \bibinfo {author} {\bibfnamefont {M.}~\bibnamefont {Lubasch}},\ }\bibfield  {title} {\bibinfo {title} {Noisy intermediate-scale quantum simulation of the one-dimensional wave equation},\ }\href {https://doi.org/10.1103/PhysRevResearch.6.043169} {\bibfield  {journal} {\bibinfo  {journal} {Phys. Rev. Research}\ }\textbf {\bibinfo {volume} {6}},\ \bibinfo {pages} {043169} (\bibinfo {year} {2024})}\BibitemShut {NoStop}%
\bibitem [{\citenamefont {Pool}\ \emph {et~al.}(2024)\citenamefont {Pool}, \citenamefont {Somoza}, \citenamefont {Keever}, \citenamefont {Lubasch},\ and\ \citenamefont {Horstmann}}]{pool2024nonlinear}%
  \BibitemOpen
  \bibfield  {author} {\bibinfo {author} {\bibfnamefont {A.~J.}\ \bibnamefont {Pool}}, \bibinfo {author} {\bibfnamefont {A.~D.}\ \bibnamefont {Somoza}}, \bibinfo {author} {\bibfnamefont {C.~M.}\ \bibnamefont {Keever}}, \bibinfo {author} {\bibfnamefont {M.}~\bibnamefont {Lubasch}},\ and\ \bibinfo {author} {\bibfnamefont {B.}~\bibnamefont {Horstmann}},\ }\bibfield  {title} {\bibinfo {title} {Nonlinear dynamics as a ground-state solution on quantum computers},\ }\href {https://doi.org/10.1103/PhysRevResearch.6.033257} {\bibfield  {journal} {\bibinfo  {journal} {Phys. Rev. Research}\ }\textbf {\bibinfo {volume} {6}},\ \bibinfo {pages} {033257} (\bibinfo {year} {2024})}\BibitemShut {NoStop}%
\bibitem [{\citenamefont {Todorova}\ and\ \citenamefont {Steijl}(2020)}]{todorova2020quantum}%
  \BibitemOpen
  \bibfield  {author} {\bibinfo {author} {\bibfnamefont {B.~N.}\ \bibnamefont {Todorova}}\ and\ \bibinfo {author} {\bibfnamefont {R.}~\bibnamefont {Steijl}},\ }\bibfield  {title} {\bibinfo {title} {Quantum algorithm for the collisionless boltzmann equation},\ }\href@noop {} {\bibfield  {journal} {\bibinfo  {journal} {J. Comput. Phys.}\ }\textbf {\bibinfo {volume} {409}},\ \bibinfo {pages} {109347} (\bibinfo {year} {2020})}\BibitemShut {NoStop}%
\bibitem [{\citenamefont {Budinski}(2021)}]{budinski2021quantum}%
  \BibitemOpen
  \bibfield  {author} {\bibinfo {author} {\bibfnamefont {L.}~\bibnamefont {Budinski}},\ }\bibfield  {title} {\bibinfo {title} {Quantum algorithm for the advection--diffusion equation simulated with the lattice boltzmann method},\ }\href@noop {} {\bibfield  {journal} {\bibinfo  {journal} {Quantum Inf. Process.}\ }\textbf {\bibinfo {volume} {20}},\ \bibinfo {pages} {57} (\bibinfo {year} {2021})}\BibitemShut {NoStop}%
\bibitem [{\citenamefont {Bakker}\ and\ \citenamefont {Watts}(2023)}]{bakker2023quantum}%
  \BibitemOpen
  \bibfield  {author} {\bibinfo {author} {\bibfnamefont {B.}~\bibnamefont {Bakker}}\ and\ \bibinfo {author} {\bibfnamefont {T.}~\bibnamefont {Watts}},\ }\bibfield  {title} {\bibinfo {title} {Quantum carleman linearization of the lattice boltzmann equation with boundary conditions},\ }\href@noop {} {\bibfield  {journal} {\bibinfo  {journal} {arXiv:2312.04781}\ } (\bibinfo {year} {2023})}\BibitemShut {NoStop}%
\bibitem [{\citenamefont {Li}\ \emph {et~al.}(2025)\citenamefont {Li}, \citenamefont {Yin}, \citenamefont {Wiebe}, \citenamefont {Chun}, \citenamefont {Schenter}, \citenamefont {Cheung},\ and\ \citenamefont {M{"u}lmenst{"a}dt}}]{li2023potential}%
  \BibitemOpen
  \bibfield  {author} {\bibinfo {author} {\bibfnamefont {X.}~\bibnamefont {Li}}, \bibinfo {author} {\bibfnamefont {X.}~\bibnamefont {Yin}}, \bibinfo {author} {\bibfnamefont {N.}~\bibnamefont {Wiebe}}, \bibinfo {author} {\bibfnamefont {J.}~\bibnamefont {Chun}}, \bibinfo {author} {\bibfnamefont {G.~K.}\ \bibnamefont {Schenter}}, \bibinfo {author} {\bibfnamefont {M.~S.}\ \bibnamefont {Cheung}},\ and\ \bibinfo {author} {\bibfnamefont {J.}~\bibnamefont {M{"u}lmenst{"a}dt}},\ }\bibfield  {title} {\bibinfo {title} {Potential quantum advantage for simulation of fluid dynamics},\ }\href {https://doi.org/10.1103/PhysRevResearch.7.013036} {\bibfield  {journal} {\bibinfo  {journal} {Phys. Rev. Research}\ }\textbf {\bibinfo {volume} {7}},\ \bibinfo {pages} {013036} (\bibinfo {year} {2025})}\BibitemShut {NoStop}%
\bibitem [{\citenamefont {Itani}\ \emph {et~al.}(2024)\citenamefont {Itani}, \citenamefont {Sreenivasan},\ and\ \citenamefont {Succi}}]{itani2024quantum}%
  \BibitemOpen
  \bibfield  {author} {\bibinfo {author} {\bibfnamefont {W.}~\bibnamefont {Itani}}, \bibinfo {author} {\bibfnamefont {K.~R.}\ \bibnamefont {Sreenivasan}},\ and\ \bibinfo {author} {\bibfnamefont {S.}~\bibnamefont {Succi}},\ }\bibfield  {title} {\bibinfo {title} {Quantum algorithm for lattice boltzmann (qalb) simulation of incompressible fluids with a nonlinear collision term},\ }\href@noop {} {\bibfield  {journal} {\bibinfo  {journal} {Phys. Fluids}\ }\textbf {\bibinfo {volume} {36}} (\bibinfo {year} {2024})}\BibitemShut {NoStop}%
\bibitem [{\citenamefont {Succi}\ \emph {et~al.}(2024{\natexlab{a}})\citenamefont {Succi}, \citenamefont {Sanavio}, \citenamefont {Scatamacchia},\ and\ \citenamefont {De~Falco}}]{succi2024three}%
  \BibitemOpen
  \bibfield  {author} {\bibinfo {author} {\bibfnamefont {S.}~\bibnamefont {Succi}}, \bibinfo {author} {\bibfnamefont {C.}~\bibnamefont {Sanavio}}, \bibinfo {author} {\bibfnamefont {R.}~\bibnamefont {Scatamacchia}},\ and\ \bibinfo {author} {\bibfnamefont {C.}~\bibnamefont {De~Falco}},\ }\bibfield  {title} {\bibinfo {title} {Three carleman routes to the quantum simulation of classical fluids},\ }\href@noop {} {\bibfield  {journal} {\bibinfo  {journal} {Phys. Fluids}\ }\textbf {\bibinfo {volume} {36}} (\bibinfo {year} {2024}{\natexlab{a}})}\BibitemShut {NoStop}%
\bibitem [{\citenamefont {Succi}\ \emph {et~al.}(2024{\natexlab{b}})\citenamefont {Succi}, \citenamefont {Itani}, \citenamefont {Sanavio}, \citenamefont {Sreenivasan},\ and\ \citenamefont {Steijl}}]{succi2024ensemble}%
  \BibitemOpen
  \bibfield  {author} {\bibinfo {author} {\bibfnamefont {S.}~\bibnamefont {Succi}}, \bibinfo {author} {\bibfnamefont {W.}~\bibnamefont {Itani}}, \bibinfo {author} {\bibfnamefont {C.}~\bibnamefont {Sanavio}}, \bibinfo {author} {\bibfnamefont {K.~R.}\ \bibnamefont {Sreenivasan}},\ and\ \bibinfo {author} {\bibfnamefont {R.}~\bibnamefont {Steijl}},\ }\bibfield  {title} {\bibinfo {title} {Ensemble fluid simulations on quantum computers},\ }\href@noop {} {\bibfield  {journal} {\bibinfo  {journal} {Comput. Fluids}\ }\textbf {\bibinfo {volume} {270}},\ \bibinfo {pages} {106148} (\bibinfo {year} {2024}{\natexlab{b}})}\BibitemShut {NoStop}%
\bibitem [{\citenamefont {Lin}\ \emph {et~al.}(2022)\citenamefont {Lin}, \citenamefont {Lowrie}, \citenamefont {Aslangil}, \citenamefont {Suba{\c{s}}{\i}},\ and\ \citenamefont {Sornborger}}]{lin2022koopman}%
  \BibitemOpen
  \bibfield  {author} {\bibinfo {author} {\bibfnamefont {Y.~T.}\ \bibnamefont {Lin}}, \bibinfo {author} {\bibfnamefont {R.~B.}\ \bibnamefont {Lowrie}}, \bibinfo {author} {\bibfnamefont {D.}~\bibnamefont {Aslangil}}, \bibinfo {author} {\bibfnamefont {Y.}~\bibnamefont {Suba{\c{s}}{\i}}},\ and\ \bibinfo {author} {\bibfnamefont {A.~T.}\ \bibnamefont {Sornborger}},\ }\bibfield  {title} {\bibinfo {title} {{K}oopman von {N}eumann mechanics and the {K}oopman representation: A perspective on solving nonlinear dynamical systems with quantum computers},\ }\href@noop {} {\bibfield  {journal} {\bibinfo  {journal} {arXiv:2202.02188}\ } (\bibinfo {year} {2022})}\BibitemShut {NoStop}%
\bibitem [{\citenamefont {Schalkers}\ and\ \citenamefont {M{\"o}ller}(2024{\natexlab{a}})}]{schalkers2024efficient}%
  \BibitemOpen
  \bibfield  {author} {\bibinfo {author} {\bibfnamefont {M.~A.}\ \bibnamefont {Schalkers}}\ and\ \bibinfo {author} {\bibfnamefont {M.}~\bibnamefont {M{\"o}ller}},\ }\bibfield  {title} {\bibinfo {title} {Efficient and fail-safe quantum algorithm for the transport equation},\ }\href@noop {} {\bibfield  {journal} {\bibinfo  {journal} {J. Comput. Phys.}\ }\textbf {\bibinfo {volume} {502}},\ \bibinfo {pages} {112816} (\bibinfo {year} {2024}{\natexlab{a}})}\BibitemShut {NoStop}%
\bibitem [{\citenamefont {Kocherla}\ \emph {et~al.}(2023)\citenamefont {Kocherla}, \citenamefont {Song}, \citenamefont {Chrit}, \citenamefont {Gard}, \citenamefont {Dumitrescu}, \citenamefont {Alexeev},\ and\ \citenamefont {Bryngelson}}]{kocherla2023fully}%
  \BibitemOpen
  \bibfield  {author} {\bibinfo {author} {\bibfnamefont {S.}~\bibnamefont {Kocherla}}, \bibinfo {author} {\bibfnamefont {Z.}~\bibnamefont {Song}}, \bibinfo {author} {\bibfnamefont {F.~E.}\ \bibnamefont {Chrit}}, \bibinfo {author} {\bibfnamefont {B.}~\bibnamefont {Gard}}, \bibinfo {author} {\bibfnamefont {E.~F.}\ \bibnamefont {Dumitrescu}}, \bibinfo {author} {\bibfnamefont {A.}~\bibnamefont {Alexeev}},\ and\ \bibinfo {author} {\bibfnamefont {S.~H.}\ \bibnamefont {Bryngelson}},\ }\bibfield  {title} {\bibinfo {title} {Fully quantum algorithm for lattice boltzmann methods with application to partial differential equations},\ }\href@noop {} {\bibfield  {journal} {\bibinfo  {journal} {arXiv:2305.07148}\ } (\bibinfo {year} {2023})}\BibitemShut {NoStop}%
\bibitem [{\citenamefont {Kocherla}\ \emph {et~al.}(2024)\citenamefont {Kocherla}, \citenamefont {Adams}, \citenamefont {Song}, \citenamefont {Alexeev},\ and\ \citenamefont {Bryngelson}}]{kocherla2024two}%
  \BibitemOpen
  \bibfield  {author} {\bibinfo {author} {\bibfnamefont {S.}~\bibnamefont {Kocherla}}, \bibinfo {author} {\bibfnamefont {A.}~\bibnamefont {Adams}}, \bibinfo {author} {\bibfnamefont {Z.}~\bibnamefont {Song}}, \bibinfo {author} {\bibfnamefont {A.}~\bibnamefont {Alexeev}},\ and\ \bibinfo {author} {\bibfnamefont {S.~H.}\ \bibnamefont {Bryngelson}},\ }\bibfield  {title} {\bibinfo {title} {A two-circuit approach to reducing quantum resources for the quantum lattice boltzmann method},\ }\href@noop {} {\bibfield  {journal} {\bibinfo  {journal} {arXiv:2401.12248}\ } (\bibinfo {year} {2024})}\BibitemShut {NoStop}%
\bibitem [{\citenamefont {Penuel}\ \emph {et~al.}(2024)\citenamefont {Penuel}, \citenamefont {Katabarwa}, \citenamefont {Johnson}, \citenamefont {Farquhar}, \citenamefont {Cao},\ and\ \citenamefont {Garrett}}]{penuel2024feasibility}%
  \BibitemOpen
  \bibfield  {author} {\bibinfo {author} {\bibfnamefont {J.}~\bibnamefont {Penuel}}, \bibinfo {author} {\bibfnamefont {A.}~\bibnamefont {Katabarwa}}, \bibinfo {author} {\bibfnamefont {P.~D.}\ \bibnamefont {Johnson}}, \bibinfo {author} {\bibfnamefont {C.}~\bibnamefont {Farquhar}}, \bibinfo {author} {\bibfnamefont {Y.}~\bibnamefont {Cao}},\ and\ \bibinfo {author} {\bibfnamefont {M.~C.}\ \bibnamefont {Garrett}},\ }\bibfield  {title} {\bibinfo {title} {Feasibility of accelerating incompressible computational fluid dynamics simulations with fault-tolerant quantum computers},\ }\href@noop {} {\bibfield  {journal} {\bibinfo  {journal} {arXiv:2406.06323}\ } (\bibinfo {year} {2024})}\BibitemShut {NoStop}%
\bibitem [{\citenamefont {Jin}\ \emph {et~al.}(2024{\natexlab{a}})\citenamefont {Jin}, \citenamefont {Liu},\ and\ \citenamefont {Yu}}]{jin2023quantum}%
  \BibitemOpen
  \bibfield  {author} {\bibinfo {author} {\bibfnamefont {S.}~\bibnamefont {Jin}}, \bibinfo {author} {\bibfnamefont {N.}~\bibnamefont {Liu}},\ and\ \bibinfo {author} {\bibfnamefont {Y.}~\bibnamefont {Yu}},\ }\bibfield  {title} {\bibinfo {title} {Quantum simulation of partial differential equations via schr{"o}dingerization},\ }\href {https://doi.org/10.1103/PhysRevLett.133.230602} {\bibfield  {journal} {\bibinfo  {journal} {Phys. Rev. Lett.}\ }\textbf {\bibinfo {volume} {133}},\ \bibinfo {pages} {230602} (\bibinfo {year} {2024}{\natexlab{a}})}\BibitemShut {NoStop}%
\bibitem [{\citenamefont {Lu}\ and\ \citenamefont {Yang}(2024)}]{lu2024quantum}%
  \BibitemOpen
  \bibfield  {author} {\bibinfo {author} {\bibfnamefont {Z.}~\bibnamefont {Lu}}\ and\ \bibinfo {author} {\bibfnamefont {Y.}~\bibnamefont {Yang}},\ }\bibfield  {title} {\bibinfo {title} {Quantum computing of reacting flows via {H}amiltonian simulation},\ }\href@noop {} {\bibfield  {journal} {\bibinfo  {journal} {Proc. Combust. Inst.}\ }\textbf {\bibinfo {volume} {40}},\ \bibinfo {pages} {105440} (\bibinfo {year} {2024})}\BibitemShut {NoStop}%
\bibitem [{\citenamefont {Meng}\ \emph {et~al.}(2023)\citenamefont {Meng}, \citenamefont {Yang} \emph {et~al.}}]{meng2023quantum}%
  \BibitemOpen
  \bibfield  {author} {\bibinfo {author} {\bibfnamefont {Z.}~\bibnamefont {Meng}}, \bibinfo {author} {\bibfnamefont {Y.}~\bibnamefont {Yang}}, \emph {et~al.},\ }\bibfield  {title} {\bibinfo {title} {Quantum computing of fluid dynamics using the hydrodynamic {S}chr{\"o}dinger equation},\ }\href@noop {} {\bibfield  {journal} {\bibinfo  {journal} {Phys. Rev. Research}\ }\textbf {\bibinfo {volume} {5}},\ \bibinfo {pages} {033182} (\bibinfo {year} {2023})}\BibitemShut {NoStop}%
\bibitem [{\citenamefont {Jin}\ \emph {et~al.}(2024{\natexlab{b}})\citenamefont {Jin}, \citenamefont {Li}, \citenamefont {Liu},\ and\ \citenamefont {Yu}}]{jin2024quantum}%
  \BibitemOpen
  \bibfield  {author} {\bibinfo {author} {\bibfnamefont {S.}~\bibnamefont {Jin}}, \bibinfo {author} {\bibfnamefont {X.}~\bibnamefont {Li}}, \bibinfo {author} {\bibfnamefont {N.}~\bibnamefont {Liu}},\ and\ \bibinfo {author} {\bibfnamefont {Y.}~\bibnamefont {Yu}},\ }\bibfield  {title} {\bibinfo {title} {Quantum simulation for partial differential equations with physical boundary or interface conditions},\ }\href@noop {} {\bibfield  {journal} {\bibinfo  {journal} {J. Comput. Phys.}\ }\textbf {\bibinfo {volume} {498}},\ \bibinfo {pages} {112707} (\bibinfo {year} {2024}{\natexlab{b}})}\BibitemShut {NoStop}%
\bibitem [{\citenamefont {K{\"o}cher}\ \emph {et~al.}(2025)\citenamefont {K{\"o}cher}, \citenamefont {Rose}, \citenamefont {Bharadwaj}, \citenamefont {Schumacher},\ and\ \citenamefont {Schumacher}}]{kocher2024numerical}%
  \BibitemOpen
  \bibfield  {author} {\bibinfo {author} {\bibfnamefont {N.}~\bibnamefont {K{\"o}cher}}, \bibinfo {author} {\bibfnamefont {H.}~\bibnamefont {Rose}}, \bibinfo {author} {\bibfnamefont {S.~S.}\ \bibnamefont {Bharadwaj}}, \bibinfo {author} {\bibfnamefont {J.}~\bibnamefont {Schumacher}},\ and\ \bibinfo {author} {\bibfnamefont {S.}~\bibnamefont {Schumacher}},\ }\bibfield  {title} {\bibinfo {title} {Numerical solution of nonlinear schr{\"o}dinger equation by a hybrid pseudospectral-variational quantum algorithm},\ }\href@noop {} {\bibfield  {journal} {\bibinfo  {journal} {Sci. Rep.}\ }\textbf {\bibinfo {volume} {15}},\ \bibinfo {pages} {23478} (\bibinfo {year} {2025})}\BibitemShut {NoStop}%
\bibitem [{\citenamefont {Bravo-Prieto}\ \emph {et~al.}(2023)\citenamefont {Bravo-Prieto}, \citenamefont {LaRose}, \citenamefont {Cerezo}, \citenamefont {Subasi}, \citenamefont {Cincio},\ and\ \citenamefont {Coles}}]{bravo2023variational}%
  \BibitemOpen
  \bibfield  {author} {\bibinfo {author} {\bibfnamefont {C.}~\bibnamefont {Bravo-Prieto}}, \bibinfo {author} {\bibfnamefont {R.}~\bibnamefont {LaRose}}, \bibinfo {author} {\bibfnamefont {M.}~\bibnamefont {Cerezo}}, \bibinfo {author} {\bibfnamefont {Y.}~\bibnamefont {Subasi}}, \bibinfo {author} {\bibfnamefont {L.}~\bibnamefont {Cincio}},\ and\ \bibinfo {author} {\bibfnamefont {P.~J.}\ \bibnamefont {Coles}},\ }\bibfield  {title} {\bibinfo {title} {Variational quantum linear solver},\ }\href@noop {} {\bibfield  {journal} {\bibinfo  {journal} {Quantum}\ }\textbf {\bibinfo {volume} {7}},\ \bibinfo {pages} {1188} (\bibinfo {year} {2023})}\BibitemShut {NoStop}%
\bibitem [{\citenamefont {Cerezo}\ \emph {et~al.}(2021)\citenamefont {Cerezo}, \citenamefont {Arrasmith}, \citenamefont {Babbush}, \citenamefont {Benjamin}, \citenamefont {Endo}, \citenamefont {Fujii}, \citenamefont {McClean}, \citenamefont {Mitarai}, \citenamefont {Yuan}, \citenamefont {Cincio} \emph {et~al.}}]{cerezo2021variational}%
  \BibitemOpen
  \bibfield  {author} {\bibinfo {author} {\bibfnamefont {M.}~\bibnamefont {Cerezo}}, \bibinfo {author} {\bibfnamefont {A.}~\bibnamefont {Arrasmith}}, \bibinfo {author} {\bibfnamefont {R.}~\bibnamefont {Babbush}}, \bibinfo {author} {\bibfnamefont {S.~C.}\ \bibnamefont {Benjamin}}, \bibinfo {author} {\bibfnamefont {S.}~\bibnamefont {Endo}}, \bibinfo {author} {\bibfnamefont {K.}~\bibnamefont {Fujii}}, \bibinfo {author} {\bibfnamefont {J.~R.}\ \bibnamefont {McClean}}, \bibinfo {author} {\bibfnamefont {K.}~\bibnamefont {Mitarai}}, \bibinfo {author} {\bibfnamefont {X.}~\bibnamefont {Yuan}}, \bibinfo {author} {\bibfnamefont {L.}~\bibnamefont {Cincio}}, \emph {et~al.},\ }\bibfield  {title} {\bibinfo {title} {Variational quantum algorithms},\ }\href@noop {} {\bibfield  {journal} {\bibinfo  {journal} {Nat. Rev. Phys.}\ }\textbf {\bibinfo {volume} {3}},\ \bibinfo {pages} {625} (\bibinfo {year} {2021})}\BibitemShut {NoStop}%
\bibitem [{\citenamefont {Wang}\ \emph {et~al.}(2021)\citenamefont {Wang}, \citenamefont {Fontana}, \citenamefont {Cerezo}, \citenamefont {Sharma}, \citenamefont {Sone}, \citenamefont {Cincio},\ and\ \citenamefont {Coles}}]{wang2021}%
  \BibitemOpen
  \bibfield  {author} {\bibinfo {author} {\bibfnamefont {S.}~\bibnamefont {Wang}}, \bibinfo {author} {\bibfnamefont {E.}~\bibnamefont {Fontana}}, \bibinfo {author} {\bibfnamefont {M.}~\bibnamefont {Cerezo}}, \bibinfo {author} {\bibfnamefont {K.}~\bibnamefont {Sharma}}, \bibinfo {author} {\bibfnamefont {A.}~\bibnamefont {Sone}}, \bibinfo {author} {\bibfnamefont {L.}~\bibnamefont {Cincio}},\ and\ \bibinfo {author} {\bibfnamefont {P.~J.}\ \bibnamefont {Coles}},\ }\bibfield  {title} {\bibinfo {title} {Noise-induced barren plateaus in variational quantum algorithms},\ }\href@noop {} {\bibfield  {journal} {\bibinfo  {journal} {Nat. Commun.}\ }\textbf {\bibinfo {volume} {12}},\ \bibinfo {pages} {6961} (\bibinfo {year} {2021})}\BibitemShut {NoStop}%
\bibitem [{\citenamefont {Holmes}\ \emph {et~al.}(2022)\citenamefont {Holmes}, \citenamefont {Sharma}, \citenamefont {Cerezo},\ and\ \citenamefont {Coles}}]{holmes2022connecting}%
  \BibitemOpen
  \bibfield  {author} {\bibinfo {author} {\bibfnamefont {Z.}~\bibnamefont {Holmes}}, \bibinfo {author} {\bibfnamefont {K.}~\bibnamefont {Sharma}}, \bibinfo {author} {\bibfnamefont {M.}~\bibnamefont {Cerezo}},\ and\ \bibinfo {author} {\bibfnamefont {P.~J.}\ \bibnamefont {Coles}},\ }\bibfield  {title} {\bibinfo {title} {Connecting ansatz expressibility to gradient magnitudes and barren plateaus},\ }\href@noop {} {\bibfield  {journal} {\bibinfo  {journal} {PRX Quantum}\ }\textbf {\bibinfo {volume} {3}},\ \bibinfo {pages} {010313} (\bibinfo {year} {2022})}\BibitemShut {NoStop}%
\bibitem [{\citenamefont {Morales}\ \emph {et~al.}(2024)\citenamefont {Morales}, \citenamefont {Pira}, \citenamefont {Schleich}, \citenamefont {Koor}, \citenamefont {Costa}, \citenamefont {An}, \citenamefont {Aspuru-Guzik}, \citenamefont {Lin}, \citenamefont {Rebentrost},\ and\ \citenamefont {Berry}}]{morales2024quantum}%
  \BibitemOpen
  \bibfield  {author} {\bibinfo {author} {\bibfnamefont {M.~E.}\ \bibnamefont {Morales}}, \bibinfo {author} {\bibfnamefont {L.}~\bibnamefont {Pira}}, \bibinfo {author} {\bibfnamefont {P.}~\bibnamefont {Schleich}}, \bibinfo {author} {\bibfnamefont {K.}~\bibnamefont {Koor}}, \bibinfo {author} {\bibfnamefont {P.}~\bibnamefont {Costa}}, \bibinfo {author} {\bibfnamefont {D.}~\bibnamefont {An}}, \bibinfo {author} {\bibfnamefont {A.}~\bibnamefont {Aspuru-Guzik}}, \bibinfo {author} {\bibfnamefont {L.}~\bibnamefont {Lin}}, \bibinfo {author} {\bibfnamefont {P.}~\bibnamefont {Rebentrost}},\ and\ \bibinfo {author} {\bibfnamefont {D.~W.}\ \bibnamefont {Berry}},\ }\bibfield  {title} {\bibinfo {title} {Quantum linear system solvers: A survey of algorithms and applications},\ }\href@noop {} {\bibfield  {journal} {\bibinfo  {journal} {arXiv preprint arXiv:2411.02522}\ } (\bibinfo {year} {2024})}\BibitemShut {NoStop}%
\bibitem [{\citenamefont {Joseph}(2020)}]{joseph2020koopman}%
  \BibitemOpen
  \bibfield  {author} {\bibinfo {author} {\bibfnamefont {I.}~\bibnamefont {Joseph}},\ }\bibfield  {title} {\bibinfo {title} {Koopman--von neumann approach to quantum simulation of nonlinear classical dynamics},\ }\href@noop {} {\bibfield  {journal} {\bibinfo  {journal} {Physical Review Research}\ }\textbf {\bibinfo {volume} {2}},\ \bibinfo {pages} {043102} (\bibinfo {year} {2020})}\BibitemShut {NoStop}%
\bibitem [{\citenamefont {Giannakis}\ \emph {et~al.}(2022)\citenamefont {Giannakis}, \citenamefont {Ourmazd}, \citenamefont {Pfeffer}, \citenamefont {Schumacher},\ and\ \citenamefont {Slawinska}}]{giannakis2022embedding}%
  \BibitemOpen
  \bibfield  {author} {\bibinfo {author} {\bibfnamefont {D.}~\bibnamefont {Giannakis}}, \bibinfo {author} {\bibfnamefont {A.}~\bibnamefont {Ourmazd}}, \bibinfo {author} {\bibfnamefont {P.}~\bibnamefont {Pfeffer}}, \bibinfo {author} {\bibfnamefont {J.}~\bibnamefont {Schumacher}},\ and\ \bibinfo {author} {\bibfnamefont {J.}~\bibnamefont {Slawinska}},\ }\bibfield  {title} {\bibinfo {title} {Embedding classical dynamics in a quantum computer},\ }\href@noop {} {\bibfield  {journal} {\bibinfo  {journal} {Phys. Rev. A}\ }\textbf {\bibinfo {volume} {105}},\ \bibinfo {pages} {052404} (\bibinfo {year} {2022})}\BibitemShut {NoStop}%
\bibitem [{\citenamefont {Giannakis}\ \emph {et~al.}(2025)\citenamefont {Giannakis}, \citenamefont {Jebelli}, \citenamefont {Montgomerry}, \citenamefont {Pfeffer}, \citenamefont {Schumacher},\ and\ \citenamefont {Slawinska}}]{giannakis2025second}%
  \BibitemOpen
  \bibfield  {author} {\bibinfo {author} {\bibfnamefont {D.}~\bibnamefont {Giannakis}}, \bibinfo {author} {\bibfnamefont {M.~J.~L.}\ \bibnamefont {Jebelli}}, \bibinfo {author} {\bibfnamefont {M.}~\bibnamefont {Montgomerry}}, \bibinfo {author} {\bibfnamefont {P.}~\bibnamefont {Pfeffer}}, \bibinfo {author} {\bibfnamefont {J.}~\bibnamefont {Schumacher}},\ and\ \bibinfo {author} {\bibfnamefont {J.}~\bibnamefont {Slawinska}},\ }\bibfield  {title} {\bibinfo {title} {Second quantization for classical nonlinear dynamics},\ }\href@noop {} {\bibfield  {journal} {\bibinfo  {journal} {arXiv preprint arXiv:2501.07419}\ } (\bibinfo {year} {2025})}\BibitemShut {NoStop}%
\bibitem [{\citenamefont {Wu}\ \emph {et~al.}(2025)\citenamefont {Wu}, \citenamefont {Wang},\ and\ \citenamefont {Li}}]{wu2025quantum}%
  \BibitemOpen
  \bibfield  {author} {\bibinfo {author} {\bibfnamefont {H.-C.}\ \bibnamefont {Wu}}, \bibinfo {author} {\bibfnamefont {J.}~\bibnamefont {Wang}},\ and\ \bibinfo {author} {\bibfnamefont {X.}~\bibnamefont {Li}},\ }\bibfield  {title} {\bibinfo {title} {Quantum algorithms for nonlinear dynamics: Revisiting carleman linearization with no dissipative conditions},\ }\href@noop {} {\bibfield  {journal} {\bibinfo  {journal} {SIAM Journal on Scientific Computing}\ }\textbf {\bibinfo {volume} {47}},\ \bibinfo {pages} {A943} (\bibinfo {year} {2025})}\BibitemShut {NoStop}%
\bibitem [{\citenamefont {Gonzalez-Conde}\ \emph {et~al.}(2024)\citenamefont {Gonzalez-Conde}, \citenamefont {Watts}, \citenamefont {Rodriguez-Grasa},\ and\ \citenamefont {Sanz}}]{gonzalez2024efficient}%
  \BibitemOpen
  \bibfield  {author} {\bibinfo {author} {\bibfnamefont {J.}~\bibnamefont {Gonzalez-Conde}}, \bibinfo {author} {\bibfnamefont {T.~W.}\ \bibnamefont {Watts}}, \bibinfo {author} {\bibfnamefont {P.}~\bibnamefont {Rodriguez-Grasa}},\ and\ \bibinfo {author} {\bibfnamefont {M.}~\bibnamefont {Sanz}},\ }\bibfield  {title} {\bibinfo {title} {Efficient quantum amplitude encoding of polynomial functions},\ }\href@noop {} {\bibfield  {journal} {\bibinfo  {journal} {Quantum}\ }\textbf {\bibinfo {volume} {8}},\ \bibinfo {pages} {1297} (\bibinfo {year} {2024})}\BibitemShut {NoStop}%
\bibitem [{\citenamefont {Pope}(2000)}]{pope2000turbulent}%
  \BibitemOpen
  \bibfield  {author} {\bibinfo {author} {\bibfnamefont {S.~B.}\ \bibnamefont {Pope}},\ }\bibfield  {title} {\bibinfo {title} {Turbulent flows},\ }\href@noop {} {\bibfield  {journal} {\bibinfo  {journal} {Cambridge University Press}\ } (\bibinfo {year} {2000})}\BibitemShut {NoStop}%
\bibitem [{\citenamefont {Williams}\ \emph {et~al.}(2015)\citenamefont {Williams}, \citenamefont {Kevrekidis},\ and\ \citenamefont {Rowley}}]{williams2015data}%
  \BibitemOpen
  \bibfield  {author} {\bibinfo {author} {\bibfnamefont {M.~O.}\ \bibnamefont {Williams}}, \bibinfo {author} {\bibfnamefont {I.~G.}\ \bibnamefont {Kevrekidis}},\ and\ \bibinfo {author} {\bibfnamefont {C.~W.}\ \bibnamefont {Rowley}},\ }\bibfield  {title} {\bibinfo {title} {A data--driven approximation of the koopman operator: Extending dynamic mode decomposition},\ }\href@noop {} {\bibfield  {journal} {\bibinfo  {journal} {J. Nonlinear Sci.}\ }\textbf {\bibinfo {volume} {25}},\ \bibinfo {pages} {1307} (\bibinfo {year} {2015})}\BibitemShut {NoStop}%
\bibitem [{\citenamefont {Valva}\ and\ \citenamefont {Giannakis}(2024)}]{valva2024physics}%
  \BibitemOpen
  \bibfield  {author} {\bibinfo {author} {\bibfnamefont {C.}~\bibnamefont {Valva}}\ and\ \bibinfo {author} {\bibfnamefont {D.}~\bibnamefont {Giannakis}},\ }\bibfield  {title} {\bibinfo {title} {Physics-informed spectral approximation of koopman operators},\ }\href@noop {} {\bibfield  {journal} {\bibinfo  {journal} {arXiv:2408.05663}\ } (\bibinfo {year} {2024})}\BibitemShut {NoStop}%
\bibitem [{\citenamefont {Liao}(2004)}]{liao2004homotopy}%
  \BibitemOpen
  \bibfield  {author} {\bibinfo {author} {\bibfnamefont {S.}~\bibnamefont {Liao}},\ }\bibfield  {title} {\bibinfo {title} {On the homotopy analysis method for nonlinear problems},\ }\href@noop {} {\bibfield  {journal} {\bibinfo  {journal} {Applied Mathematics and Computation}\ }\textbf {\bibinfo {volume} {147}},\ \bibinfo {pages} {499} (\bibinfo {year} {2004})}\BibitemShut {NoStop}%
\bibitem [{\citenamefont {Liao}(1992)}]{liao1992proposed}%
  \BibitemOpen
  \bibfield  {author} {\bibinfo {author} {\bibfnamefont {S.-J.}\ \bibnamefont {Liao}},\ }\emph {\bibinfo {title} {The proposed homotopy analysis technique for the solution of nonlinear problems}},\ \href@noop {} {Ph.D. thesis},\ \bibinfo  {school} {PhD thesis, Shanghai Jiao Tong University} (\bibinfo {year} {1992})\BibitemShut {NoStop}%
\bibitem [{\citenamefont {He}(1999)}]{he1999homotopy}%
  \BibitemOpen
  \bibfield  {author} {\bibinfo {author} {\bibfnamefont {J.-H.}\ \bibnamefont {He}},\ }\bibfield  {title} {\bibinfo {title} {Homotopy perturbation technique},\ }\href@noop {} {\bibfield  {journal} {\bibinfo  {journal} {Comput. Methods Appl. Mech. Eng.}\ }\textbf {\bibinfo {volume} {178}},\ \bibinfo {pages} {257} (\bibinfo {year} {1999})}\BibitemShut {NoStop}%
\bibitem [{\citenamefont {Xue}\ \emph {et~al.}(2021)\citenamefont {Xue}, \citenamefont {Wu},\ and\ \citenamefont {Guo}}]{xue2021quantum}%
  \BibitemOpen
  \bibfield  {author} {\bibinfo {author} {\bibfnamefont {C.}~\bibnamefont {Xue}}, \bibinfo {author} {\bibfnamefont {Y.-C.}\ \bibnamefont {Wu}},\ and\ \bibinfo {author} {\bibfnamefont {G.-P.}\ \bibnamefont {Guo}},\ }\bibfield  {title} {\bibinfo {title} {Quantum homotopy perturbation method for nonlinear dissipative ordinary differential equations},\ }\href@noop {} {\bibfield  {journal} {\bibinfo  {journal} {New Journal of Physics}\ }\textbf {\bibinfo {volume} {23}},\ \bibinfo {pages} {123035} (\bibinfo {year} {2021})}\BibitemShut {NoStop}%
\bibitem [{\citenamefont {Xue}\ \emph {et~al.}(2022)\citenamefont {Xue}, \citenamefont {Xu}, \citenamefont {Wu},\ and\ \citenamefont {Guo}}]{xue2022quantum}%
  \BibitemOpen
  \bibfield  {author} {\bibinfo {author} {\bibfnamefont {C.}~\bibnamefont {Xue}}, \bibinfo {author} {\bibfnamefont {X.-F.}\ \bibnamefont {Xu}}, \bibinfo {author} {\bibfnamefont {Y.-C.}\ \bibnamefont {Wu}},\ and\ \bibinfo {author} {\bibfnamefont {G.-P.}\ \bibnamefont {Guo}},\ }\bibfield  {title} {\bibinfo {title} {Quantum algorithm for solving a quadratic nonlinear system of equations},\ }\href@noop {} {\bibfield  {journal} {\bibinfo  {journal} {Physical Review A}\ }\textbf {\bibinfo {volume} {106}},\ \bibinfo {pages} {032427} (\bibinfo {year} {2022})}\BibitemShut {NoStop}%
\bibitem [{\citenamefont {Liao}(2003{\natexlab{a}})}]{liao2003beyond}%
  \BibitemOpen
  \bibfield  {author} {\bibinfo {author} {\bibfnamefont {S.}~\bibnamefont {Liao}},\ }\href@noop {} {\emph {\bibinfo {title} {Beyond perturbation: introduction to the homotopy analysis method}}}\ (\bibinfo  {publisher} {Chapman and Hall/CRC},\ \bibinfo {year} {2003})\BibitemShut {NoStop}%
\bibitem [{\citenamefont {Liang}\ and\ \citenamefont {Jeffrey}(2009)}]{liang2009comparison}%
  \BibitemOpen
  \bibfield  {author} {\bibinfo {author} {\bibfnamefont {S.}~\bibnamefont {Liang}}\ and\ \bibinfo {author} {\bibfnamefont {D.~J.}\ \bibnamefont {Jeffrey}},\ }\bibfield  {title} {\bibinfo {title} {Comparison of homotopy analysis method and homotopy perturbation method through an evolution equation},\ }\href@noop {} {\bibfield  {journal} {\bibinfo  {journal} {Commun. Nonlinear Sci. Numer.}\ }\textbf {\bibinfo {volume} {14}},\ \bibinfo {pages} {4057} (\bibinfo {year} {2009})}\BibitemShut {NoStop}%
\bibitem [{\citenamefont {Liao}(2012)}]{liao2012homotopy}%
  \BibitemOpen
  \bibfield  {author} {\bibinfo {author} {\bibfnamefont {S.}~\bibnamefont {Liao}},\ }\href@noop {} {\emph {\bibinfo {title} {Homotopy analysis method in nonlinear differential equations}}}\ (\bibinfo  {publisher} {Springer},\ \bibinfo {year} {2012})\BibitemShut {NoStop}%
\bibitem [{\citenamefont {Gupta}\ and\ \citenamefont {Ray}(2014)}]{gupta2014comparison}%
  \BibitemOpen
  \bibfield  {author} {\bibinfo {author} {\bibfnamefont {A.}~\bibnamefont {Gupta}}\ and\ \bibinfo {author} {\bibfnamefont {S.~S.}\ \bibnamefont {Ray}},\ }\bibfield  {title} {\bibinfo {title} {Comparison between homotopy perturbation method and optimal homotopy asymptotic method for the soliton solutions of boussinesq--burger equations},\ }\href@noop {} {\bibfield  {journal} {\bibinfo  {journal} {Comp. Fluids}\ }\textbf {\bibinfo {volume} {103}},\ \bibinfo {pages} {34} (\bibinfo {year} {2014})}\BibitemShut {NoStop}%
\bibitem [{\citenamefont {Bharadwaj}\ \emph {et~al.}(2023)\citenamefont {Bharadwaj}, \citenamefont {Nadiga}, \citenamefont {Eidenbenz},\ and\ \citenamefont {Sreenivasan}}]{bharadwaj2023quantum}%
  \BibitemOpen
  \bibfield  {author} {\bibinfo {author} {\bibfnamefont {S.~S.}\ \bibnamefont {Bharadwaj}}, \bibinfo {author} {\bibfnamefont {B.}~\bibnamefont {Nadiga}}, \bibinfo {author} {\bibfnamefont {S.}~\bibnamefont {Eidenbenz}},\ and\ \bibinfo {author} {\bibfnamefont {K.~R.}\ \bibnamefont {Sreenivasan}},\ }\bibfield  {title} {\bibinfo {title} {Quantum boundary integral algorithm for linear and nonlinear pdes in quantum computation of fluid dynamics {A}70.00008},\ }\href@noop {} {\bibfield  {journal} {\bibinfo  {journal} {Bull. Am. Phys. Soc.}\ } (\bibinfo {year} {2023})}\BibitemShut {NoStop}%
\bibitem [{\citenamefont {Aaronson}(2015)}]{aaronson2015read}%
  \BibitemOpen
  \bibfield  {author} {\bibinfo {author} {\bibfnamefont {S.}~\bibnamefont {Aaronson}},\ }\bibfield  {title} {\bibinfo {title} {Read the fine print},\ }\href@noop {} {\bibfield  {journal} {\bibinfo  {journal} {Nat. Phys.}\ }\textbf {\bibinfo {volume} {11}},\ \bibinfo {pages} {291} (\bibinfo {year} {2015})}\BibitemShut {NoStop}%
\bibitem [{\citenamefont {Berry}\ \emph {et~al.}(2007)\citenamefont {Berry}, \citenamefont {Ahokas}, \citenamefont {Cleve},\ and\ \citenamefont {Sanders}}]{berry2007efficient}%
  \BibitemOpen
  \bibfield  {author} {\bibinfo {author} {\bibfnamefont {D.~W.}\ \bibnamefont {Berry}}, \bibinfo {author} {\bibfnamefont {G.}~\bibnamefont {Ahokas}}, \bibinfo {author} {\bibfnamefont {R.}~\bibnamefont {Cleve}},\ and\ \bibinfo {author} {\bibfnamefont {B.~C.}\ \bibnamefont {Sanders}},\ }\bibfield  {title} {\bibinfo {title} {Efficient quantum algorithms for simulating sparse hamiltonians},\ }\href@noop {} {\bibfield  {journal} {\bibinfo  {journal} {Communications in Mathematical Physics}\ }\textbf {\bibinfo {volume} {270}},\ \bibinfo {pages} {359} (\bibinfo {year} {2007})}\BibitemShut {NoStop}%
\bibitem [{\citenamefont {Berry}\ \emph {et~al.}(2014)\citenamefont {Berry}, \citenamefont {Childs}, \citenamefont {Cleve}, \citenamefont {Kothari},\ and\ \citenamefont {Somma}}]{berry2014exponential}%
  \BibitemOpen
  \bibfield  {author} {\bibinfo {author} {\bibfnamefont {D.~W.}\ \bibnamefont {Berry}}, \bibinfo {author} {\bibfnamefont {A.~M.}\ \bibnamefont {Childs}}, \bibinfo {author} {\bibfnamefont {R.}~\bibnamefont {Cleve}}, \bibinfo {author} {\bibfnamefont {R.}~\bibnamefont {Kothari}},\ and\ \bibinfo {author} {\bibfnamefont {R.~D.}\ \bibnamefont {Somma}},\ }\bibfield  {title} {\bibinfo {title} {Exponential improvement in precision for simulating sparse hamiltonians},\ }in\ \href@noop {} {\emph {\bibinfo {booktitle} {Proceedings of the forty-sixth annual ACM symposium on Theory of computing}}}\ (\bibinfo {year} {2014})\ p.\ \bibinfo {pages} {283}\BibitemShut {NoStop}%
\bibitem [{\citenamefont {Berry}\ \emph {et~al.}(2015{\natexlab{a}})\citenamefont {Berry}, \citenamefont {Childs}, \citenamefont {Cleve}, \citenamefont {Kothari},\ and\ \citenamefont {Somma}}]{berry2015simulating}%
  \BibitemOpen
  \bibfield  {author} {\bibinfo {author} {\bibfnamefont {D.~W.}\ \bibnamefont {Berry}}, \bibinfo {author} {\bibfnamefont {A.~M.}\ \bibnamefont {Childs}}, \bibinfo {author} {\bibfnamefont {R.}~\bibnamefont {Cleve}}, \bibinfo {author} {\bibfnamefont {R.}~\bibnamefont {Kothari}},\ and\ \bibinfo {author} {\bibfnamefont {R.~D.}\ \bibnamefont {Somma}},\ }\bibfield  {title} {\bibinfo {title} {Simulating hamiltonian dynamics with a truncated taylor series},\ }\href@noop {} {\bibfield  {journal} {\bibinfo  {journal} {Phys. Rev. Lett.}\ }\textbf {\bibinfo {volume} {114}},\ \bibinfo {pages} {090502} (\bibinfo {year} {2015}{\natexlab{a}})}\BibitemShut {NoStop}%
\bibitem [{\citenamefont {Berry}\ \emph {et~al.}(2015{\natexlab{b}})\citenamefont {Berry}, \citenamefont {Childs},\ and\ \citenamefont {Kothari}}]{berry2015hamiltonian}%
  \BibitemOpen
  \bibfield  {author} {\bibinfo {author} {\bibfnamefont {D.~W.}\ \bibnamefont {Berry}}, \bibinfo {author} {\bibfnamefont {A.~M.}\ \bibnamefont {Childs}},\ and\ \bibinfo {author} {\bibfnamefont {R.}~\bibnamefont {Kothari}},\ }\bibfield  {title} {\bibinfo {title} {Hamiltonian simulation with nearly optimal dependence on all parameters},\ }in\ \href@noop {} {\emph {\bibinfo {booktitle} {2015 IEEE 56th annual symposium on foundations of computer science}}}\ (\bibinfo {organization} {IEEE},\ \bibinfo {year} {2015})\ p.\ \bibinfo {pages} {792}\BibitemShut {NoStop}%
\bibitem [{\citenamefont {Low}\ and\ \citenamefont {Chuang}(2017)}]{low2017optimal}%
  \BibitemOpen
  \bibfield  {author} {\bibinfo {author} {\bibfnamefont {G.~H.}\ \bibnamefont {Low}}\ and\ \bibinfo {author} {\bibfnamefont {I.~L.}\ \bibnamefont {Chuang}},\ }\bibfield  {title} {\bibinfo {title} {Optimal hamiltonian simulation by quantum signal processing},\ }\href@noop {} {\bibfield  {journal} {\bibinfo  {journal} {Phys. Rev. Lett.}\ }\textbf {\bibinfo {volume} {118}},\ \bibinfo {pages} {010501} (\bibinfo {year} {2017})}\BibitemShut {NoStop}%
\bibitem [{\citenamefont {Berry}\ \emph {et~al.}(2020)\citenamefont {Berry}, \citenamefont {Childs}, \citenamefont {Su}, \citenamefont {Wang},\ and\ \citenamefont {Wiebe}}]{berry2020time}%
  \BibitemOpen
  \bibfield  {author} {\bibinfo {author} {\bibfnamefont {D.~W.}\ \bibnamefont {Berry}}, \bibinfo {author} {\bibfnamefont {A.~M.}\ \bibnamefont {Childs}}, \bibinfo {author} {\bibfnamefont {Y.}~\bibnamefont {Su}}, \bibinfo {author} {\bibfnamefont {X.}~\bibnamefont {Wang}},\ and\ \bibinfo {author} {\bibfnamefont {N.}~\bibnamefont {Wiebe}},\ }\bibfield  {title} {\bibinfo {title} {Time-dependent hamiltonian simulation with $ l 1$-norm scaling},\ }\href@noop {} {\bibfield  {journal} {\bibinfo  {journal} {Quantum}\ }\textbf {\bibinfo {volume} {4}},\ \bibinfo {pages} {254} (\bibinfo {year} {2020})}\BibitemShut {NoStop}%
\bibitem [{\citenamefont {An}\ \emph {et~al.}(2021)\citenamefont {An}, \citenamefont {Fang},\ and\ \citenamefont {Lin}}]{an2021time}%
  \BibitemOpen
  \bibfield  {author} {\bibinfo {author} {\bibfnamefont {D.}~\bibnamefont {An}}, \bibinfo {author} {\bibfnamefont {D.}~\bibnamefont {Fang}},\ and\ \bibinfo {author} {\bibfnamefont {L.}~\bibnamefont {Lin}},\ }\bibfield  {title} {\bibinfo {title} {Time-dependent unbounded hamiltonian simulation with vector norm scaling},\ }\href@noop {} {\bibfield  {journal} {\bibinfo  {journal} {Quantum}\ }\textbf {\bibinfo {volume} {5}},\ \bibinfo {pages} {459} (\bibinfo {year} {2021})}\BibitemShut {NoStop}%
\bibitem [{\citenamefont {Childs}\ and\ \citenamefont {Weibe}(2012)}]{childs2012hamiltonian}%
  \BibitemOpen
  \bibfield  {author} {\bibinfo {author} {\bibfnamefont {A.}~\bibnamefont {Childs}}\ and\ \bibinfo {author} {\bibfnamefont {N.}~\bibnamefont {Weibe}},\ }\bibfield  {title} {\bibinfo {title} {Hamiltonian simulation using linear combinations of unitary operations},\ }\href@noop {} {\bibfield  {journal} {\bibinfo  {journal} {Quantum Info. Comput.}\ }\textbf {\bibinfo {volume} {12}},\ \bibinfo {pages} {910} (\bibinfo {year} {2012})}\BibitemShut {NoStop}%
\bibitem [{\citenamefont {Childs}\ \emph {et~al.}(2017)\citenamefont {Childs}, \citenamefont {Kothari},\ and\ \citenamefont {Somma}}]{childs2017quantum}%
  \BibitemOpen
  \bibfield  {author} {\bibinfo {author} {\bibfnamefont {A.~M.}\ \bibnamefont {Childs}}, \bibinfo {author} {\bibfnamefont {R.}~\bibnamefont {Kothari}},\ and\ \bibinfo {author} {\bibfnamefont {R.~D.}\ \bibnamefont {Somma}},\ }\bibfield  {title} {\bibinfo {title} {Quantum algorithm for systems of linear equations with exponentially improved dependence on precision},\ }\href@noop {} {\bibfield  {journal} {\bibinfo  {journal} {SIAM J. Comput.}\ }\textbf {\bibinfo {volume} {46}},\ \bibinfo {pages} {1920} (\bibinfo {year} {2017})}\BibitemShut {NoStop}%
\bibitem [{\citenamefont {Schalkers}\ and\ \citenamefont {M{\"o}ller}(2024{\natexlab{b}})}]{schalkers2024importance}%
  \BibitemOpen
  \bibfield  {author} {\bibinfo {author} {\bibfnamefont {M.~A.}\ \bibnamefont {Schalkers}}\ and\ \bibinfo {author} {\bibfnamefont {M.}~\bibnamefont {M{\"o}ller}},\ }\bibfield  {title} {\bibinfo {title} {On the importance of data encoding in quantum boltzmann methods},\ }\href@noop {} {\bibfield  {journal} {\bibinfo  {journal} {Quantum Inf. Process.}\ }\textbf {\bibinfo {volume} {23}},\ \bibinfo {pages} {20} (\bibinfo {year} {2024}{\natexlab{b}})}\BibitemShut {NoStop}%
\bibitem [{\citenamefont {Wang}\ \emph {et~al.}(2025)\citenamefont {Wang}, \citenamefont {Meng}, \citenamefont {Zhao},\ and\ \citenamefont {Yang}}]{wang2025quantum}%
  \BibitemOpen
  \bibfield  {author} {\bibinfo {author} {\bibfnamefont {B.}~\bibnamefont {Wang}}, \bibinfo {author} {\bibfnamefont {Z.}~\bibnamefont {Meng}}, \bibinfo {author} {\bibfnamefont {Y.}~\bibnamefont {Zhao}},\ and\ \bibinfo {author} {\bibfnamefont {Y.}~\bibnamefont {Yang}},\ }\bibfield  {title} {\bibinfo {title} {Quantum lattice boltzmann method for simulating nonlinear fluid dynamics},\ }\href {https://doi.org/10.1038/s41534-025-01142-6} {\bibfield  {journal} {\bibinfo  {journal} {npj Quantum Inf.}\ }\textbf {\bibinfo {volume} {11}},\ \bibinfo {pages} {196} (\bibinfo {year} {2025})}\BibitemShut {NoStop}%
\bibitem [{\citenamefont {Meng}\ \emph {et~al.}(2024)\citenamefont {Meng}, \citenamefont {Zhong}, \citenamefont {Xu}, \citenamefont {Wang}, \citenamefont {Chen}, \citenamefont {Jin}, \citenamefont {Zhu}, \citenamefont {Gao}, \citenamefont {Wu}, \citenamefont {Zhang} \emph {et~al.}}]{meng2024simulating}%
  \BibitemOpen
  \bibfield  {author} {\bibinfo {author} {\bibfnamefont {Z.}~\bibnamefont {Meng}}, \bibinfo {author} {\bibfnamefont {J.}~\bibnamefont {Zhong}}, \bibinfo {author} {\bibfnamefont {S.}~\bibnamefont {Xu}}, \bibinfo {author} {\bibfnamefont {K.}~\bibnamefont {Wang}}, \bibinfo {author} {\bibfnamefont {J.}~\bibnamefont {Chen}}, \bibinfo {author} {\bibfnamefont {F.}~\bibnamefont {Jin}}, \bibinfo {author} {\bibfnamefont {X.}~\bibnamefont {Zhu}}, \bibinfo {author} {\bibfnamefont {Y.}~\bibnamefont {Gao}}, \bibinfo {author} {\bibfnamefont {Y.}~\bibnamefont {Wu}}, \bibinfo {author} {\bibfnamefont {C.}~\bibnamefont {Zhang}}, \emph {et~al.},\ }\bibfield  {title} {\bibinfo {title} {Simulating unsteady flows on a superconducting quantum processor},\ }\href@noop {} {\bibfield  {journal} {\bibinfo  {journal} {Commun. Phys.}\ }\textbf {\bibinfo {volume} {7}},\ \bibinfo {pages} {349} (\bibinfo {year} {2024})}\BibitemShut {NoStop}%
\bibitem [{\citenamefont {Hu}\ \emph {et~al.}(2024)\citenamefont {Hu}, \citenamefont {Jin}, \citenamefont {Liu},\ and\ \citenamefont {Zhang}}]{hu2024quantum}%
  \BibitemOpen
  \bibfield  {author} {\bibinfo {author} {\bibfnamefont {J.}~\bibnamefont {Hu}}, \bibinfo {author} {\bibfnamefont {S.}~\bibnamefont {Jin}}, \bibinfo {author} {\bibfnamefont {N.}~\bibnamefont {Liu}},\ and\ \bibinfo {author} {\bibfnamefont {L.}~\bibnamefont {Zhang}},\ }\bibfield  {title} {\bibinfo {title} {Quantum circuits for partial differential equations via schr{"o}dingerisation},\ }\href {https://doi.org/10.22331/q-2024-12-12-1563} {\bibfield  {journal} {\bibinfo  {journal} {Quantum}\ }\textbf {\bibinfo {volume} {8}},\ \bibinfo {pages} {1563} (\bibinfo {year} {2024})}\BibitemShut {NoStop}%
\bibitem [{\citenamefont {Bharadwaj}(2024{\natexlab{a}})}]{bharadwaj2024qflows}%
  \BibitemOpen
  \bibfield  {author} {\bibinfo {author} {\bibfnamefont {S.~S.}\ \bibnamefont {Bharadwaj}},\ }\bibfield  {title} {\bibinfo {title} {{Q}{F}low{S}: Quantum simulator for fluid flows},\ }\href@noop {} {\bibfield  {journal} {\bibinfo  {journal} {Phys. Fluids}\ }\textbf {\bibinfo {volume} {36}},\ \bibinfo {pages} {107112} (\bibinfo {year} {2024}{\natexlab{a}})}\BibitemShut {NoStop}%
\bibitem [{\citenamefont {Liao}(1997)}]{liao1997boundary}%
  \BibitemOpen
  \bibfield  {author} {\bibinfo {author} {\bibfnamefont {S.-J.}\ \bibnamefont {Liao}},\ }\bibfield  {title} {\bibinfo {title} {Boundary element method for general nonlinear differential operators},\ }\href@noop {} {\bibfield  {journal} {\bibinfo  {journal} {Eng. Anal. Bound. Elem.}\ }\textbf {\bibinfo {volume} {20}},\ \bibinfo {pages} {91} (\bibinfo {year} {1997})}\BibitemShut {NoStop}%
\bibitem [{\citenamefont {Liao}(1999)}]{liao1999blasius}%
  \BibitemOpen
  \bibfield  {author} {\bibinfo {author} {\bibfnamefont {S.-J.}\ \bibnamefont {Liao}},\ }\bibfield  {title} {\bibinfo {title} {An explicit, totally analytic approximate solution for blasius’ viscous flow problems},\ }\href@noop {} {\bibfield  {journal} {\bibinfo  {journal} {Int. J. Non-Linear Mech.}\ }\textbf {\bibinfo {volume} {34}},\ \bibinfo {pages} {759} (\bibinfo {year} {1999})}\BibitemShut {NoStop}%
\bibitem [{\citenamefont {Liao}\ and\ \citenamefont {Campo}(2002)}]{liao2002blasius2}%
  \BibitemOpen
  \bibfield  {author} {\bibinfo {author} {\bibfnamefont {S.}~\bibnamefont {Liao}}\ and\ \bibinfo {author} {\bibfnamefont {A.}~\bibnamefont {Campo}},\ }\bibfield  {title} {\bibinfo {title} {Analytic solutions of the temperature distribution in blasius viscous flow problems},\ }\href@noop {} {\bibfield  {journal} {\bibinfo  {journal} {J. Fluid Mech.}\ }\textbf {\bibinfo {volume} {453}},\ \bibinfo {pages} {411} (\bibinfo {year} {2002})}\BibitemShut {NoStop}%
\bibitem [{\citenamefont {Liao}(2003{\natexlab{b}})}]{liao2003mhd}%
  \BibitemOpen
  \bibfield  {author} {\bibinfo {author} {\bibfnamefont {S.-J.}\ \bibnamefont {Liao}},\ }\bibfield  {title} {\bibinfo {title} {On the analytic solution of magnetohydrodynamic flows of non-newtonian fluids over a stretching sheet},\ }\href@noop {} {\bibfield  {journal} {\bibinfo  {journal} {J. Fluid Mech.}\ }\textbf {\bibinfo {volume} {488}},\ \bibinfo {pages} {189} (\bibinfo {year} {2003}{\natexlab{b}})}\BibitemShut {NoStop}%
\bibitem [{\citenamefont {Xu}\ \emph {et~al.}(2010)\citenamefont {Xu}, \citenamefont {Lin}, \citenamefont {Liao}, \citenamefont {Wu},\ and\ \citenamefont {Majdalani}}]{xu2010porous}%
  \BibitemOpen
  \bibfield  {author} {\bibinfo {author} {\bibfnamefont {H.}~\bibnamefont {Xu}}, \bibinfo {author} {\bibfnamefont {Z.-L.}\ \bibnamefont {Lin}}, \bibinfo {author} {\bibfnamefont {S.-J.}\ \bibnamefont {Liao}}, \bibinfo {author} {\bibfnamefont {J.-Z.}\ \bibnamefont {Wu}},\ and\ \bibinfo {author} {\bibfnamefont {J.}~\bibnamefont {Majdalani}},\ }\bibfield  {title} {\bibinfo {title} {Homotopy based solutions of the navier--stokes equations for a porous channel with orthogonally moving walls},\ }\href@noop {} {\bibfield  {journal} {\bibinfo  {journal} {Phys. Fluids}\ }\textbf {\bibinfo {volume} {22}} (\bibinfo {year} {2010})}\BibitemShut {NoStop}%
\bibitem [{\citenamefont {Bharadwaj}(2024{\natexlab{b}})}]{bharadwaj2024thesis}%
  \BibitemOpen
  \bibfield  {author} {\bibinfo {author} {\bibfnamefont {S.~S.}\ \bibnamefont {Bharadwaj}},\ }\bibfield  {title} {\bibinfo {title} {Quantum computation of fluid dynamics (phd thesis)},\ }\href@noop {} {\bibfield  {journal} {\bibinfo  {journal} {ProQuest}\ }\textbf {\bibinfo {volume} {ISBN: 9798384450764}} (\bibinfo {year} {2024}{\natexlab{b}})}\BibitemShut {NoStop}%
\bibitem [{\citenamefont {Xue}\ \emph {et~al.}(2024)\citenamefont {Xue}, \citenamefont {Xu}, \citenamefont {Zhuang}, \citenamefont {Sun}, \citenamefont {Wang}, \citenamefont {Tan}, \citenamefont {Ye}, \citenamefont {Liu}, \citenamefont {Wu}, \citenamefont {Chen} \emph {et~al.}}]{xue2024quantum}%
  \BibitemOpen
  \bibfield  {author} {\bibinfo {author} {\bibfnamefont {C.}~\bibnamefont {Xue}}, \bibinfo {author} {\bibfnamefont {X.-F.}\ \bibnamefont {Xu}}, \bibinfo {author} {\bibfnamefont {X.-N.}\ \bibnamefont {Zhuang}}, \bibinfo {author} {\bibfnamefont {T.-P.}\ \bibnamefont {Sun}}, \bibinfo {author} {\bibfnamefont {Y.-J.}\ \bibnamefont {Wang}}, \bibinfo {author} {\bibfnamefont {M.-Y.}\ \bibnamefont {Tan}}, \bibinfo {author} {\bibfnamefont {C.-C.}\ \bibnamefont {Ye}}, \bibinfo {author} {\bibfnamefont {H.-Y.}\ \bibnamefont {Liu}}, \bibinfo {author} {\bibfnamefont {Y.-C.}\ \bibnamefont {Wu}}, \bibinfo {author} {\bibfnamefont {Z.-Y.}\ \bibnamefont {Chen}}, \emph {et~al.},\ }\bibfield  {title} {\bibinfo {title} {Quantum homotopy analysis method with secondary linearization for nonlinear partial differential equations},\ }\href@noop {} {\bibfield  {journal} {\bibinfo  {journal} {arXiv:2411.06759}\ } (\bibinfo {year} {2024})}\BibitemShut {NoStop}%
\bibitem [{\citenamefont {Odibat}(2010)}]{odibat2010study}%
  \BibitemOpen
  \bibfield  {author} {\bibinfo {author} {\bibfnamefont {Z.~M.}\ \bibnamefont {Odibat}},\ }\bibfield  {title} {\bibinfo {title} {A study on the convergence of homotopy analysis method},\ }\href@noop {} {\bibfield  {journal} {\bibinfo  {journal} {Appl. Math. Comput.}\ }\textbf {\bibinfo {volume} {217}},\ \bibinfo {pages} {782} (\bibinfo {year} {2010})}\BibitemShut {NoStop}%
\bibitem [{\citenamefont {Schlimgen}\ \emph {et~al.}(2021)\citenamefont {Schlimgen}, \citenamefont {Head-Marsden}, \citenamefont {Sager}, \citenamefont {Narang},\ and\ \citenamefont {Mazziotti}}]{schlimgen2021quantum}%
  \BibitemOpen
  \bibfield  {author} {\bibinfo {author} {\bibfnamefont {A.~W.}\ \bibnamefont {Schlimgen}}, \bibinfo {author} {\bibfnamefont {K.}~\bibnamefont {Head-Marsden}}, \bibinfo {author} {\bibfnamefont {L.~M.}\ \bibnamefont {Sager}}, \bibinfo {author} {\bibfnamefont {P.}~\bibnamefont {Narang}},\ and\ \bibinfo {author} {\bibfnamefont {D.~A.}\ \bibnamefont {Mazziotti}},\ }\bibfield  {title} {\bibinfo {title} {Quantum simulation of open quantum systems using a unitary decomposition of operators},\ }\href@noop {} {\bibfield  {journal} {\bibinfo  {journal} {Phys. Rev. Lett.}\ }\textbf {\bibinfo {volume} {127}},\ \bibinfo {pages} {270503} (\bibinfo {year} {2021})}\BibitemShut {NoStop}%
\bibitem [{\citenamefont {Lewis}\ \emph {et~al.}(2024)\citenamefont {Lewis}, \citenamefont {Eidenbenz}, \citenamefont {Nadiga},\ and\ \citenamefont {Suba{\c{s}}{\i}}}]{lewis2024limitations}%
  \BibitemOpen
  \bibfield  {author} {\bibinfo {author} {\bibfnamefont {D.}~\bibnamefont {Lewis}}, \bibinfo {author} {\bibfnamefont {S.}~\bibnamefont {Eidenbenz}}, \bibinfo {author} {\bibfnamefont {B.}~\bibnamefont {Nadiga}},\ and\ \bibinfo {author} {\bibfnamefont {Y.}~\bibnamefont {Suba{\c{s}}{\i}}},\ }\bibfield  {title} {\bibinfo {title} {Limitations for quantum algorithms to solve turbulent and chaotic systems},\ }\href@noop {} {\bibfield  {journal} {\bibinfo  {journal} {Quantum}\ }\textbf {\bibinfo {volume} {8}},\ \bibinfo {pages} {1509} (\bibinfo {year} {2024})}\BibitemShut {NoStop}%
\bibitem [{\citenamefont {Liao}(2002)}]{liao2002direct}%
  \BibitemOpen
  \bibfield  {author} {\bibinfo {author} {\bibfnamefont {S.-J.}\ \bibnamefont {Liao}},\ }\bibfield  {title} {\bibinfo {title} {A direct boundary element approach for unsteady non-linear heat transfer problems},\ }\href@noop {} {\bibfield  {journal} {\bibinfo  {journal} {Eng. Anal. Bound. Elem.}\ }\textbf {\bibinfo {volume} {26}},\ \bibinfo {pages} {55} (\bibinfo {year} {2002})}\BibitemShut {NoStop}%
\bibitem [{\citenamefont {Varah}(1975)}]{varah1975lower}%
  \BibitemOpen
  \bibfield  {author} {\bibinfo {author} {\bibfnamefont {J.~M.}\ \bibnamefont {Varah}},\ }\bibfield  {title} {\bibinfo {title} {A lower bound for the smallest singular value of a matrix},\ }\href@noop {} {\bibfield  {journal} {\bibinfo  {journal} {Linear Algebra Appl.}\ }\textbf {\bibinfo {volume} {11}},\ \bibinfo {pages} {3} (\bibinfo {year} {1975})}\BibitemShut {NoStop}%
\bibitem [{\citenamefont {Richardson}\ and\ \citenamefont {Gaunt}(1927)}]{richardson1927viii}%
  \BibitemOpen
  \bibfield  {author} {\bibinfo {author} {\bibfnamefont {L.~F.}\ \bibnamefont {Richardson}}\ and\ \bibinfo {author} {\bibfnamefont {J.~A.}\ \bibnamefont {Gaunt}},\ }\bibfield  {title} {\bibinfo {title} {Viii. the deferred approach to the limit},\ }\href@noop {} {\bibfield  {journal} {\bibinfo  {journal} {Philos. Trans. R. Soc. A}\ }\textbf {\bibinfo {volume} {226}},\ \bibinfo {pages} {299} (\bibinfo {year} {1927})}\BibitemShut {NoStop}%
\bibitem [{\citenamefont {Fisher}(1937)}]{fisher1937wave}%
  \BibitemOpen
  \bibfield  {author} {\bibinfo {author} {\bibfnamefont {R.~A.}\ \bibnamefont {Fisher}},\ }\bibfield  {title} {\bibinfo {title} {The wave of advance of advantageous genes},\ }\href@noop {} {\bibfield  {journal} {\bibinfo  {journal} {Annals of Eugenics}\ }\textbf {\bibinfo {volume} {7}},\ \bibinfo {pages} {355} (\bibinfo {year} {1937})}\BibitemShut {NoStop}%
\bibitem [{\citenamefont {Kolmogorov}\ \emph {et~al.}(1937)\citenamefont {Kolmogorov}, \citenamefont {Petrovsky},\ and\ \citenamefont {Piskunov}}]{kolmogorov1937study}%
  \BibitemOpen
  \bibfield  {author} {\bibinfo {author} {\bibfnamefont {A.~N.}\ \bibnamefont {Kolmogorov}}, \bibinfo {author} {\bibfnamefont {I.~G.}\ \bibnamefont {Petrovsky}},\ and\ \bibinfo {author} {\bibfnamefont {N.~S.}\ \bibnamefont {Piskunov}},\ }\bibfield  {title} {\bibinfo {title} {Study of the diffusion equation with growth of the quantity of matter and its application to a biological problem},\ }\href@noop {} {\bibfield  {journal} {\bibinfo  {journal} {Bulletin of Moscow University, Mathematics and Mechanics}\ }\textbf {\bibinfo {volume} {1}},\ \bibinfo {pages} {1} (\bibinfo {year} {1937})}\BibitemShut {NoStop}%
\bibitem [{\citenamefont {Aronson}\ and\ \citenamefont {Weinberger}(1975)}]{aronson1975nonlinear}%
  \BibitemOpen
  \bibfield  {author} {\bibinfo {author} {\bibfnamefont {D.~G.}\ \bibnamefont {Aronson}}\ and\ \bibinfo {author} {\bibfnamefont {H.~F.}\ \bibnamefont {Weinberger}},\ }\bibfield  {title} {\bibinfo {title} {Nonlinear diffusion in population genetics, combustion, and nerve pulse propagation},\ }in\ \href {https://doi.org/10.1007/BFb0070595} {\emph {\bibinfo {booktitle} {Partial Differential Equations and Related Topics}}},\ \bibinfo {series} {Lecture Notes in Mathematics}, Vol.\ \bibinfo {volume} {446}\ (\bibinfo  {publisher} {Springer},\ \bibinfo {address} {Berlin, Heidelberg},\ \bibinfo {year} {1975})\ pp.\ \bibinfo {pages} {5--49}\BibitemShut {NoStop}%
\bibitem [{\citenamefont {Kuramoto}\ and\ \citenamefont {Tsuzuki}(1976)}]{kuramoto1976persistent}%
  \BibitemOpen
  \bibfield  {author} {\bibinfo {author} {\bibfnamefont {Y.}~\bibnamefont {Kuramoto}}\ and\ \bibinfo {author} {\bibfnamefont {T.}~\bibnamefont {Tsuzuki}},\ }\bibfield  {title} {\bibinfo {title} {Persistent propagation of concentration waves in dissipative media far from thermal equilibrium},\ }\href {https://doi.org/10.1143/PTP.55.356} {\bibfield  {journal} {\bibinfo  {journal} {Progress of Theoretical Physics}\ }\textbf {\bibinfo {volume} {55}},\ \bibinfo {pages} {356} (\bibinfo {year} {1976})}\BibitemShut {NoStop}%
\bibitem [{\citenamefont {Sivashinsky}(1977)}]{sivashinsky1977nonlinear}%
  \BibitemOpen
  \bibfield  {author} {\bibinfo {author} {\bibfnamefont {G.~I.}\ \bibnamefont {Sivashinsky}},\ }\bibfield  {title} {\bibinfo {title} {Nonlinear analysis of hydrodynamic instability in laminar flames--i. derivation of basic equations},\ }\href {https://doi.org/10.1016/0094-5765(77)90096-0} {\bibfield  {journal} {\bibinfo  {journal} {Acta Astronautica}\ }\textbf {\bibinfo {volume} {4}},\ \bibinfo {pages} {1177} (\bibinfo {year} {1977})}\BibitemShut {NoStop}%
\bibitem [{\citenamefont {Sivashinsky}\ and\ \citenamefont {Michelson}(1980)}]{sivashinsky1980irregular}%
  \BibitemOpen
  \bibfield  {author} {\bibinfo {author} {\bibfnamefont {G.~I.}\ \bibnamefont {Sivashinsky}}\ and\ \bibinfo {author} {\bibfnamefont {D.~M.}\ \bibnamefont {Michelson}},\ }\bibfield  {title} {\bibinfo {title} {On irregular wavy flow of a liquid film down a vertical plane},\ }\href {https://doi.org/10.1143/PTP.63.2112} {\bibfield  {journal} {\bibinfo  {journal} {Progress of Theoretical Physics}\ }\textbf {\bibinfo {volume} {63}},\ \bibinfo {pages} {2112} (\bibinfo {year} {1980})}\BibitemShut {NoStop}%
\bibitem [{\citenamefont {Allen}\ and\ \citenamefont {Cahn}(1979)}]{allen1979microscopic}%
  \BibitemOpen
  \bibfield  {author} {\bibinfo {author} {\bibfnamefont {S.~M.}\ \bibnamefont {Allen}}\ and\ \bibinfo {author} {\bibfnamefont {J.~W.}\ \bibnamefont {Cahn}},\ }\bibfield  {title} {\bibinfo {title} {A microscopic theory for antiphase boundary motion and its application to antiphase domain coarsening},\ }\href {https://doi.org/10.1016/0001-6160(79)90196-2} {\bibfield  {journal} {\bibinfo  {journal} {Acta Metallurgica}\ }\textbf {\bibinfo {volume} {27}},\ \bibinfo {pages} {1085} (\bibinfo {year} {1979})}\BibitemShut {NoStop}%
\bibitem [{\citenamefont {Bray}(1994)}]{bray1994theory}%
  \BibitemOpen
  \bibfield  {author} {\bibinfo {author} {\bibfnamefont {A.~J.}\ \bibnamefont {Bray}},\ }\bibfield  {title} {\bibinfo {title} {Theory of phase-ordering kinetics},\ }\href {https://doi.org/10.1080/00018739400101505} {\bibfield  {journal} {\bibinfo  {journal} {Advances in Physics}\ }\textbf {\bibinfo {volume} {43}},\ \bibinfo {pages} {357} (\bibinfo {year} {1994})}\BibitemShut {NoStop}%
\bibitem [{\citenamefont {Cross}\ and\ \citenamefont {Hohenberg}(1993)}]{cross1993pattern}%
  \BibitemOpen
  \bibfield  {author} {\bibinfo {author} {\bibfnamefont {M.~C.}\ \bibnamefont {Cross}}\ and\ \bibinfo {author} {\bibfnamefont {P.~C.}\ \bibnamefont {Hohenberg}},\ }\bibfield  {title} {\bibinfo {title} {Pattern formation outside of equilibrium},\ }\href {https://doi.org/10.1103/RevModPhys.65.851} {\bibfield  {journal} {\bibinfo  {journal} {Reviews of Modern Physics}\ }\textbf {\bibinfo {volume} {65}},\ \bibinfo {pages} {851} (\bibinfo {year} {1993})}\BibitemShut {NoStop}%
\end{thebibliography}%
\end{document}